\documentclass[preprint,12pt,numberedappendix,appendixfloats]{emulateapj}

\slugcomment{accepted to ApJ}  
\shorttitle{Inverted Metallicity Gradients in High-$z$ Dwarf Galaxies}
\shortauthors{Wang et al. (2019)}

\usepackage[english]{babel}
\usepackage[utf8]{inputenc} 
\usepackage[T1]{fontenc}
\usefont{T1}{tnr}{m}{sl}
\usepackage{fancyhdr}       
\usepackage{pgf,pgfarrows,pgfnodes,pgfautomata,pgfheaps}
\usepackage{graphicx}       
\usepackage{natbib}         
\usepackage{url}            
\usepackage{grffile}        
\usepackage{mathtools}      
\usepackage{bbold}          

\usepackage{bbm}            

\usepackage{multirow}       
\usepackage{xspace}         
\usepackage{epstopdf}
\usepackage{amsmath,amssymb,amsxtra,amsfonts}   
\usepackage{txfonts}       
\newcommand{\kms}{\ensuremath{\rm km~s^{-1}}\xspace}

 \usepackage{color}         
 \definecolor{gold}{rgb}{1,0.80,0}
 \definecolor{orange}{rgb}{1,0.5,0}
 \definecolor{midgray}{gray}{0.3}
 \definecolor{lblue}{rgb}{0,0.2,0.6}
 \definecolor{dgreen}{rgb}{0.1,0.6,0.3}

\usepackage[colorlinks=true,citecolor=lblue,linkcolor=dgreen]{hyperref}    
\newcommand{\be}{\begin{equation}}
\newcommand{\ee}{\end{equation}}
\newcommand{\non}{\nonumber}
\newcommand{\ba}{\begin{align}}
\newcommand{\ea}{\end{align}}

\newcommand{\defeq}{\vcentcolon=}

\newcommand{\Msun}{\ensuremath{M_\odot}\xspace}

\newcommand{\chisq}{\ensuremath{\chi^2}\xspace}

\newcommand{\Mstar}{\ensuremath{M_\ast}\xspace}
\newcommand{\Lstar}{\ensuremath{L_\ast}\xspace}
\newcommand{\Sstar}{\ensuremath{\Sigma_\ast}\xspace}
\newcommand{\oh}{\ensuremath{12+\log({\rm O/H})}\xspace}
\newcommand{\Av}{\ensuremath{A_{\rm V}}\xspace}
\newcommand{\Rv}{\ensuremath{R_{\rm V}}\xspace}

\newcommand{\SFR}{\ensuremath{{\rm SFR}}\xspace}
\newcommand{\Mgas}{\ensuremath{M_{\rm gas}}\xspace}
\newcommand{\Sgas}{\ensuremath{\Sigma_{\rm gas}}\xspace}
\newcommand{\fgas}{\ensuremath{f_{\rm gas}}\xspace}
\newcommand{\Zgas}{\ensuremath{Z_{\rm gas}}\xspace}
\newcommand{\tage}{\ensuremath{t_{\rm age}}\xspace}

\newcommand{\Dn}{\ensuremath{{\rm D}_n(4000)}\xspace}
\newcommand{\HdA}{\ensuremath{{\rm H}\delta_A}\xspace}
\newcommand{\scrit}{\ensuremath{\sigma_{\rm crit}}\xspace}

\newcommand{\pc}{\ensuremath{\rm pc}\xspace}
\newcommand{\kpc}{\ensuremath{\rm kpc}\xspace}
\newcommand{\Mpc}{\ensuremath{\rm Mpc}\xspace}

\newcommand{\Hunit}{\ensuremath{\rm km~s^{-1}~Mpc^{-1}}\xspace}
\newcommand{\Funit}{\ensuremath{\rm erg~s^{-1}~cm^{-2}}\xspace}

\newcommand{\SBunit}{\ensuremath{\rm erg~s^{-1}~cm^{-2}~arcsec^{-2}}\xspace}

\newcommand{\Msunyr}{\ensuremath{\Msun~\mathrm{yr}^{-1}}\xspace}
\newcommand{\yr}{\ensuremath{\rm yr}\xspace}
\newcommand{\Myr}{\ensuremath{\rm Myr}\xspace}
\newcommand{\Gyr}{\ensuremath{\rm Gyr}\xspace}
\def\micron{\ensuremath{\mu\textrm{m}}\xspace}  

\newcommand\ionp[2]{#1$\;${\scshape{#2}}}      
\newcommand{\Ha}{\textrm{H}\ensuremath{\alpha}\xspace}
\newcommand{\Hb}{\textrm{H}\ensuremath{\beta}\xspace}
\newcommand{\Hg}{\textrm{H}\ensuremath{\gamma}\xspace}
\newcommand{\Hd}{\textrm{H}\ensuremath{\delta}\xspace}
\newcommand{\HII}{\textrm{H}\textsc{ii}\xspace}

\newcommand{\OII}{[\textrm{O}~\textsc{ii}]\xspace}
\newcommand{\OIII}{[\textrm{O}~\textsc{iii}]\xspace}

\newcommand{\NII}{[\textrm{N}~\textsc{ii}]\xspace}

\newcommand{\NeIII}{[\textrm{Ne}~\textsc{iii}]\xspace}
\newcommand{\sersic}{S\'{e}rsic\xspace}
\def\B{\ensuremath{B_{435}}\xspace}

\def\I{\ensuremath{I_{814}}\xspace}

\def\H{\ensuremath{H_{160}}\xspace}

\newcommand{\clyi}{MACS1149.6+2223\xspace}

\newcommand{\clsan}{Abell 370\xspace}

\newcommand{\clba}{MACS0744.9+3927\xspace}

\newcommand{\sex}{\textsc{SExtractor}\xspace}
\newcommand{\emc}{\textsc{Emcee}\xspace}

\newcommand{\adriz}{\textsc{AstroDrizzle}\xspace}

\newcommand{\fast}{\textsc{FAST}\xspace}

\newcommand{\SJ}{\textsc{Sharon \& Johnson}\xspace}
\newcommand{\grzl}{\textsc{Grzili}\xspace}
\newcommand{\burst}{\textsc{Starburst99}\xspace}

\newcommand{\hst}{\textit{HST}\xspace}
\newcommand{\jwst}{\textit{JWST}\xspace}

\newcommand{\glass}{\textit{GLASS}\xspace}
\newcommand{\clash}{\textit{CLASH}\xspace}
\newcommand{\hff}{\textit{HFF}\xspace}

\def\clash{\textit{CLASH}\xspace}

\newcommand{\osiris}{\textit{OSIRIS}\xspace}
\newcommand{\keck}{\textit{Keck}\xspace}

\def\etal{et al.\xspace}
\def\ie{i.e.\xspace}
\def\eg{e.g.\xspace}
\def\etc{etc.\xspace}

\def\vsv{vis-\'a-vis\xspace}
\renewcommand\({\left(}
\renewcommand\){\right)}

\newcommand\mgs{metallicity gradients\xspace}

\newcommand\gpm{gas-phase metallicity\xspace}
\def\sf{star-forming\xspace}

\newcommand{\el}[1]{\ensuremath{\textrm{EL}_{#1}}}

\newcommand{\Om} {\ensuremath{\Omega_{\rm{m}}}\xspace}

\newcommand{\Ol} {\ensuremath{\Omega_{\Lambda}}\xspace}

\newcommand{\taueq}{\ensuremath{\tau_{\rm eq}}\xspace}

\usepackage{etoolbox}
\makeatletter
\patchcmd{\NAT@citex}
  {\@citea\NAT@hyper@{\NAT@nmfmt{\NAT@nm}\NAT@date}}
  {\@citea\NAT@nmfmt{\NAT@nm}\NAT@hyper@{\NAT@date}}
  {}
  {}
\patchcmd{\NAT@citex}
  {\@citea\NAT@hyper@{%
     \NAT@nmfmt{\NAT@nm}%
     \hyper@natlinkbreak{\NAT@aysep\NAT@spacechar}{\@citeb\@extra@b@citeb}%
     \NAT@date}}
  {\@citea\NAT@nmfmt{\NAT@nm}%
   \NAT@aysep\NAT@spacechar%
   \NAT@hyper@{\NAT@date}}
  {}
  {}
\patchcmd{\NAT@citex}
  {\@citea\NAT@hyper@{%
     \NAT@nmfmt{\NAT@nm}%
     \hyper@natlinkbreak{\NAT@spacechar\NAT@@open\if*#1*\else#1\NAT@spacechar\fi}%
       {\@citeb\@extra@b@citeb}%
     \NAT@date}}
  {\@citea\NAT@nmfmt{\NAT@nm}%
   \NAT@spacechar\NAT@@open\if*#1*\else#1\NAT@spacechar\fi%
   \NAT@hyper@{\NAT@date}}
  {}
  {}
\makeatother

\begin{document}


\title{Discovery of Strongly Inverted Metallicity Gradients in Dwarf Galaxies at $z$$\sim$2}

\author{
    Xin~Wang$^{1}$\thanks{E-mail: xwang@astro.ucla.edu},
    Tucker~A.~Jones$^{2}$,
    Tommaso~Treu$^{1}$,
    Jessie~Hirtenstein$^{2}$,
    Gabriel~B.~Brammer$^{3}$,
    Emanuele~Daddi$^{4}$,
    Xiao-Lei~Meng$^{5}$,
    Takahiro~Morishita$^{3}$,
    Louis~E.~Abramson$^{1}$,
    Alaina~L.~Henry$^{6}$,
    Ying-jie~Peng$^{7}$,
    Kasper~B.~Schmidt$^{8}$,
    Keren~Sharon$^{9}$,
    Michele~Trenti$^{10,11}$,
    Benedetta~Vulcani$^{12}$
}
\affil{$^{1}$  Department of Physics and Astronomy, University of California, Los Angeles, CA, USA 90095-1547}
\affil{$^{2}$  University of California Davis, 1 Shields Avenue, Davis, CA 95616, USA}
\affil{$^{3}$  Space Telescope Science Institute, 3700 San Martin Drive, Baltimore, MD, 21218, USA}
\affil{$^{4}$  Laboratoire AIM, CEA/DSM-CNRS-Universit\'e Paris Diderot, IRFU/Service d'Astrophysique, B\^at. 709, CEA Saclay, F-91191 Gif-sur-Yvette Cedex, France}
\affil{$^{5}$  Physics Department and Tsinghua Centre for Astrophysics, Tsinghua University, Beijing 100084, China}
\affil{$^{6}$  Space Telescope Science Institute, 3700 San Martin Drive, Baltimore, MD, 21218, USA}
\affil{$^{7}$  Kavli Institute for Astronomy and Astrophysics, Peking University, Beijing 100871, China}
\affil{$^{8}$  Leibniz-Institut f\"ur Astrophysik Potsdam (AIP), An der Sternwarte 16, D-14482 Potsdam, Germany}
\affil{$^{9}$  Department of Astronomy, University of Michigan, 1085 S. University Avenue, Ann Arbor, MI 48109, USA}
\affil{$^{10}$ School of Physics, University of Melbourne, VIC 3010, Australia}
\affil{$^{11}$ ARC Centre of Excellence for All-Sky Astrophysics in 3-Dimensions, Australia}
\affil{$^{12}$ INAF- Osservatorio Astronomico di Padova, Vicolo Osservatorio 5, I-35122 Padova, Italy 0000-0003-0980-1499}
\email{xwang@astro.ucla.edu}

\begin{abstract}
    We report the first sub-kiloparsec spatial resolution measurements of strongly inverted gas-phase 
    metallicity gradients in two dwarf galaxies at $z$$\sim$2.
    The galaxies have stellar masses $\sim$$10^9$\Msun, specific star-formation rate $\sim$20 \Gyr$^{-1}$, and 
    global metallicity $\oh\sim8.1$ (1/4 solar), assuming the \citet{2008A&A...488..463M}
    strong line calibrations of \OIII/\Hb and \OII/\Hb.
    Their metallicity radial gradients are measured to be highly inverted, \ie, 0.122$\pm$0.008 and 
    0.111$\pm$0.017 dex/kpc, which is hitherto unseen at such small masses in similar redshift ranges.
    From the Hubble Space Telescope observations of the source nebular emission and stellar continuum, we 
    present the 2-dimensional spatial maps of star-formation rate surface density, stellar population age, and 
    gas fraction, which show that our galaxies are currently undergoing rapid mass assembly via disk 
    inside-out growth.  More importantly, using a simple chemical evolution model, we find that the gas 
    fractions for different metallicity regions cannot be explained by pure gas accretion.  Our spatially 
    resolved analysis based on a more advanced gas regulator model results in a spatial map of net gaseous 
    outflows, triggered by active central starbursts, that potentially play a significant role in shaping the 
    spatial distribution of metallicity by effectively transporting stellar nucleosynthesis yields outwards.  
    The relation between wind mass loading factors and stellar surface densities measured in different regions 
    of our galaxies shows that a single type of wind mechanism, driven by either energy or momentum 
    conservation, cannot explain the entire galaxy.  These sources present a unique constraint on the effects 
    of gas flows on the early phase of disk growth from the perspective of spatially resolved chemical 
    evolution within individual systems.
\end{abstract}

\keywords{galaxies: abundances --- galaxies: evolution --- galaxies: formation --- galaxies: high-redshift --- gravitational lensing: strong}

\section{Introduction}\label{sect:intro}

Galaxy formation models require inflows and outflows of gas to regulate star formation 
\citep{2008MNRAS.385.2181F,Recchi:2008gw,Bouche:2010kh,2012MNRAS.421...98D,Dayal:2013im,Dekel:2013id,Lilly:2013ko,Dekel:2014jm,Peng:2014hn,Pipino:2014it}, 
yet this ``baryon cycle'' is not quantitatively understood. The interstellar medium (ISM) oxygen abundance 
(\ie metallicity\footnote{Throughout the paper, we refer to \gpm as metallicity for simplicity.}) and its 
spatial distribution
is fortunately a key observational probe of this process
\citep{Tremonti:2004ed,Erb:2006kn,2008A&A...488..463M,Bresolin:2009hh,2010MNRAS.408.2115M,Mannucci:2011be,Zahid:2011bb,Yates:2012kx,Zahid:2012cd,
Henry:2013cn,2013ApJ...765...48J,2014A&A...563A..49S,TheUniversalRelati:2014kx,Bresolin:2015fk,Ho:2015gq,Sanders:2015gk,Strom:2016vn}.
``Inside-out'' galaxy growth implies that initially steep radial gradients of metallicity flatten at later 
times (higher masses)
as disks grow larger, yet other scenarios suggest metallicities are initially well mixed by strong galactic feedback, and then
locked into negative gradients as winds lose the power to disrupt massive gas disks
\citep{Prantzos:2000gb,Hou:2000tq,Molla:2005eq,Kobayashi:2011cr,Few:2012jl,Pilkington:2012ib,Gibson:2013jw,2017MNRAS.466.4780M}.
What in common between these scenarios is that none of them predict the existence of a steep positive (\ie 
inverted) radial gradient such that metallicity increases with galacto-centric radius.

However, there is growing evidence of such phenomenon in both the local and distant Universe
\citep{Cresci:2010hr,Queyrel:2012hw,2014MNRAS.443.2695S,Metallicityevolutio:2014kg,2014A&A...563A..49S,PerezMontero:2016hs,2016ApJ...827...74W,Belfiore:2017bv,Carton:2018kv}.
The key reason for local galaxies possessing inverted gradients is gas re-distribution by tidal force in strongly interacting
systems \citep{Kewley:2006gb,Kewley:2010eg,Rupke:2010cg,AnIntegralFieldSt:2012hn,Torrey:2012kf}.
At high redshifts, inverted gradients are often attributed to the inflows of metal-poor gas from the filaments of cosmic web,
infalling directly onto galaxy centers, diluting central metallicities and hence creating positive gradients
\citep{Cresci:2010hr,Mott:2013bt}.
Given most of the high-$z$ observations are conducted from the ground with natural seeing, the targets are usually
super-\Lstar galaxies with stellar mass (\Mstar) $\gtrsim$$10^{10}$\Msun \citep[see \eg,][]{Metallicityevolutio:2014kg}.

These high-$z$ inverted gradients are in concert with the ``cold-mode'' gas accretion which has long been recognized to play a
crucial role in galaxies getting their baryonic mass supply
\citep{Birnboim:2003fo,Keres:2005gb,Dekel:2006cn,Dekel:2009fz,2009MNRAS.395..160K}.
Instead of being shock-heated to dark matter (DM) halo virial temperature ($\sim$10$^6$K for a $M_{\rm 
h}$$\sim$10$^{12}$\Msun halo) and then radiate away the thermal energy to condense and form stars (\vsv 
``hot-mode'' accretion),
gas streams can remain relatively cold (<10$^5$K) while being steadily accreted onto galaxy 
disks\footnote{Note however that
cold-mode accretion does not necessarily enforce that gas has to reach galaxy center first
given the large dynamic range of the scales of galaxy disks ($\sim$\kpc) and cosmic web ($\sim$\Mpc).}.
This cold accretion dominates the growth of galaxies forming in low-mass halos irrespective of redshifts since a hot permeating
halo of virialized gas can only manifest in halos above 2-3$\times10^{11}$\Msun, at $z\lesssim2$
\citep{Birnboim:2003fo,Keres:2005gb}.

A question thus arises: if cold-mode gas accretion dominates in low-mass systems (with \Mstar less than a few
$10^{10}$\Msun) and is thought to lead to inverted gradients under the condition that the incoming gas streams 
are centrally directed, can we observe this phenomenon in dwarf galaxies (with $\Mstar\lesssim10^9$) at high 
redshifts?
The answer is not straightforward since the effect of ejective feedback (\eg galactic winds driven by supernovae) is more
pronounced in lower mass galaxies, given their shallower gravitational potential wells and higher specific
star-formation rate (sSFR) \citep[see \eg][]{GalaxiesonFIREFe:2014dn,2014Natur.509..177V}.
On one hand, galactic winds can bring about kinematic turbulence that prevents a smooth accretion of 
filamentary gas streams directly onto galaxy center, resulting in rapid formation of in-situ clumps 
\citep{Dekel:2009bn}.
On the other hand, metal-enriched outflows triggered by these powerful winds can help remove stellar nucleosynthesis yields from
galaxy center \citep{Tremonti:2004ed,Erb:2006kn}.
Therefore the existence of strongly inverted gradients in dwarf galaxies at high redshifts, if any, presents a sensitive test of 
the relative strength of feedback-induced radial gas flows, in the early phase of the disk mass assembly process.
There have not been any attempts to investigate such existence, primarily due to the small sizes of these 
dwarf galaxies and sub-kiloparsec (sub-kpc) spatial resolution required to yield accurate gradient 
measurements \citep{2013ApJ...767..106Y}.
In this work, we present the first effort to secure two robustly measured inverted metallicity gradients in $z\sim2$ star-forming
dwarf galaxies from the Hubble Space Telescope (\hst) near-infrared (NIR) grism slitless spectroscopy, aided 
with galaxy cluster lensing magnification.
The details of data and sample galaxies are presented in Section~\ref{sect:data}. We describe our analysis methods alongside main 
results in Section~\ref{sect:rslt}, and conclude in Section~\ref{sect:conclu}.
Throughout this paper, a flat $\Lambda$CDM cosmology (\Om=0.3, \Ol=0.7, $H_0$=70\Hunit) is assumed.

\section{Data and galaxy sample}\label{sect:data}

The two galaxies with exceptional inverted gradients are selected from a comprehensive study of $\sim$300 galaxies with 
metallicity measurements at $1.2\lesssim z\lesssim2.3$ (Wang et al. 2017; Wang et al. in prep.). Before we 
discuss the two systems in detail in Section~\ref{subsect:galaxy_sample}, we give for convenience a brief 
summary of the spectroscopic data (Section~\ref{subsect:grism_data}), a concise description of the data 
reduction procedure (Section~\ref{subsect:grism_reduce}), and the ancillary imaging used in this work 
(Section~\ref{subsect:imaging_data}).

\subsection{\hst grism slitless spectroscopy}\label{subsect:grism_data}

We use the diffraction-limited spatially resolved slitless spectroscopy, obtained using the \hst wide-field 
camera 3 (WFC3) NIR grisms (G102 and G141), acquired by the Grism Lens-Amplified Survey from Space
\citep[\glass,][]{2014ApJ...782L..36S,2015ApJ...812..114T}.
\glass observes the distant Universe through 10 massive galaxy clusters as natural telescopes, exposing 10 orbits of G102
(0.8-1.15\micron, $R$$\sim$210) and 4 orbits of G141 (1.1-1.7\micron, $R$$\sim$130) per sightline.
This amounts to a sum of $\sim$22 kiloseconds of G102 and $\sim$9 kiloseconds of G141, as well as $\sim$7 
kiloseconds of
F140W+F105W direct imaging for astrometric alignment and wavelength/flux calibrations per field.
These exposures distributed over two separate pointings per cluster with nearly orthogonal orientations,
designed to help disentangle spectral contamination from neighboring objects.
So for each source, two sets of G102+G141 spectrum are obtained, covering an uninterrupted wavelength range of 0.8-1.7\micron
with almost unchanging sensitivity, reaching a 1-$\sigma$ surface brightness of $3\times10^{-16}~\SBunit$ across the entire
spectral range.
The \glass collaboration has made the catalogs of their redshift identifications in the 10 fields, based on visual inspections of
emission line (EL) features, publicly available at \url{https://archive.stsci.edu/prepds/glass/}.

\subsection{Grism data reduction}\label{subsect:grism_reduce}

To explore the chemical properties of galaxies at the peak epoch of cosmic chemical enrichment, we select from these
catalogs, a parent sample consisting of $\sim$300 galaxies with secure redshifts (\ie redshift quality $\geq$3 in 
the publicly available catalogs as described by \citet{2015ApJ...812..114T}) in the range of $1.2\lesssim 
z\lesssim2.3$.
This range is chosen for the detection of multiple nebular ELs\footnote{The names of the forbidden lines are
simplified as usual, if presented without wavelength numbers: $\OIII\lambda5008\defeq\OIII$,
$\OII\lambda\lambda3727,3730\defeq\OII$.} --- in particular the Balmer lines, \OIII, and \OII --- enabling the 
metallicity measurements, as in our earlier work \citep{2015AJ....149..107J,Wang:2016um}.
The \glass data for these $\sim$300 galaxies are reduced using the Grism Redshift and Line analysis
(\grzl\footnote{\url{https://github.com/gbrammer/grizli/}}; G. Brammer et al. in prep) software.
\grzl presents an end-to-end processing of the paired grism and direct exposures. The procedure includes five steps: 1)
pre-processing of the raw grism exposures, 2) full field-of-view (FoV) grism model construction, 3) 1D/2D spectrum extraction, and
4) solving for best-fit redshift from spectral template fitting (see Appendix~\ref{sect:grismspec} for more 
details), 5) refining full FoV grism model and extractions of source 1D/2D spectrum and EL stamps.
In step 1), the pre-processing consists of hot-pixel/persistence masking, cosmic ray flagging, flat fielding, astrometric
alignment, sky background subtraction, and extraction of visit-level source catalogs and segmentation maps.
In step 5), the EL stamps are drizzled onto a grid with a pixel scale of $0\farcs06$, Nyquist sampling the WFC3 point spread
function (PSF).
We apply an additional step on the \grzl output products to obtain pure 2D maps of \OIII$\lambda$5008 and \Hb, clean from the
partial contamination of \OIII$\lambda$4960, due to the limited grism spectral resolution and extended source morphology.
Our procedure properly combines EL maps at multiple orientations, preserving angular resolution and accounting for EL
blending.

\subsection{\hst imaging: estimating \Mstar from SED fitting}\label{subsect:imaging_data}

In addition to the deep NIR spectroscopy, there exists a wealth of ancillary imaging data with equally high spatial resolution on
the 10 \glass fields, which encompass all 6 Hubble Frontier Field \citep[\hff,][]{Lotz:2016ca} clusters and 4 from the Cluster
Lensing And Supernova Survey with Hubble \citep[\clash,][]{Postman:2012ca}.
These broad-band photometry, covering observed wavelengths of $\sim$0.4-1.7 \micron, can help constrain 
stellar population
properties (especially \Mstar) of our selected $\sim$300 galaxies at sufficient confidence.
We use the images sampled with $0\farcs06$ pixel size, and apply kernel convolutions to match the angular resolution of all images
to that of the F160W filter. We subtract contamination from intracluster light using established procedures 
\citep{Morishita:2016wu}.
Since our targets have rest-frame optical ELs with high equivalent widths (EWs), we subtract the {\it nebular} 
flux contribution from the broad-band photometry to obtain the {\it stellar} continuum flux.
As given in Appendix~\ref{sect:grismspec}, we model the nebular ELs as Gaussian profiles centered at the 
corresponding wavelengths.
The nebular contribution to the broad-band photometry is thereby estimated by convolving the filter throughput 
with the best-fit Gaussian profiles, and then subtracted off.
\footnote{
For the two sources selected in Section~\ref{subsect:galaxy_sample}, the reduction of broad-band fluxes after 
subtracting nebular ELs measured in grism data is as follows. For ID03751, the F140W and F105W fluxes are reduced 
by factors of 0.7 and 0.94, respectively. For ID01203, the F140W and F105W fluxes are reduced by 0.64 and 0.93, 
respectively.}
We then fit the {\it stellar} continuum spectral energy distribution (SED) with the \citet{Bruzual:2003ck} (BC03) 
stellar population synthesis models using the software \fast\citep{Kriek:2009eo}.
We assume a \citet{Chabrier:2003ki} initial mass function (IMF), constant star formation history, stellar dust attenuation in the
range $\Av^{\rm S}$=0-4 with a \citet{Calzetti:2000iy} extinction curve, and age
ranging from 5 Myr to the Hubble time at the redshifts of our targets. Stellar metallicity is fixed to 1/5 solar and we verify that this
assumption affects the results by no more than <0.05 dex on \Mstar.

\subsection{Two dwarf galaxies with strongly inverted metallicity gradients}\label{subsect:galaxy_sample}

Out of the parent sample of $\sim$300 galaxies, we are able to secure accurate (\ie at sub-kpc resolution) 
radial
metallicity gradients on 81 sources with suitable spatial extent and high signal-to-noise ratio (SNR) nebular 
emission.
These extended sources typically have half-light radii $R_{50}\gtrsim0\farcs15$.
In a range of $7\lesssim\log\left(\Mstar/\Msun\right)\lesssim10$ given by the analyses in Section~\ref{subsect:imaging_data}, our 
sample probes much lower \Mstar than other surveys of spatially resolved line emission at similar redshifts 
\citep{2016ApJ...827...74W,ForsterSchreiber:2018uq}, thanks to the enhanced resolution from lensing 
magnification, and high sensitivity of the \hst NIR grisms.
We have previously described the properties of 10 galaxies in our
sample from the cluster \clyi \citep{Wang:2016um}; results for the full sample are in preparation.

In most cases, we find that metallicity gradients are approximately flat (\ie consistent with zero given the typical
$\sigma=0.03$ dex~kpc$^{-1}$) or slightly negative. A minority (10/81) of our sample shows positive (\ie ``inverted'') gradients,
which are of interest as they pose a challenge to standard galactic chemical evolution models 
\citep[\eg,][]{Molla:2005eq}.
We have selected the two best examples with strongly inverted gradients for further study in this paper.
The two sources are ID03751 ($z=1.96,~\Mstar=1.12\times10^9\Msun$) in the prime field of \clsan, and ID01203
($z=1.65,~\Mstar=2.55\times10^9\Msun$) in the prime field of \clba.

\begin{figure*}
    \centering
    \includegraphics[width=.16\textwidth]{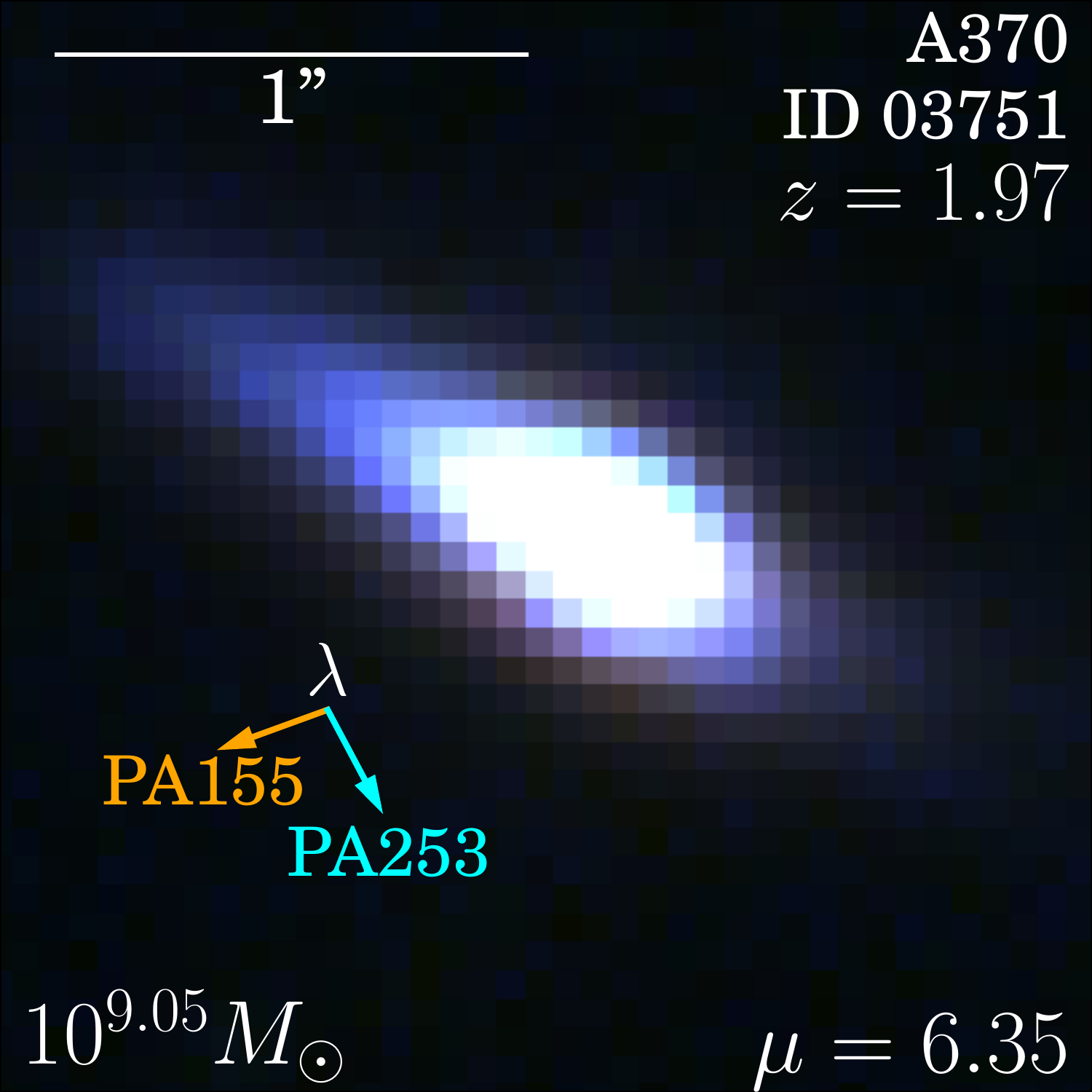}
    \includegraphics[width=.16\textwidth]{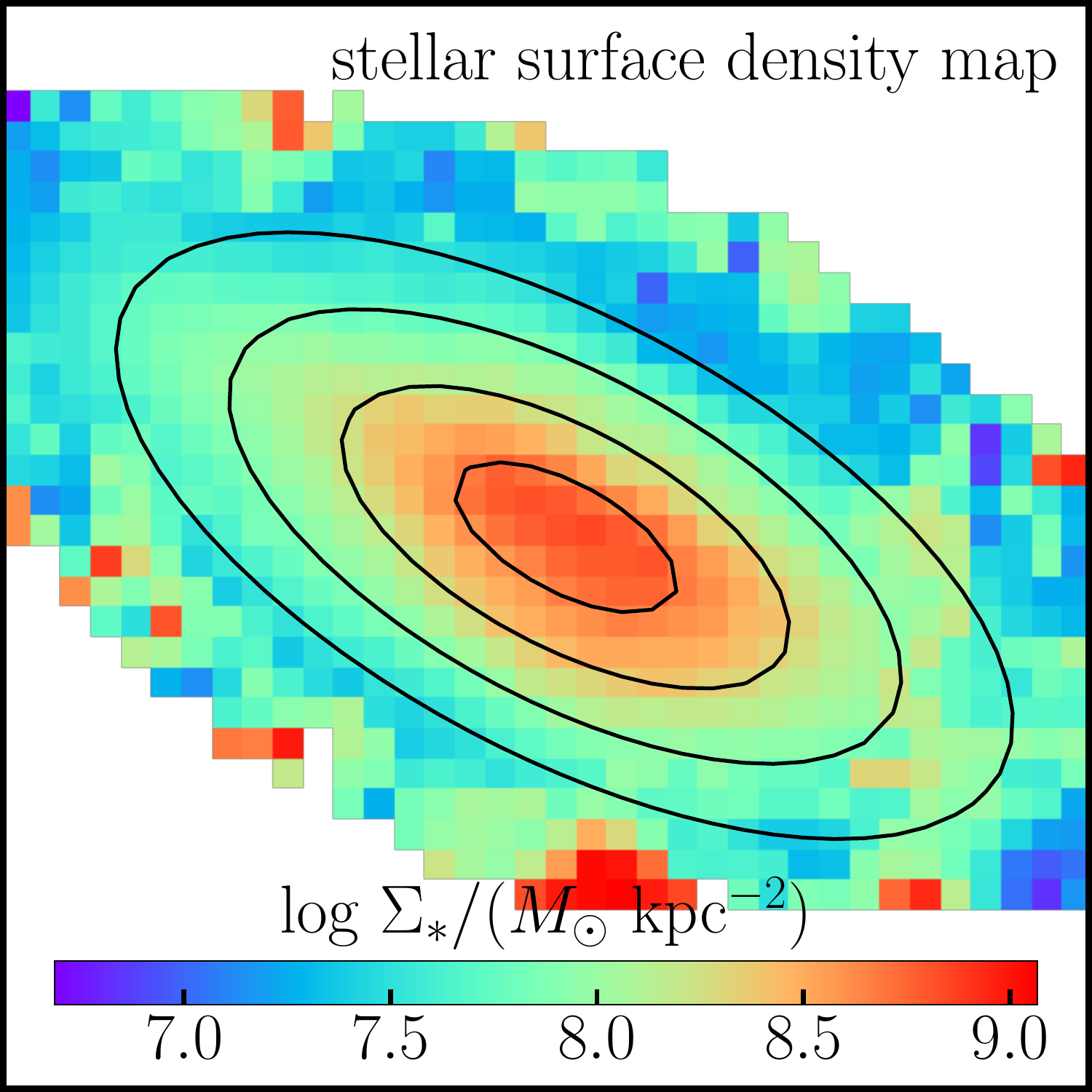}
    \includegraphics[width=.16\textwidth]{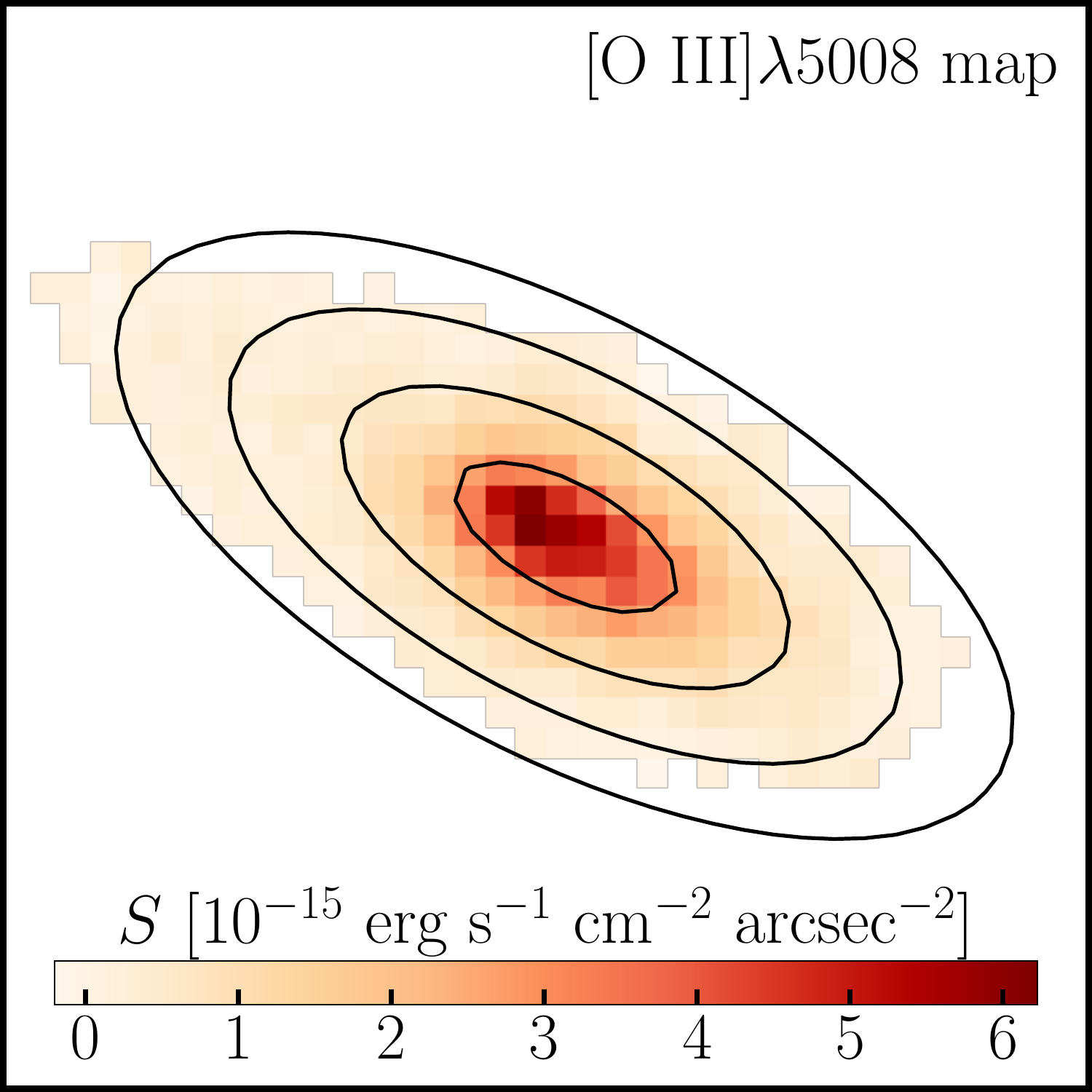}
    \includegraphics[width=.16\textwidth]{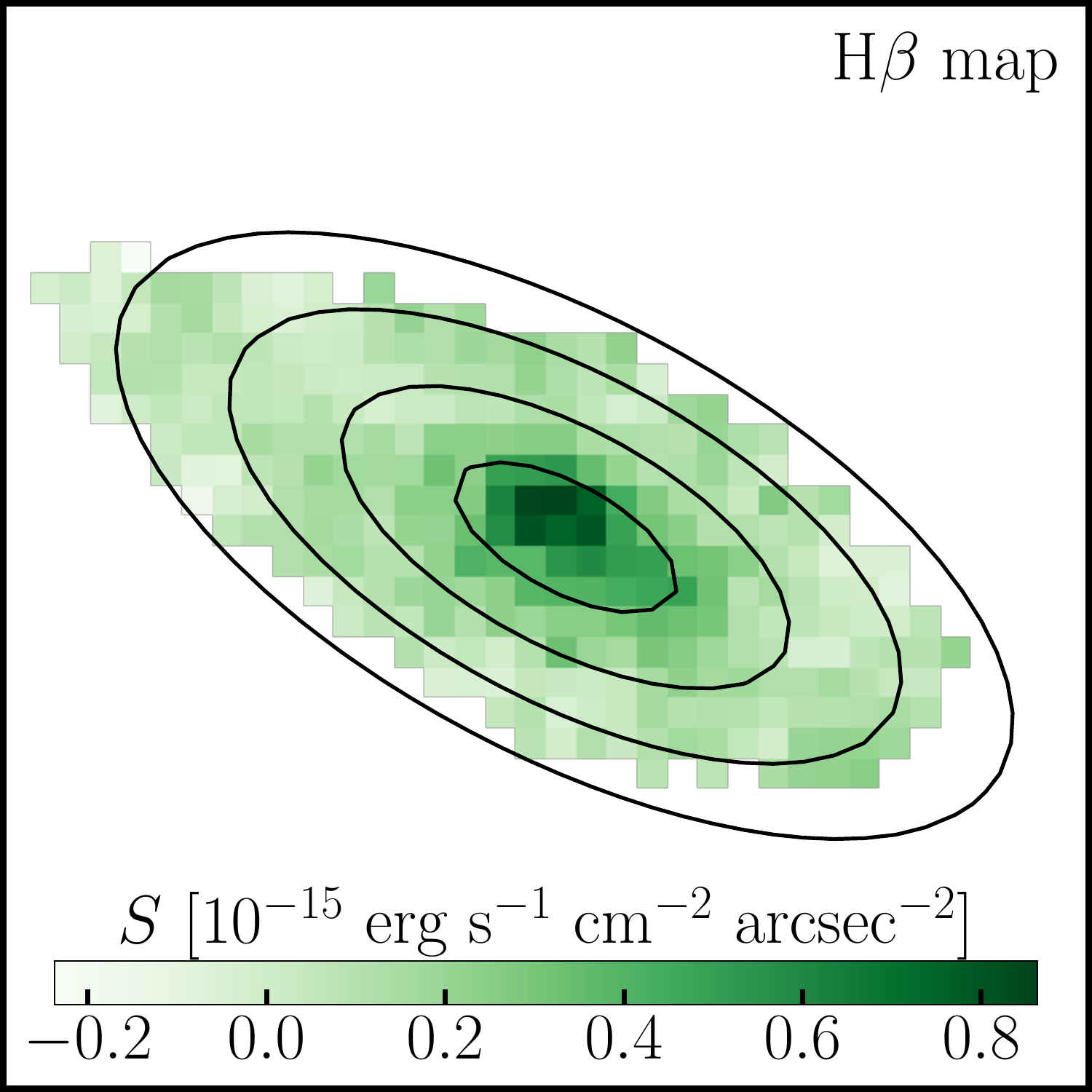}
    \includegraphics[width=.16\textwidth]{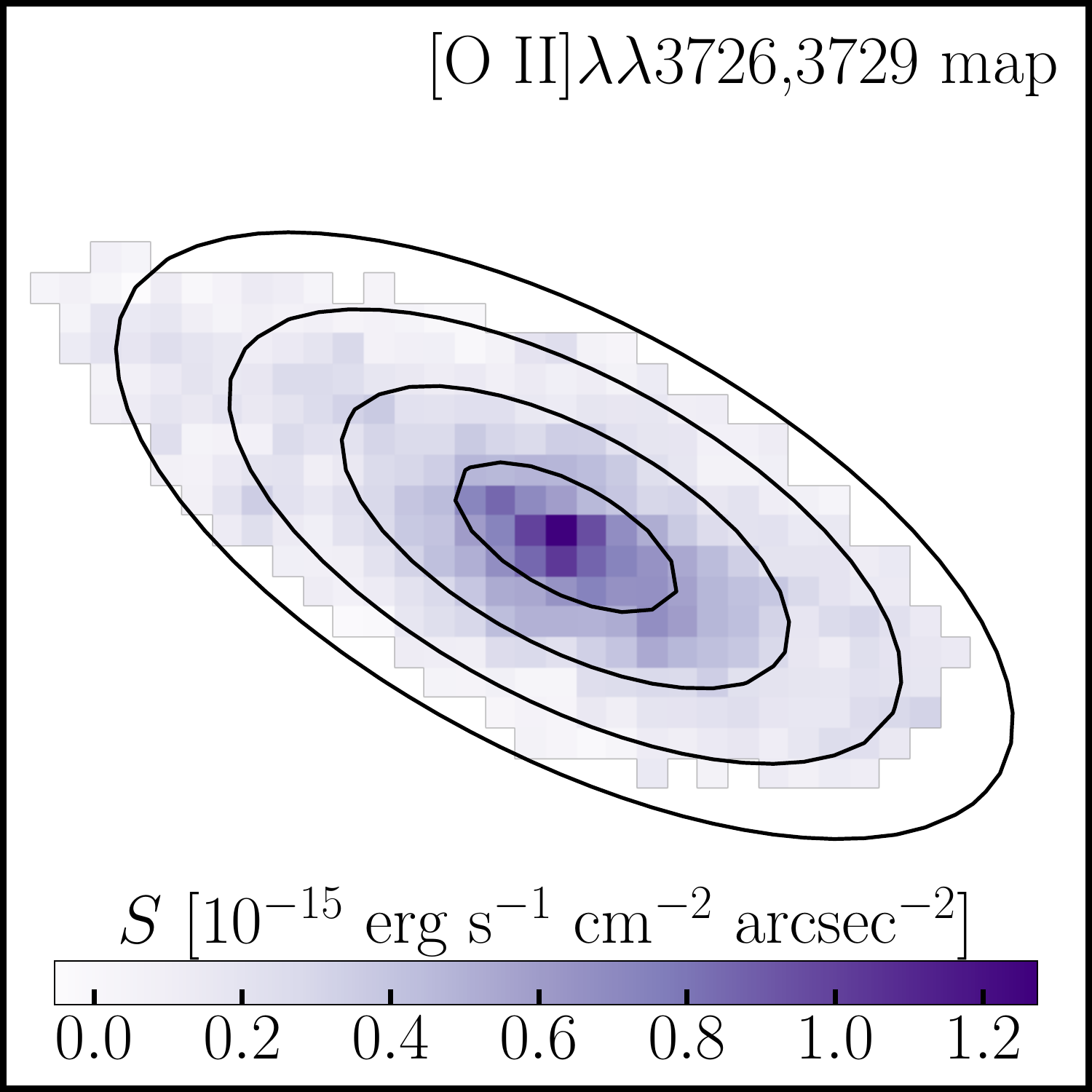}
    \includegraphics[width=.16\textwidth]{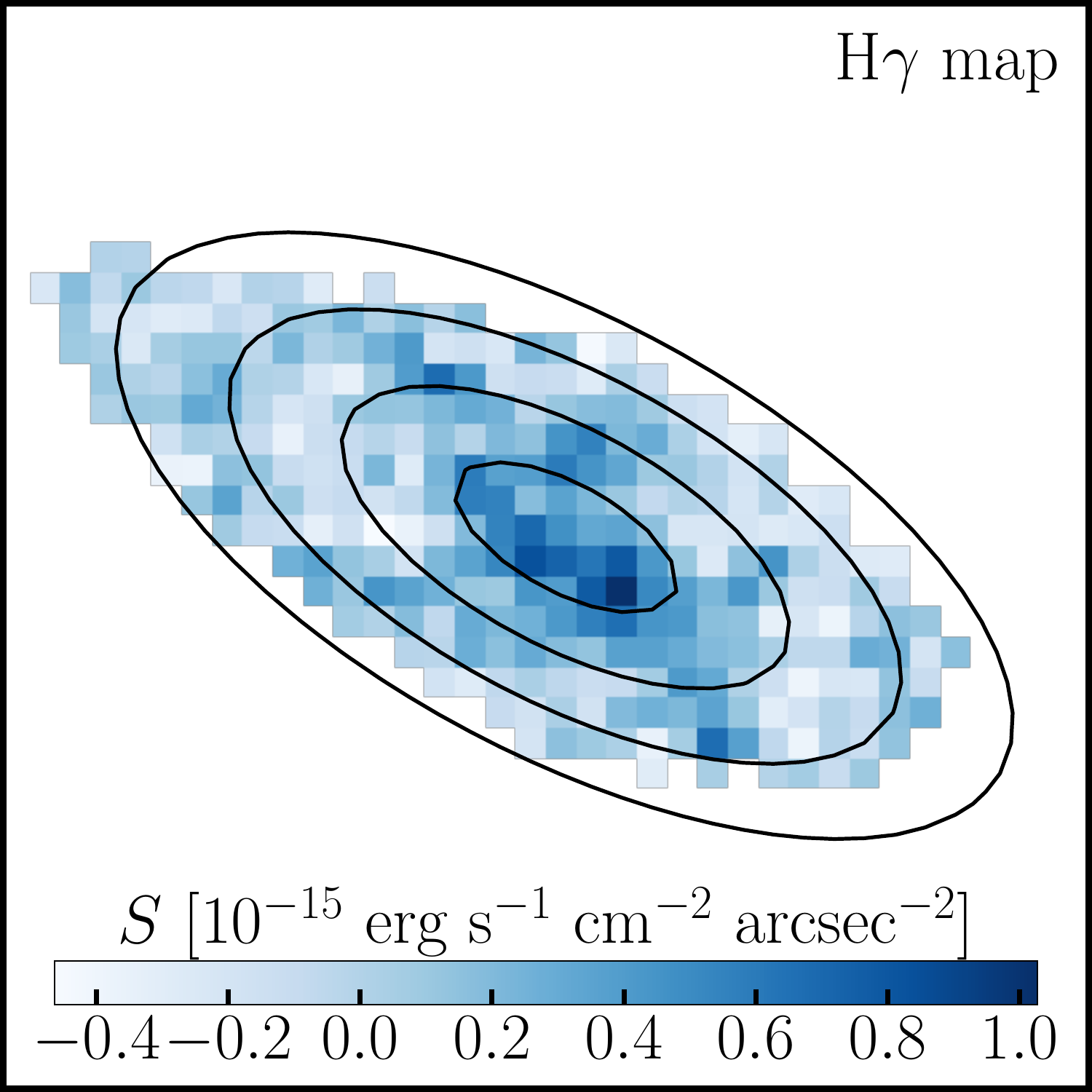}\\
    \includegraphics[width=.16\textwidth]{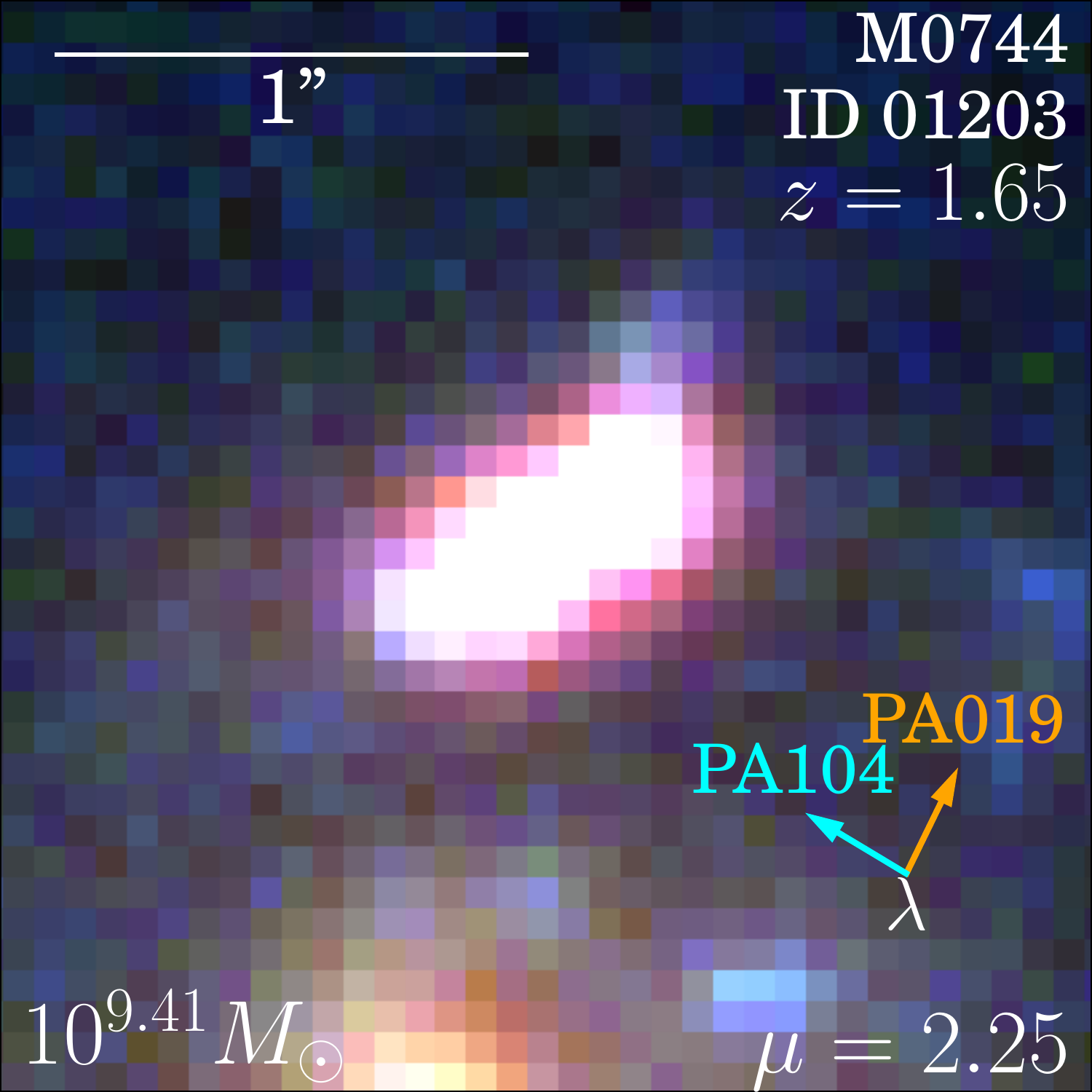}
    \includegraphics[width=.16\textwidth]{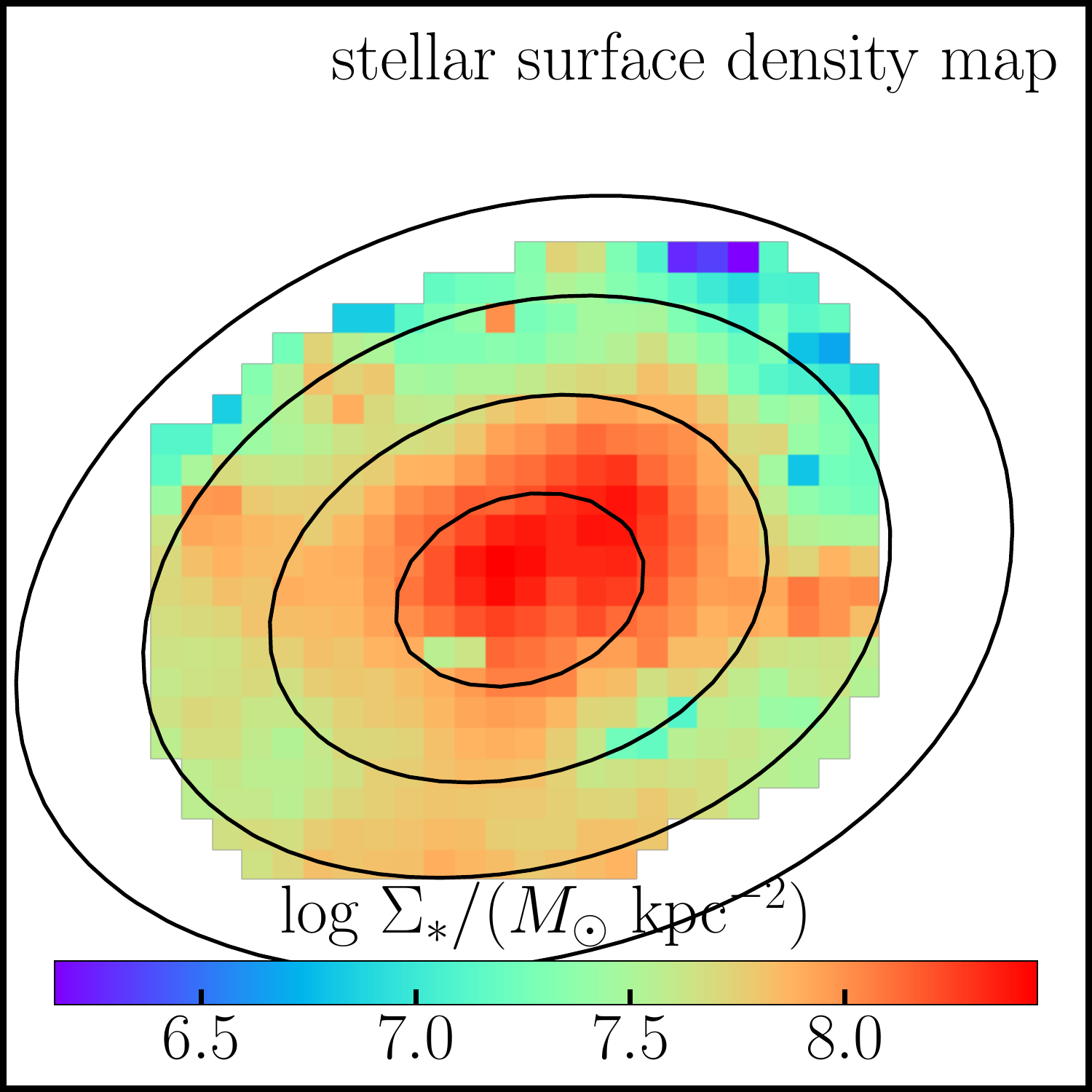}
    \includegraphics[width=.16\textwidth]{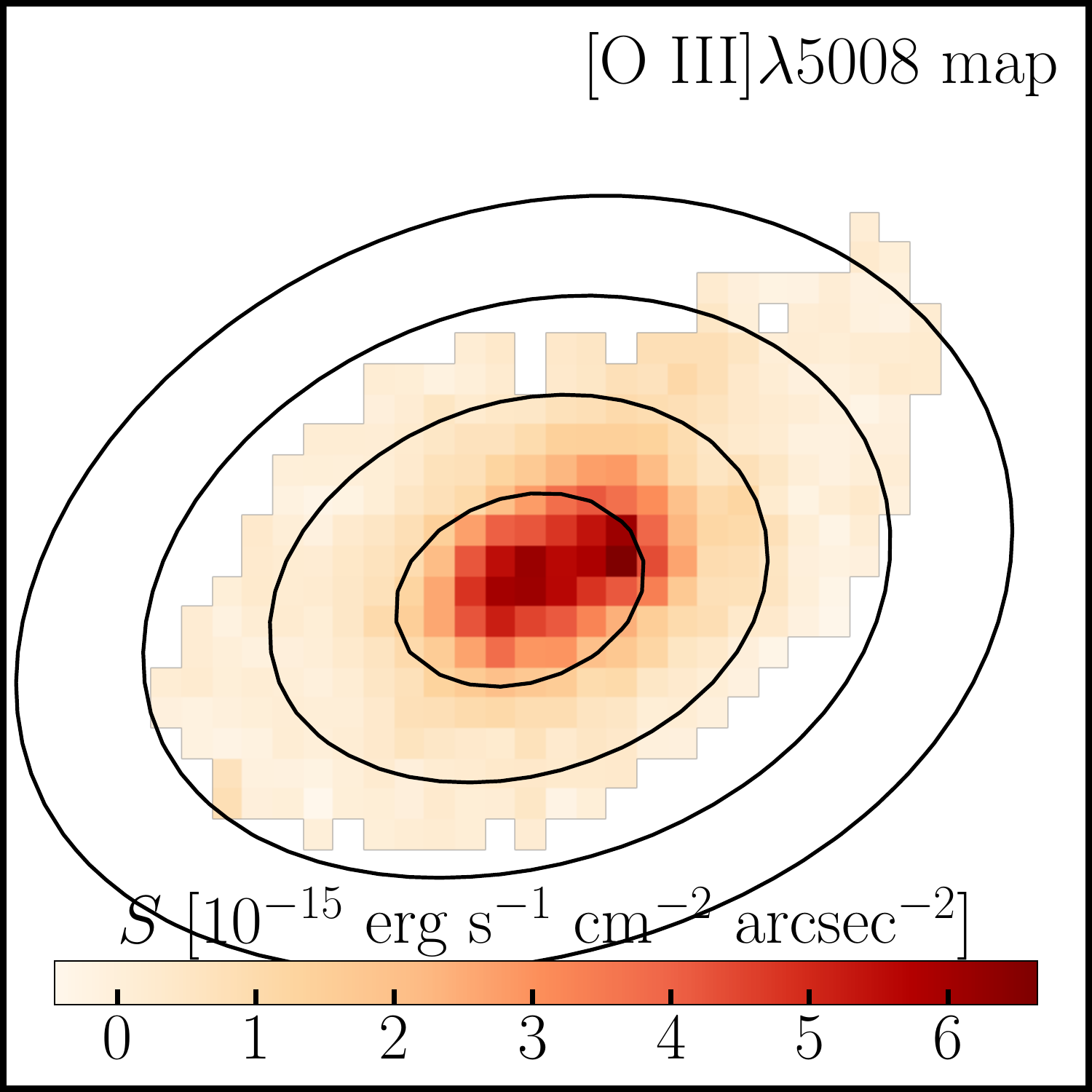}
    \includegraphics[width=.16\textwidth]{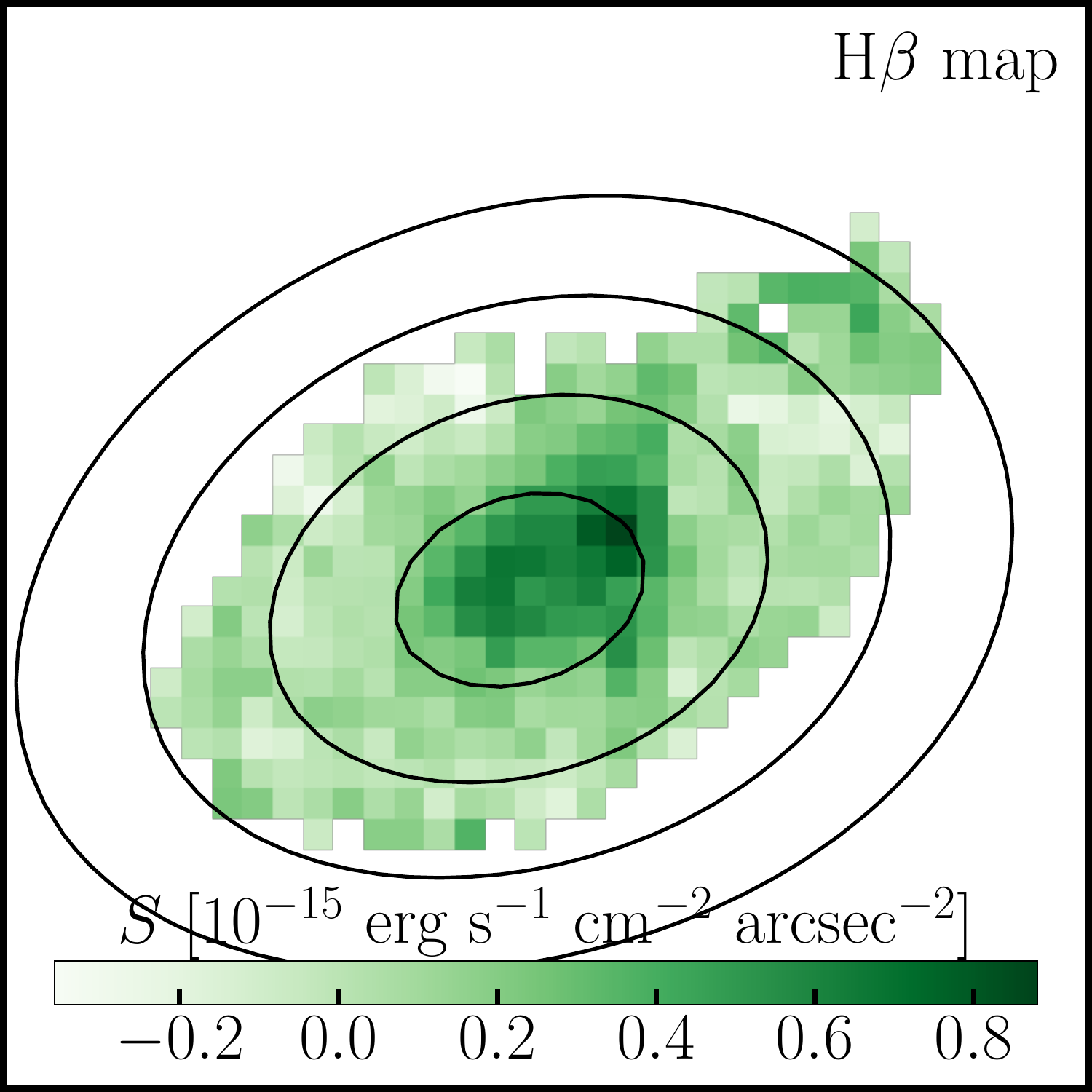}
    \includegraphics[width=.16\textwidth]{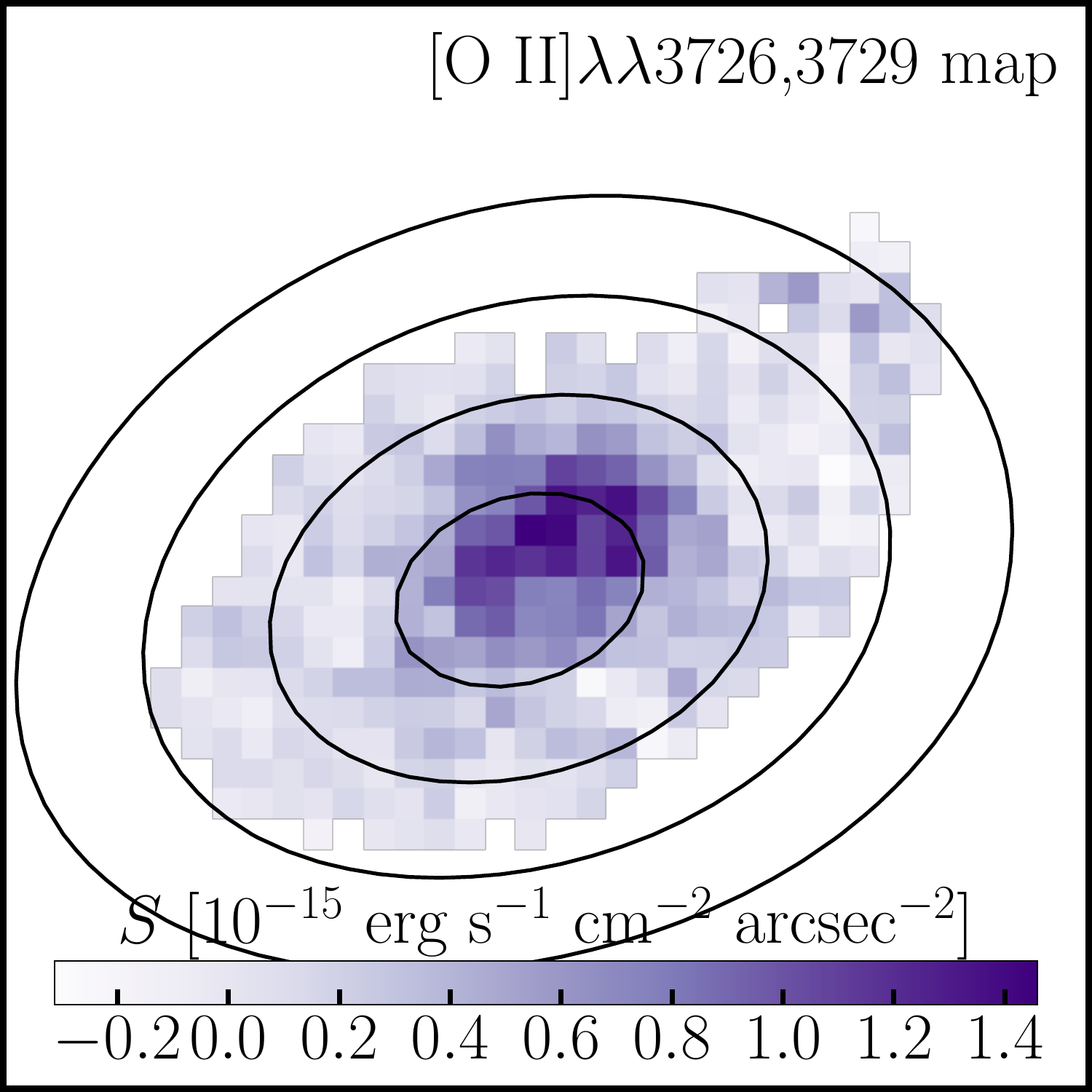}
    \includegraphics[width=.16\textwidth]{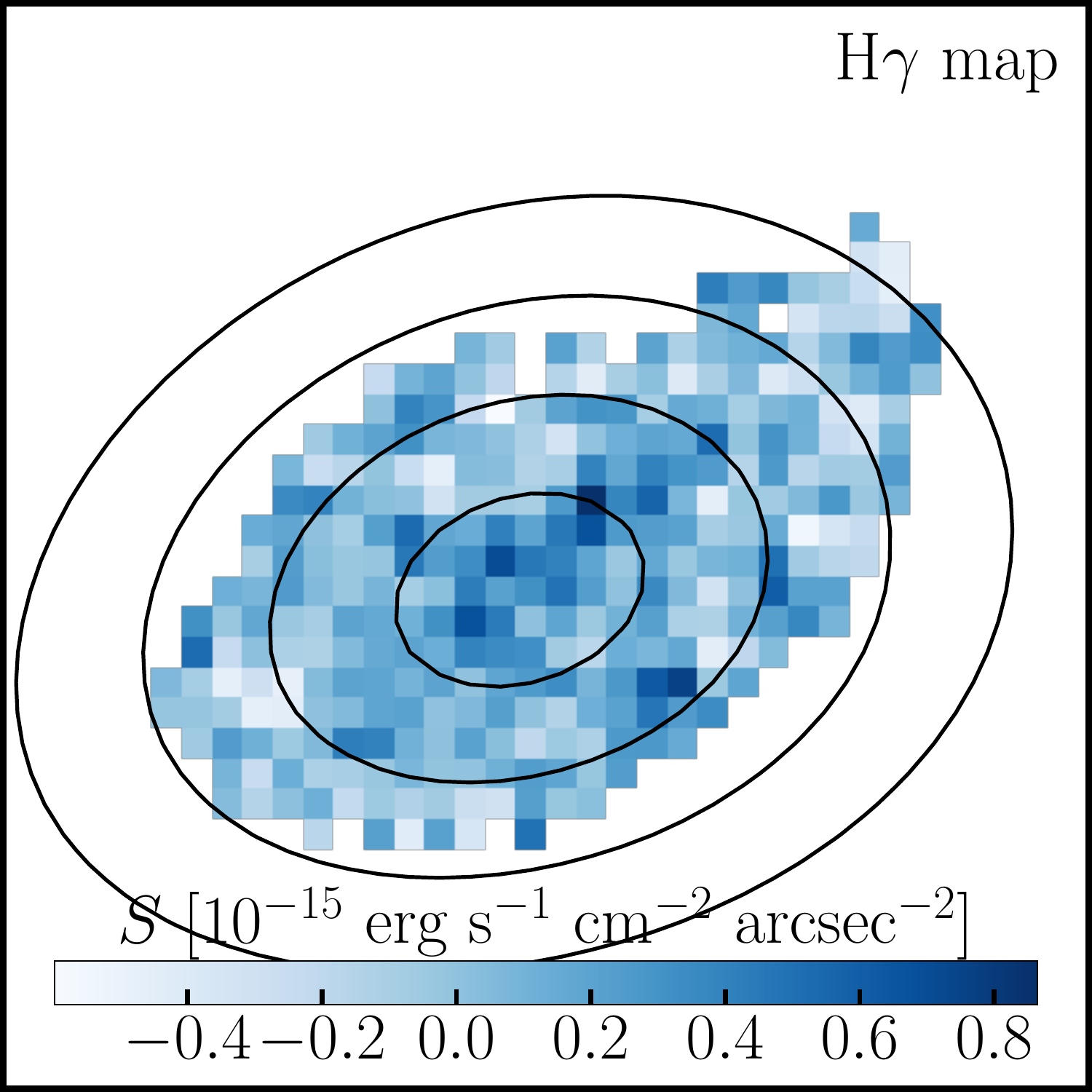}
    \caption{Two star-forming dwarf galaxies at $z$$\sim$2 displaying unusually strong inverted metallicity gradients, securely
    determined at sub-kpc spatial resolution. For each source we show, from left to right: color composite image (created from
    \hst broad-band photometry), stellar surface density map (obtained from SED fitting to \hst photometry), and surface
    brightness map of ELs \OIII, \Hb, \OII, and \Hg. The black contours overlaid represent the source plane de-projected
    galacto-centric radii with 1 kpc interval. The two light-dispersion directions for the grism exposures are 
    denoted by the orange and cyan arrows (see Figures~\ref{fig:3751spec} and \ref{fig:1203spec} for the 
    corresponding spectra).
    The spatial extent and orientation are unchanged for the two sources in all 2D
    stamps throughout. North is up and east is to the left.
    \label{fig:combELmap}}
\end{figure*}

Table~\ref{tab:srcprop} presents their properties.
Figure~\ref{fig:combELmap} shows the color-composite \hst images of these two galaxies and their 2D spatially
resolved maps of the nebular ELs.
Remarkably, they have \Mstar considerably lower --- by one order of magnitude --- than those of previous positive gradients 
measured at similar redshifts \citep[see \eg,][]{Cresci:2010hr,Queyrel:2012hw,2014MNRAS.443.2695S,Metallicityevolutio:2014kg}.
To complement the low dispersion grism spectra, we have obtained adaptive optics (AO) assisted kinematic data on our sources using 
ground-based integral-field unit (IFU) spectrograph when available.
The observation of source ID01203 is presented in Appendix~\ref{sect:kinem}.
The full data analysis is presented in \citet{Hirtenstein:2018tn} in detail.

\begin{deluxetable}{lcccccccccccc}
    \tablecolumns{13}
    \tablewidth{0pt}
    \tablecaption{Measured quantities of the two dwarf galaxies\label{tab:srcprop}}
\tablehead{
    \colhead{ID}  &     \colhead{03751}     &       \colhead{01203}
}
\startdata
    cluster         &       \clsan      &       \clba      \\
    R.A. (deg.)     &       39.977361   &       116.197585  \\
    Decl. (deg.)    &       -1.591636   &       39.456698   \\
    $z_{\rm spec}$  &       1.96        &       1.65       \\
    $\mu$~\tablenotemark{a}     &      6.35$_{-0.58}^{+0.40}$      &       2.25$_{-0.03}^{+0.04}$  \\
    \hline\noalign{\smallskip}
    \multicolumn{3}{c}{Observed emission line fluxes}   \\
    $f_{\OIII}$~[$10^{-17}$\Funit]     &  111.41$\pm$0.84   &   117.66$\pm$1.17 \\
    $f_{\Hb}$~[$10^{-17}$\Funit]       &   17.68$\pm$0.68    &   17.46$\pm$1.06    \\
    $f_{\OII}$~[$10^{-17}$\Funit]      &   29.57$\pm$0.51    &   34.00$\pm$0.96    \\
    $f_{\Hg}$~[$10^{-17}$\Funit]       &    7.21$\pm$0.67     &   7.06$\pm$1.00   \\
    \hline\noalign{\smallskip}
    \multicolumn{3}{c}{Restframe equivalent widths}   \\
    EW$_{\OIII}$~[\AA]     &  466.22$\pm$3.52    &   797.14$\pm$7.95   \\
    EW$_{\Hb}$~[\AA]       &   73.98$\pm$2.83    &   118.29$\pm$7.18   \\
    EW$_{\OII}$~[\AA]      &   79.14$\pm$1.37    &   123.91$\pm$3.50   \\
    EW$_{\Hg}$~[\AA]       &   30.18$\pm$2.82    &    25.73$\pm$3.68    \\
    \hline\noalign{\smallskip}
    \multicolumn{3}{c}{Estimated physical parameters}   \\
    \Mstar[10$^9$\Msun]\tablenotemark{b} &      1.12$_{-0.14}^{+0.14}$      &       2.55$_{-0.04}^{+0.04}$  \\
    \oh\tablenotemark{c}   &       8.08$_{-0.12}^{+0.11}$  &       8.10$_{-0.11}^{+0.11}$          \\
    $\Delta\log({\rm O/H})/\Delta r$~[dex/kpc]    &    0.122$\pm$0.008    &       0.111$\pm$0.017     \\
    \SFR[\Msunyr]\tablenotemark{b}       &   25.39$\pm$2.19      &       48.86$\pm$3.04     \\
    \Av             &       0.84$\pm$0.13           &       0.90$\pm$0.16                   \\
    \tage[$10^7$ yrs]      &   7.93$\pm$0.88     &       3.98$\pm$0.51     \\
    \Mgas[10$^{9}$\Msun]\tablenotemark{b}    &   4.07$\pm$1.27   &       23.85$\pm$7.33       \\
    \fgas\tablenotemark{d}           &       0.56$\pm$0.24   &       0.86$\pm$0.35       \\
    $B/T$\tablenotemark{e}  &       0.36$\pm$0.14   &       0.14$\pm$0.07       \\
    $R_{\rm eff}$ [kpc]\tablenotemark{b}  &   1.53$\pm$0.12    &   1.66$\pm$0.17     \\
    \hline\noalign{\smallskip}
    \multicolumn{3}{c}{Gas kinematics}   \\
    $\sigma$~[km/s]     &   \nodata\tablenotemark{f}     &   73$\pm$3    \\
    $V/\sigma$      &   \nodata\tablenotemark{f}     &   1.3$\pm$0.1 \\
    \hline\noalign{\smallskip}
    \multicolumn{3}{c}{Measurements of the gaseous outflows at the central 1\kpc}   \\
    $\lambda$  &  49.9$\pm$14.7   &   52.1$\pm$20.2     \\
    $\Psi$~[\Msunyr]   &  311.3$\pm$96.2   &   1700.9$\pm$681.9
\enddata
    \tablenotetext{a}{The magnification estimates are obtained from the \SJ version 4 model of \clsan\citep{Johnson:2014cf} and the
    Zitrin PIEMD+eNFW version 2 model of \clba\citep{2015ApJ...801...44Z}, for the two sources respectively.}
    \tablenotetext{b}{Values presented here are corrected for lensing magnification.}
    \tablenotetext{c}{Values represent global metallicity, inferred from integrated line fluxes.}
    \tablenotetext{d}{Here the gas fraction is calculated according to Eq.~\ref{eq:fgas}, using
    surface densities of the stellar and gas components, the latter of which is given by inverting the extended
    Schmidt law (Eq.~\ref{eq:KSlaw}) and the former from spatially resolved SED fitting.
    We caution that stellar mass estimates from resolved photometry can be systematically higher (by factors of
    up to 5) than spatially unresolved photometry, as elaborated in \citet{Sorba:2018hd}.
    In our cases, the total stellar masses derived from adding up stellar surface densities given by
    resolved photometry amount to (3.20$\pm$1.92)$\times10^9$\Msun and (3.88$\pm$2.76)$\times10^9$\Msun, for ID03751 and ID01203, respectively.
    }
    \tablenotetext{e}{In the bulge-disk decomposition, we fix the \sersic index $n=4$ (\ie de Vaucouleurs) for the bulge
    component, and $n=1$ (\ie exponential) for the disk component.}
    \tablenotetext{f}{Ground-based \keck OSIRIS follow-up observations targetting \Ha or \OIII gas kinematics for this source are 
    not feasible due to significantly low atmospheric transmission at the corresponding wavelengths.}
\end{deluxetable}


\section{Methods and results}\label{sect:rslt}

In this section, we describe our key methods used to derive radial metallicity gradients (Section~\ref{sect:metalgrad}), 2D maps 
of \SFR, average stellar population age, and gas fraction (Section~\ref{sect:physprop}), as well as spatial 
distributions of net gaseous outflow rate and mass loading factor (Section~\ref{sect:regulator}).
The main results are presented alongside the corresponding methods.

\subsection{Radial metallicity gradients}\label{sect:metalgrad}

\begin{figure*}
    \centering
    \includegraphics[width=\textwidth]{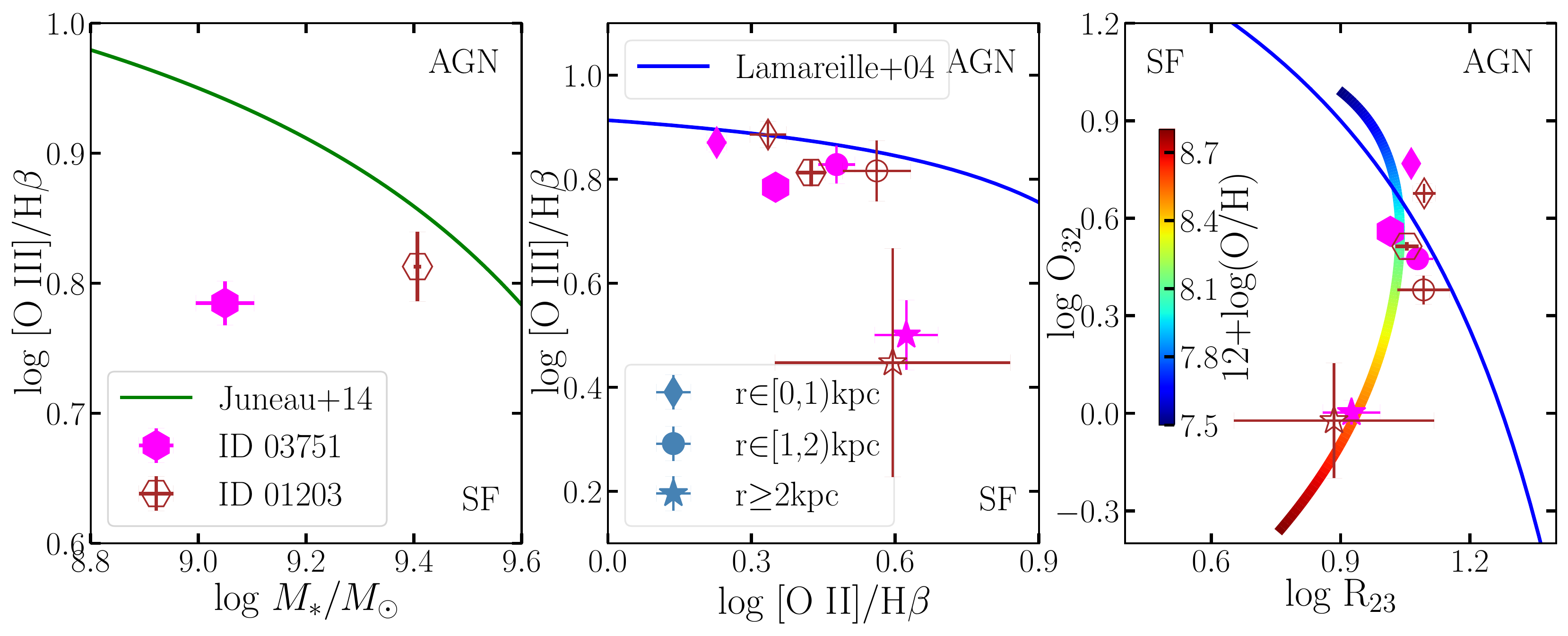}
    \caption{The diagnostic diagrams for our sources.
    On the left, we show the mass-excitation diagram with the demarcation scheme between the loci of AGNs and 
    star-forming (SF) galaxies proposed in \citet{Juneau:2014ca}. Our galaxies can be safely classified as the 
    latter.
    In the middle and right panels, we show the ``blue'' diagrams with the boundaries described by \citet{Lamareille:2004jk}.
    Following the conventions in the left panel, we use filled symbols to represent measurements for ID03751 
    whilst empty ones for ID01203.
    Furthermore, the hexagons correspond to the measurements integrated over the entire galaxies (as in the left 
    panel), whereas the other symbols denote results obtained in different radial annuli, as explained in the 
    legend of the middle panel.
    Again we see that the contamination from AGN ionization is minimum for our sources, even in the central 
    regions. In the right panel,
    we also show the gas-phase metallicities given by the \citet{2008A&A...488..463M} calibrations involving 
    R$_{23}$ and O$_{32}$.
    The two galaxies presented in this work share similar evolutionary trends in their excitation and ionization 
    states with respect to galacto-centric radius, strongly indicating that their metallicity increases with 
    galacto-centric radius.
    All the line flux ratios are corrected for dust extinction following the \citet{1989ApJ...345..245C} law with 
    nebular attenuation estimated from the observed Balmer decrement of $f_{\Hg}/f_{\Hb}$ under the assumption of 
    Case B recombination conditions.
    \label{fig:bluediagram}}
\end{figure*}

Since we infer metallicity from strong line flux ratio diagnostics, calibrated by either empirical methods, or 
theoretical methods, or a hybrid of both, it is essential to make sure that the line emission is not 
contaminated by active galactic nucleus (AGN) ionization or shock excitation.
As shown in Figure~\ref{fig:bluediagram}, we verify that our targets have a low probability (<\%10) of being 
classified as AGNs according to the mass-excitation diagram \citep{Juneau:2014ca}.
Their individual radial annuli also have excitation and ionization states compatible with \HII regions, as 
revealed in their loci in the ``blue'' diagnostic diagrams of $f_{\OIII}/f_{\Hb}$ versus $f_{\OII}/f_{\Hb}$ (the 
middle panel of Figure~\ref{fig:bluediagram}) and O$_{32}$=$(f_{\OIII\lambda5008}+f_{\OIII\lambda4960})/f_{\OII}$ 
versus R$_{23}$=$(f_{\OIII\lambda5008} + f_{\OIII\lambda4960} + f_{\OII})/f_{\Hb}$ (the right panel of 
Figure~\ref{fig:bluediagram}) \citep{Lamareille:2004jk,Rodrigues:2012dr,2015ApJ...813..126J}.
Notably, the intrinsic ratios of R$_{23}$ and O$_{32}$ decrease with galacto-centric radius for both of our 
sources, indicative of positive radial gradients of metallicities, opposite the normal trend,
which is confirmed with direct metallicity measurements in nearby disk galaxies
\citep[see \eg,][]{Bresolin:2009hh,2015ApJ...806...16B,Croxall:2015fl,Croxall:2016ew}.
In the source ID01203 covered by our follow-up \osiris observations (see Appendix~\ref{sect:kinem}), its 
integrated $f_{\NII}/f_{\Ha}$ ($\lesssim0.1$ at 3-$\sigma$) also shows no sign of AGN or shocked gas emission.

Our measurements of radial \mgs largely follow the procedures described in our previous work \citep{Wang:2016um}.
We use a Bayesian approach to jointly infer metallicity (\oh), nebular dust extinction ($\Av^{\rm N}$), and de-reddened \Hb flux
($f_{\Hb}$).  We explore the parameter space using the Markov Chain Monte Carlo sampler \emc\citep{ForemanMackey:2013io}.
The likelihood function is given by $\mathrm{L}\propto\exp(-\chisq/2)$ with
\begin{align}\label{eq:calibr}
    \chisq = \sum_i \frac{\(f_{\el{i}} - R_i \cdot f_{\Hb}\)^2}
        {\(\sigma_{\el{i}}\)^2 + \(f_{\Hb}\)^2\cdot\(\sigma_{R_i}\)^2},
\end{align}
where \el{i} corresponds to each available EL: \OIII, \Hb, \OII, and \Hg.
As shown in Figures~\ref{fig:3751spec} and \ref{fig:1203spec} in Appendix~\ref{sect:grismspec}, \Hd is only 
weakly detected for both of our galaxies, with SNR$\lesssim$5 from their spatially integrated grism spectra 
combined from two orientations, and thereby not included in our Bayesian analysis.
On the other hand, our source spectra shows marked detection of the EL $\NeIII\lambda3869$.
However, due to the relatively low wavelength dispersion of the \hst NIR grism channels (\ie $R_{\rm 
G141}\sim130$ and $R_{\rm G102}\sim210$),
$\NeIII\lambda3869$ is heavily blended with \ionp{He}{i}$\lambda3889$+H8 on the red side and H9 on the blue side.
As a result, we exclude \NeIII in our subsequent analyses as well.
$f_{\el{i}}$ and $\sigma_{\el{i}}$ are the flux and uncertainty of \el{i}. $R_i$ is the flux ratio between \el{i} 
and \Hb, with $\sigma_{R_i}$ being the intrinsic scatter at fixed physical properties.
In the case $\el{i}=\Hg$, $R_i$ is given by the Balmer decrement $f_{\Hg}/f_{\Hb}=0.47$.
For $\el{i} \in \{\OII,~\OIII\}$, $R_i$ and $\sigma_{R_i}$ are given by the strong line metallicity diagnostics 
($f_{\OIII}/f_{\Hb}$ and $f_{\OII}/f_{\Hb}$) calibrated by \citet{2008A&A...488..463M}.
The \citet{2008A&A...488..463M} calibrations combine the direct electron temperature measurements from the Sloan 
Digital Sky Survey (SDSS) in the low-metallicity (\oh$\lesssim$~8.35) branch \citep{2006A&A...459...85N} and the 
photoionization model predictions in the high-metallicity (\oh$\gtrsim$~8.35) branch \citep{Kewley:2002ep}, 
providing a continuous and coherent recipe over a wide metallicity range.

Our metallicity inference presented here primarily originates from the flux ratios between the oxygen 
collisionally excited lines (CELs) (\ie \OIII, \OII) and \Hb \citep[see the recent review by][for pros and cons 
of all strong line ratio diagnostics in the literature]{Maiolino:2018vq}.
Despite their bi-modal relationship with metallicity, R$_{23}$ and $f_{\OIII}/f_{\Hb}$ are found to produce the 
best accuracy among all the strong line ratios used in $z$$\sim$2 studies \citep{Patricio:2018up}.
Notably, multiple independent analyses have shown that there is a clean metallicity sequence in the diagnostic 
diagram spanned by $f_{\OIII}/f_{\Hb}$ and $f_{\OII}/f_{\Hb}$ \citep[][also see 
Figure~\ref{fig:bluediagram}]{Andrews:2013dn,2015ApJ...813..126J,Gebhardt:2015ui,Curti:2017fn}.
Since \OIII and \OII are among the brightest and thus most accessible strong ELs in high-$z$ galaxies, the flux 
ratios involving oxygen CELs will remain the most promising venue for
conversion into O/H, since they trace directly the emissivities and abundances of the oxygen ions.

In Appendix~\ref{sect:compare_calib}, we perform a comparative study of measuring metallicity gradients with
the pure empirical calibrations by \citet{2015ApJ...813..126J} and \citet{Curti:2017fn},
based upon the electron temperature metallicity measurements from the DEEP2 \citep[at 
$z\sim0.8$,][]{Newman:2013ha} and the SDSS \citep[at $0.027<z<0.25$,][]{Abazajian:2009ef} datasets.
We find that the systematic differences between the absolute gradient slopes derived assuming different strong 
line calibrations can be as high as 0.06 dex/kpc.
However, a positive radial gradient slope can always be recovered in each of our sources at high statistical 
significance, regardless of the calibrations used (see Appendix~\ref{sect:compare_calib} for more details).
We thus verify that there is no significant bias from the strong line calibrations adopted in our gradient 
measurements.
The same process is applied to both galaxy-integrated fluxes and to fluxes contained in individual spatial pixels 
(spaxels).

To obtain the correct intrinsic de-projected distance scale for each spaxel, we conducted full source plane morphological
reconstruction of our sources.
We ray-trace the image of each galaxy to its source plane using up-to-date lens models for each cluster: the macroscopic
model of \SJ version 4 for \clsan \citep{Johnson:2014cf}, and the \textsc{Zitrin} version 2 model for \clba
\citep{2015ApJ...801...44Z}. Other lens models are available for these clusters 
\cite[\eg][]{Diego:2016ww,Strait:2018ul} and we verified that the morphology of each source
is robust to the choice of model.

For each source, we fit the SED of individual spaxels using the procedures described in Section~\ref{sect:data}, 
obtaining the 2D stellar surface density (\Sstar) map shown in Figure~\ref{fig:combELmap}. Then we reconstruct 
\Sstar map in the source plane by de-lensing the surface densities according to the deflection field given by the 
macroscopic lens models.
To minimize the stochasticity in stellar population synthesis \citep{Fouesneau:2010ea,Eldridge:2012ds}, we make 
sure that the source plane resolution elements during this reconstruction contain enough stellar masses ($\gtrsim 
10^5$ \Msun) to be representative of complete stellar populations.
The axis ratios, inclinations, and major axis orientations are determined from an elliptical Gaussian fit. This 
procedure provides the intrinsic lensing-corrected morphology, and in particular, the galacto-centric radius at 
each point of the observed images.
The radial scale as black contours in all figures is used to establish the absolute metallicity gradient slope (i.e., in units of 
dex per proper kpc).
From the source reconstructed morphology, we measure their effective radius where the enclosed mass reaches half 
the total mass of the source. The measurements are represented by $R_{\rm eff}$ in Table~\ref{tab:srcprop}.

\begin{figure*}
    \centering
    \includegraphics[width=\textwidth]{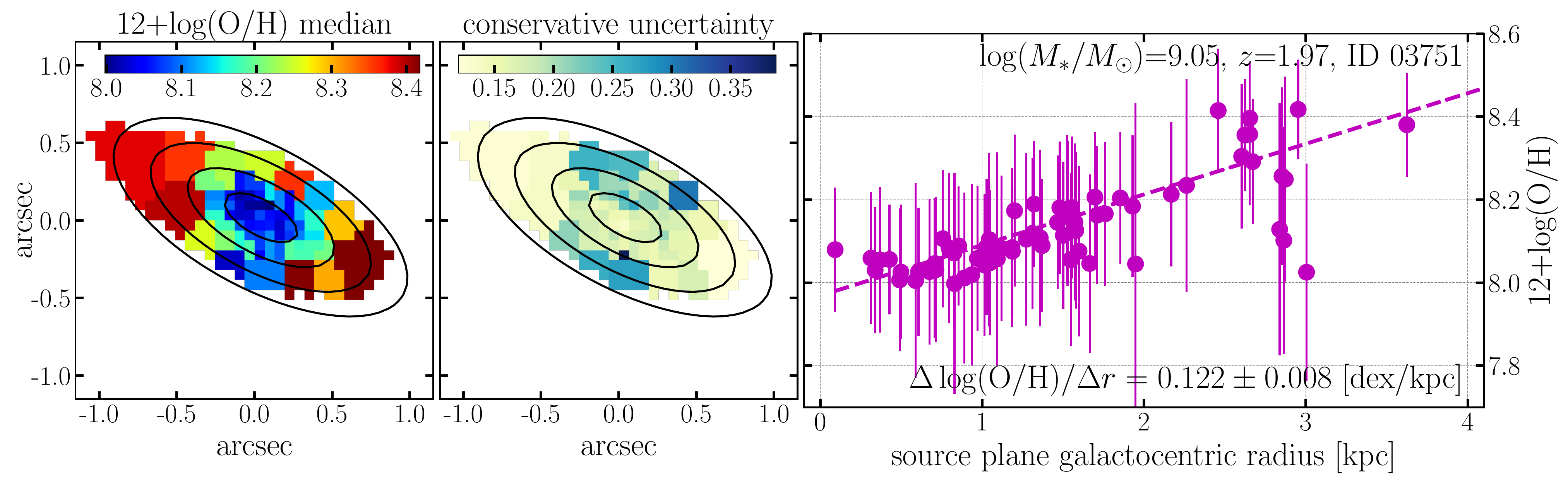}
    \includegraphics[width=\textwidth]{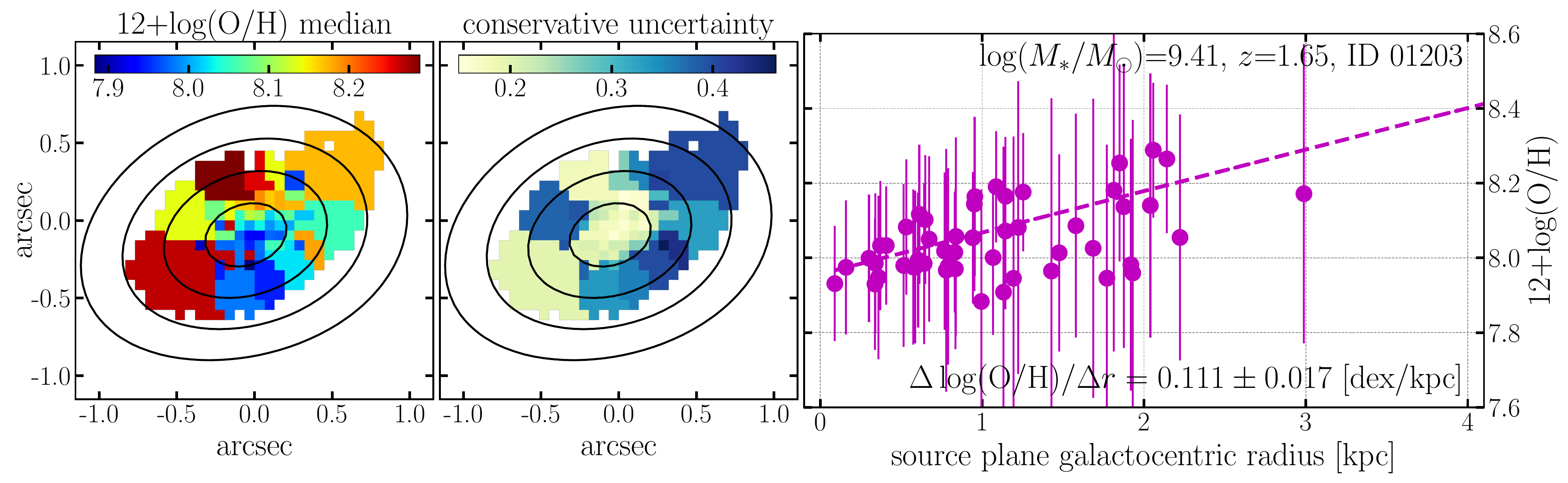}
    \caption{Metallicity maps and radial gradient measurements of the two galaxies. The left panels show the 2D maps of the median
    value estimates of metallicity, and the central panels show their conservative uncertainties (\ie the larger side of the
    asymmetric 1-$\sigma$ error bars). The right panels show the corresponding radial gradients measurements.
    The black contours again mark the source plane de-projected galacto-centric distances as in 
    Figure~\ref{fig:combELmap}.
    We adopted weighted Voronoi tessellation \citep{Cappellari:2003eu,Diehl:2006cz},
    with a SNR of 10 on \OIII for the binned metallicity maps.
    In the right column, these bins are plotted as individual data points.
    The dashed line denotes the linear regression from these points, with the measured radial slope shown at the bottom of each
    panel.  For both galaxies, the radial gradient is strongly positive (\ie inverted).
    \label{fig:oh12grad}}
\end{figure*}

Figure~\ref{fig:oh12grad} shows the 2D maps of metallicity of our selected two dwarf galaxies at $z\sim2$.
Clearly, the outskirts of our galaxies display highly elevated oxygen abundance ratios.
In particular, the outskirts of ID03751 are more metal enriched by $\sim$0.4 dex (\ie a factor of 2.5) than its center, and more 
metal-rich by $\sim$0.2 dex than the value inferred based on the fundamental metallicity relation (FMR) given its 
integrated \Mstar \cite{2010MNRAS.408.2115M,Mannucci:2011be}.
Note that our metallicity measurements extend beyond the source effective radius to cover large enough dynamic 
range, but not into the region where a plateau/flattening in metallicity \citep[\ie, at $R>2-2.5 R_{\rm 
eff}$,][]{2014A&A...563A..49S,SanchezMenguiano:2016gj} is likely to occur, which might bias the overall gradient 
determination.
For the first time, we are able to detect strongly inverted metallicity gradients in $z$$\sim$2 dwarf galaxies
at unprecedentedly high confidence: 0.122$\pm$0.008 dex/kpc for ID03751 ($\sim15.2\sigma$), and
0.111$\pm$0.017 dex/kpc for ID01203 ($\sim6.5\sigma$).

The question is thus what caused these dwarf galaxies to have such strongly inverted gradients?
First of all, our sources show no evidence of major mergers, supported by their regular morphology displayed in the 2D maps of 
\Mstar and EL surface brightness in Figure~\ref{fig:combELmap}.
For source ID01203 with \osiris data, this statement is further strengthened by the kinematic evidence of disk orderly rotation.
Secondly, the fact that the outskirts of our sources show elevated metallicity as compared to the FMR 
expectations indicates that there are more metals in the outer regions than could be produced by the stars in those 
regions.
This discourages any explanations involving solely low-metallicity gas inflows, not limited to those induced by mergers.
In the subsequent sections, we thus gather all available pieces of observational evidence to further investigate the possible 
cause.

\subsection{\SFR, stellar population age, and gas fraction}\label{sect:physprop}

To understand the cause of the strongly inverted metallicity gradients seen in these dwarf galaxies,
we combine their EL maps with \hst broad-band photometry to derive 2D maps of \Mstar, \SFR,
stellar population age, and gas surface density for each galaxy.
The \SFR is derived from extinction-corrected Balmer emission line flux. Maps of \Hb and \Hg emission are shown in
Figure~\ref{fig:combELmap}. The \Hb/\Hg line ratio provides a measurement of nebular extinction although it is limited by the
modest signal-to-noise of \Hg. We obtain more precise results from \hst photometry, by converting \B-\I color maps to spatial
distributions of stellar reddening $E_{\rm S}(B-V)$ \citep{Daddi:2004hj}. Nebular reddening $E_{\rm N}(B-V)$ is then calculated
following \citet{Valentino:2017by}.  The nebular reddening maps of both our galaxies show lower dust attenuation in centers than
that in outskirts, consistent with the inverted metallicity gradients shown in Figure~\ref{fig:oh12grad}.

We calculate extinction in \Hb adopting a \citet{1989ApJ...345..245C} dust extinction law
(with \Rv=3.1) and assuming Case B recombination with Balmer ratios appropriate for fiducial \HII region properties (\ie,
\Ha/\Hb~=~2.86).  Finally, we convert intrinsic \Ha luminosity to \SFR through the commonly used calibration
\citep{Kennicutt:1998ki},
\begin{align}
    {\rm SFR} = 4.6\times10^{-42}~ \frac{L(\Ha)}{\rm erg/s} \quad [\Msun/\textrm{yr}],
\end{align}
appropriate for the \citet{Chabrier:2003ki} IMF.
This provides the instantaneous star formation rate on $\sim$10 Myr time scales; we note that the ultraviolet continuum probed by
\hst photometry is sensitive to recent \SFR over a longer time span ($\sim$100-300 \Myr). The short timescales probed by Balmer
emission are most relevant for determining outflow physical properties, which are highly dynamic on small spatial scales, \eg, at 
sub-kpc level.

Next we derive average stellar age maps, using the spatial distribution of EL EW as the primary constraint. We
calculate \Hb rest-frame EWs from our maps of the emission line flux and stellar continuum flux density. Stellar continuum maps
are corrected for emission line contamination as described in Section~\ref{sect:data}. We correct for stellar Balmer absorption
which we estimate to be rest-frame EW~$\sim$3 \AA\ in \Hb based on the derived galaxy properties \citep{Kashino:2013ev}.  Maps of
\Hb EW are then converted to average stellar age using a series of \burst stellar population synthesis
models \citep{Leitherer:1999jt,Zanella:2015ej} assuming 1/5 solar metallicity and constant star formation history.

We also compare the age estimates given by our SED fitting (Section~\ref{sect:data}) and \Hb rest-frame EW using the method
described above. The median values given by the former practice are systematically larger than those of the latter by $\sim$0.5
dex, but we note that the uncertainties by the SED fitting are usually much larger due to the absence of prominent continuum
spectral age indicators, \eg, \Dn and \HdA \citep{Kauffmann:2003cu}. Hence, we adopt the results from \Hb rest-frame EW as the
average age for stellar populations throughout our paper, as we consider this a more reliable estimate.

Finally, we calculate the gas fraction defined as
\begin{align}
    \fgas=\Sgas/\left(\Sgas+\Sstar\right).
    \label{eq:fgas}
\end{align}
Since we do not directly observe the bulk of interstellar gas, we instead estimate gas surface density \Sgas by 
inverting the
Kennicutt-Schmidt (KS) law \citep{Schmidt:1959bp,Kennicutt:1998id}, \ie, $\Sigma_{\SFR}\propto\Sgas^{N}$ together with our
measurements of $\Sigma_{\SFR}$ described above.  We adopt the more robust extended version of the KS law developed by 
\citet{Shi:2011ck,Shi:2018wf} which is especially useful in low density regimes:
\begin{align}
    \frac{\Sigma_{\SFR}}{\Msun/\yr/\kpc^2} = 10^{-4.76} \left(\frac{\Sstar}{\Msun/\pc^2}\right)^{0.545}
    \left(\frac{\Sgas}{\Msun/\pc^2}\right)^{1.09}.
    \label{eq:KSlaw}
\end{align}
This extended KS law has been tested in numerous ensembles of galaxies as well as low surface brightness regions in individual
galaxies, and is shown to have relatively small scatter ($\sim$0.3 dex) over a large dynamic range of gas and \SFR surface
densities. We have combined in quadrature this systematic uncertainty of 0.3 dex in our estimates of \Sgas.

\begin{figure*}
    \centering
    \includegraphics[width=.33\textwidth]{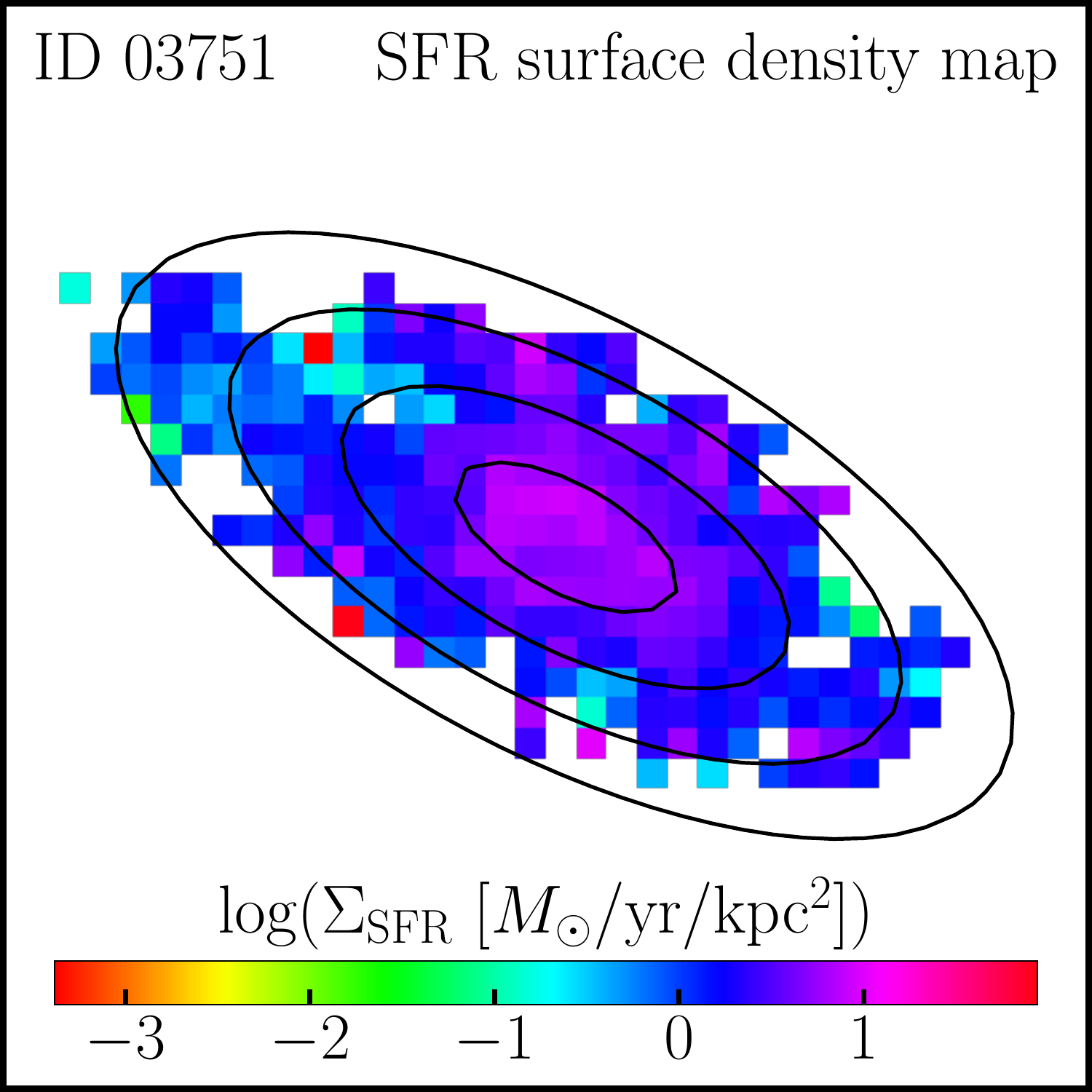}
    \includegraphics[width=.33\textwidth]{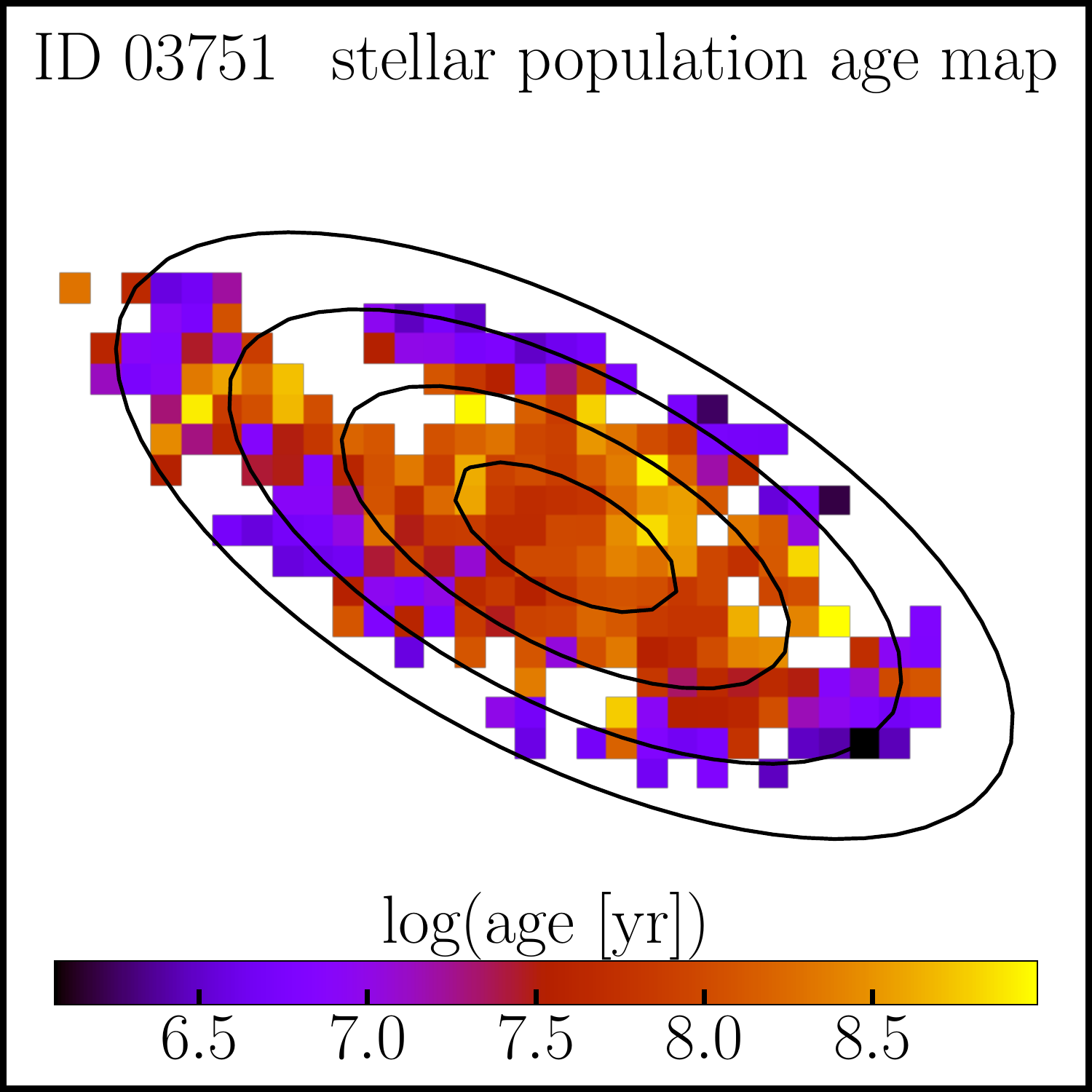}
    \includegraphics[width=.33\textwidth]{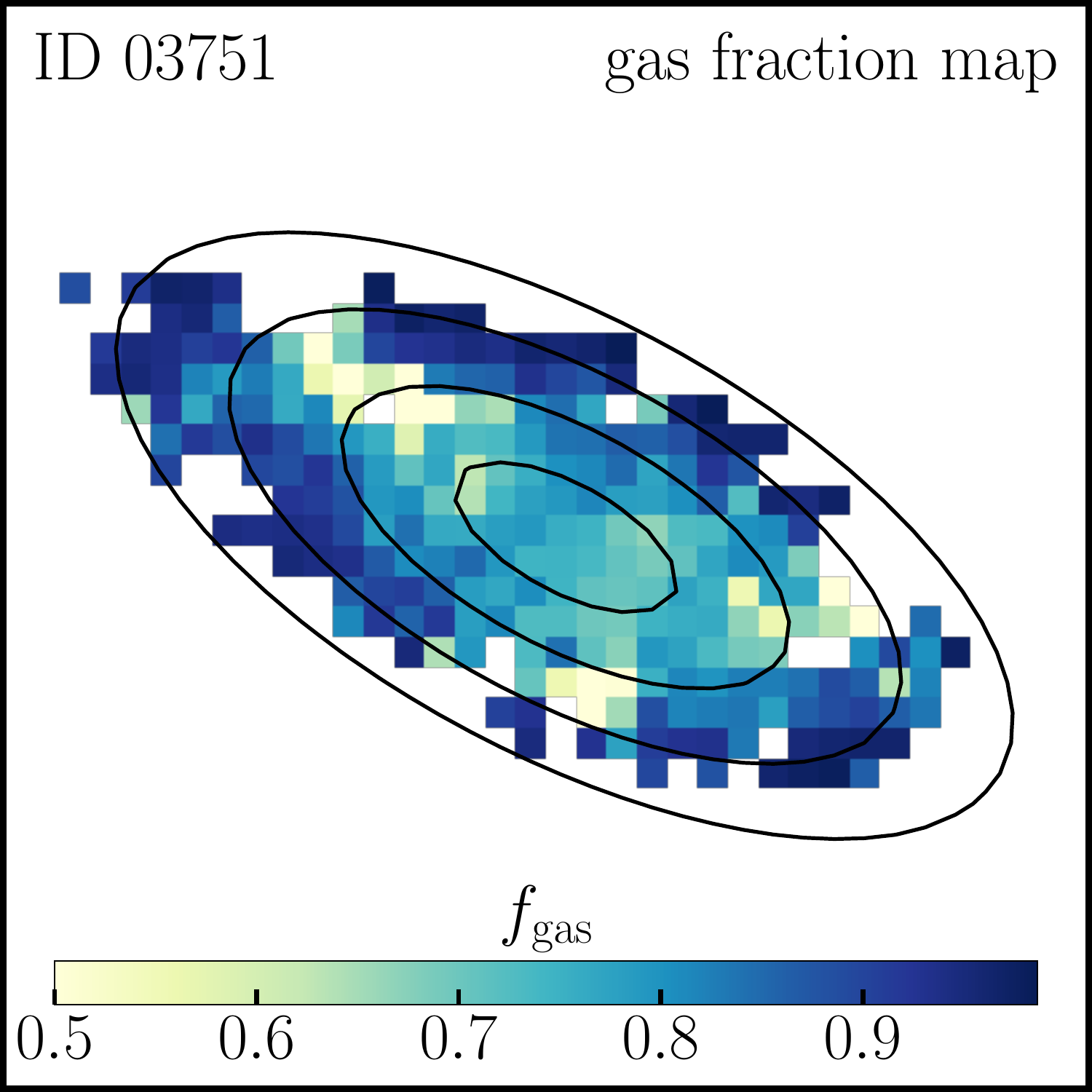}\\
    \includegraphics[width=.33\textwidth]{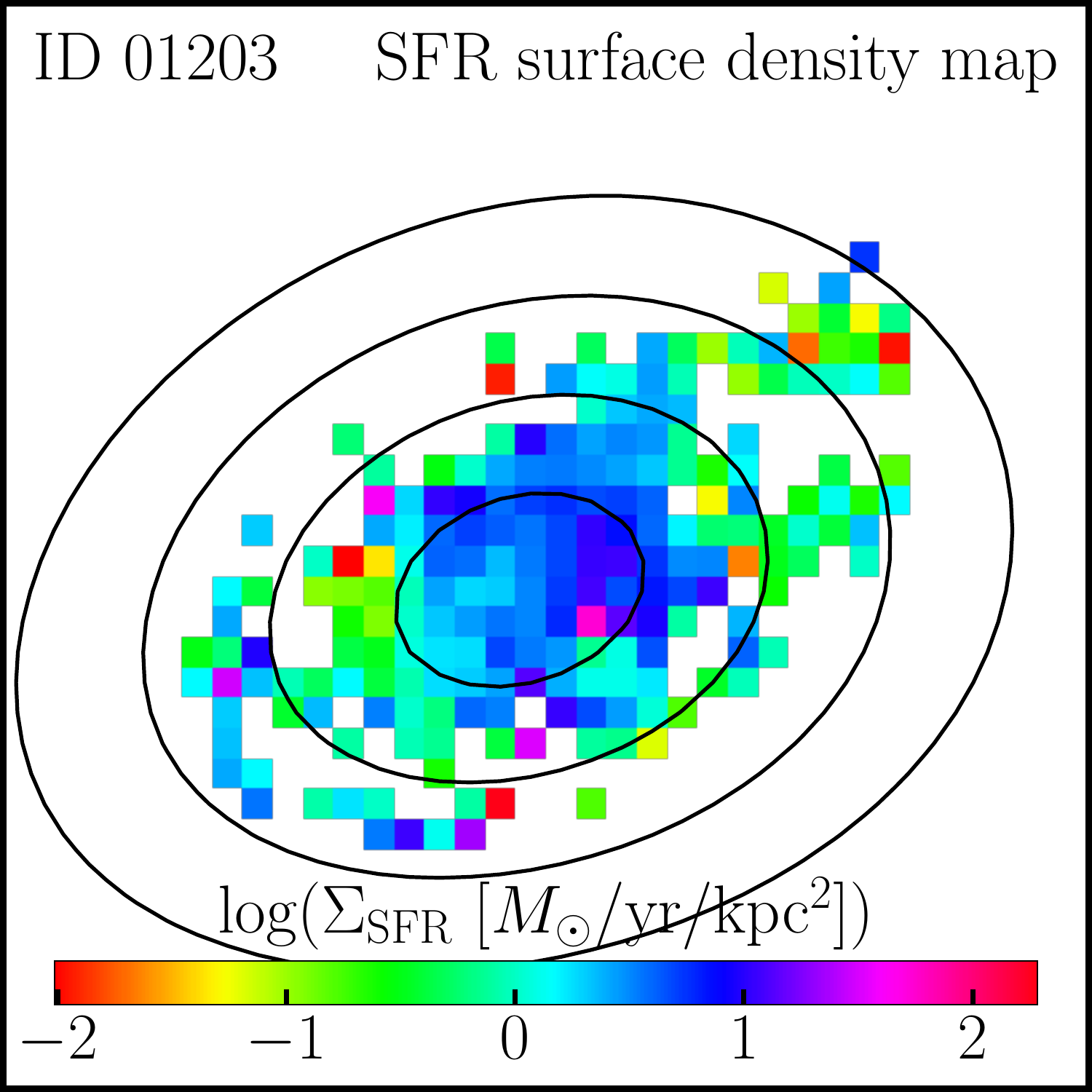}
    \includegraphics[width=.33\textwidth]{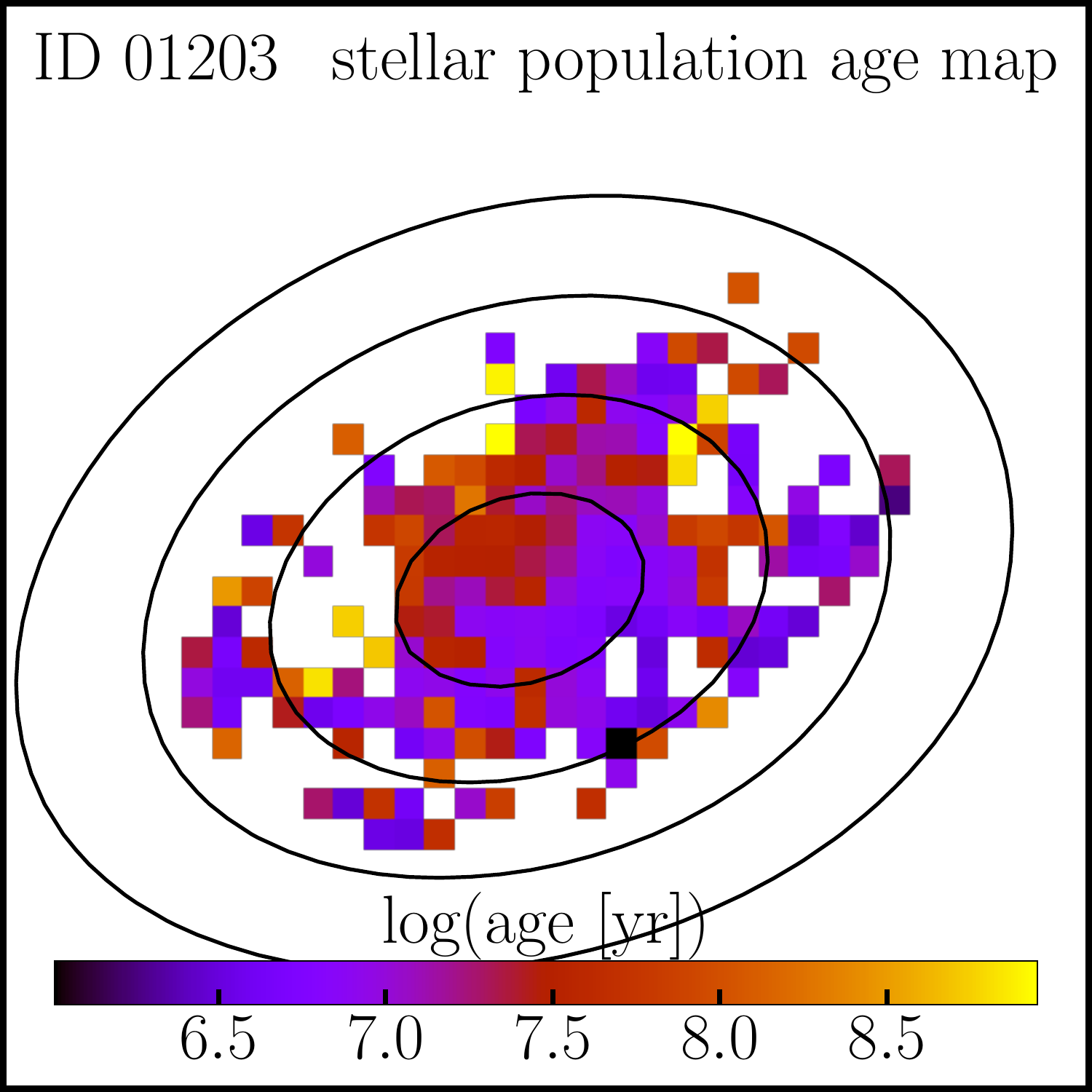}
    \includegraphics[width=.33\textwidth]{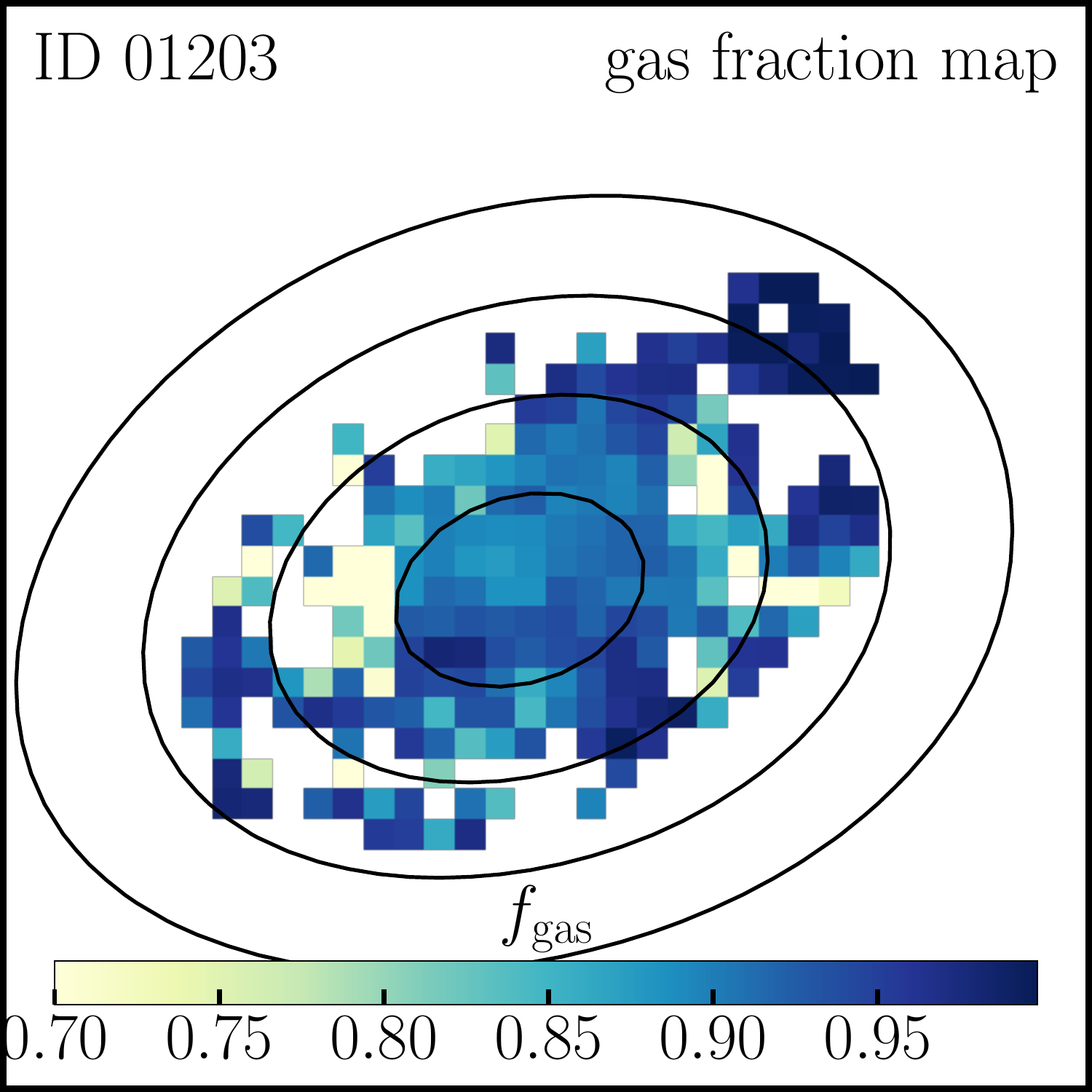}
    \caption{Maps of SFR surface density, average stellar population age, and gas fraction for our galaxies, derived from our
    spatially resolved analysis of stellar continuum and nebular emission.
    The spatial extent and orientation follows that in Figure~\ref{fig:combELmap}.
    We see that for both sources compared with their outskirts, their central regions have more active star formation, older 
    stellar population, and lower gas fraction.
    \label{fig:physmap}}
\end{figure*}

Figure~\ref{fig:physmap} shows the derived 2D maps of \SFR, average stellar age, and gas fraction.
In general, we observe centrally concentrated star formation, with the most actively star-forming regions having surface densities 
$\gtrsim10~\Msun/\yr/\kpc$.
On average, the central regions also have older stellar populations and smaller gas fractions than the outskirts, indicating that
the outer regions are still in the early stages of converting their gas into stars.
These features together indicate that we are witnessing the rapid build-up of galactic disks through in-situ star formation and 
strongly support an inside-out mode of galaxy growth \citep{Nelson:2014is,2013ApJ...765...48J}.

\begin{figure*}
    \centering
    \includegraphics[width=.48\textwidth]{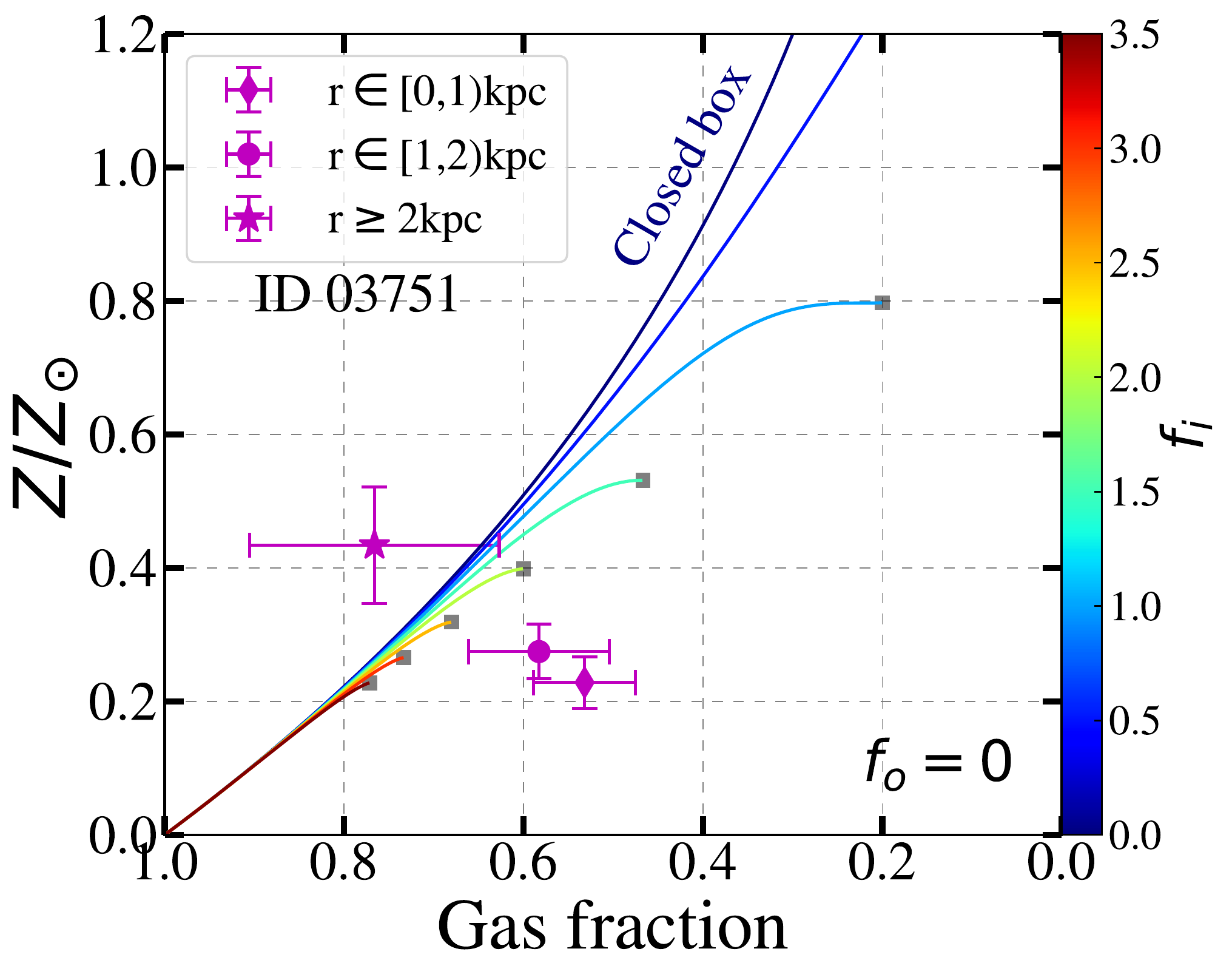}
    \includegraphics[width=.48\textwidth]{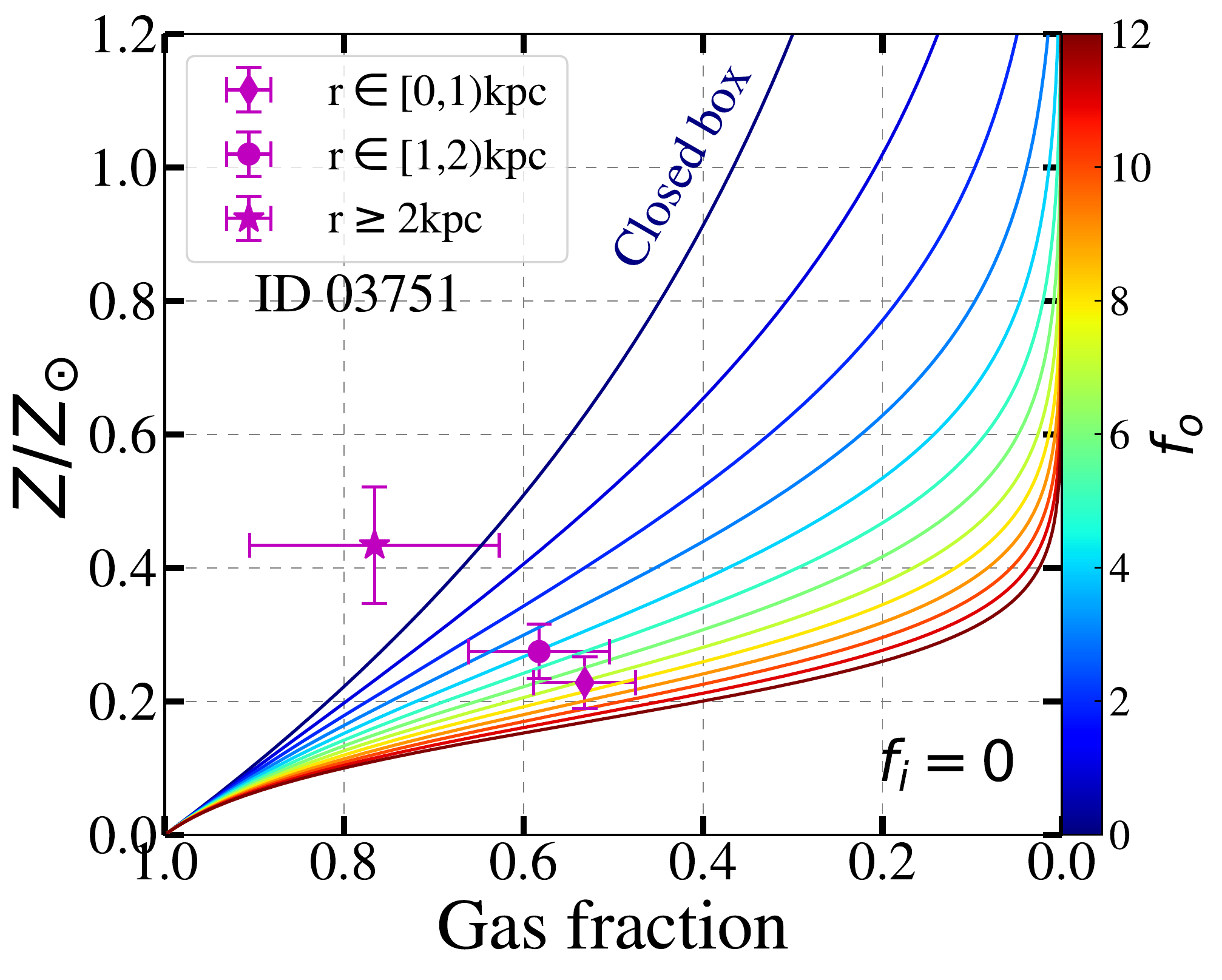}
    \caption{Gas fraction and metallicity estimated in different radial annuli for galaxy ID03751.
    The diamond, circle, and star symbols represent measurements derived at a galacto-centric radius of $r\in[0,1)$\kpc,
    $r\in[1,2)$\kpc, and $r\gtrsim2$\kpc, respectively.
    We also overlay the curves calculated from a simple chemical evolution model \cite{Erb:2008di} under extreme conditions, \ie,
    pure gas inflow ($f_o=0$; left) and pure gas outflow ($f_i=0$; right).
    Note that the trajectories of pure gas inflow cases cease at the grey squares for high infall rate ($f_i\gtrsim1$) conditions;
    any extensions from those grey squares toward low gas fraction (while fixing metallicity) are unphysical.
    This simple comparison shows that purely gas accretion does not suffice to explain the strong inverted gradients seen in our
    galaxies.
    \label{fig:Zerb_fgas}}
\end{figure*}

We compare our radially averaged \fgas and metallicity measurements against the predictions from
the simple chemical evolution model developed by \citet{Erb:2008di}.
To separate the effects of gas inflows and outflows, we compute two extreme sets of models, one being pure gas 
accretion (\ie with no outflows, $f_o=\Psi/\SFR=0$) and the other corresponding to the leaky box model (\ie with 
no inflows, $f_i=\Phi/\SFR=0$).
The results are shown in Figure~\ref{fig:Zerb_fgas}.
We note that for the pure gas accretion scenario, \fgas cannot decrease beyond a certain value, \ie,
\begin{align}
    \fgas^{\rm min} = 1-\frac{1-R}{f_i - f_o},
\end{align}
where $f_o=0$ and $R$ is the instantaneous return fraction. This $\fgas^{\rm min}$, implicitly imposed by 
Eq.~(11) of \citet{Erb:2008di}, physically indicates that galaxies cannot exhaust their gas reservoir to below a 
certain amount without the help of outflows,
under the equilibrium condition with steady gas accretion (see Section~\ref{sect:regulator} when this equilibrium assumption is 
relaxed).
Therefore, the pure gas accretion scenario cannot explain the observed gas fractions in our source central regions (at $\lesssim$ 
2\kpc) where metallicities are also lower.
The leaky box model, on the other hand, provides a plausible explanation for our observation such that the outflow rate tends to 
increase towards galaxy center.
However, we stress that in reality both gas outflows and inflows are acting together to re-distribute 
metallicity. This test using simple chemical evolution models just clearly shows that using gas accretions 
\emph{alone} cannot explain our spatially resolved measurements.

\subsection{Spatially resolved gaseous outflows}\label{sect:regulator}

\begin{figure*}
    \centering
    \includegraphics[width=.49\textwidth]{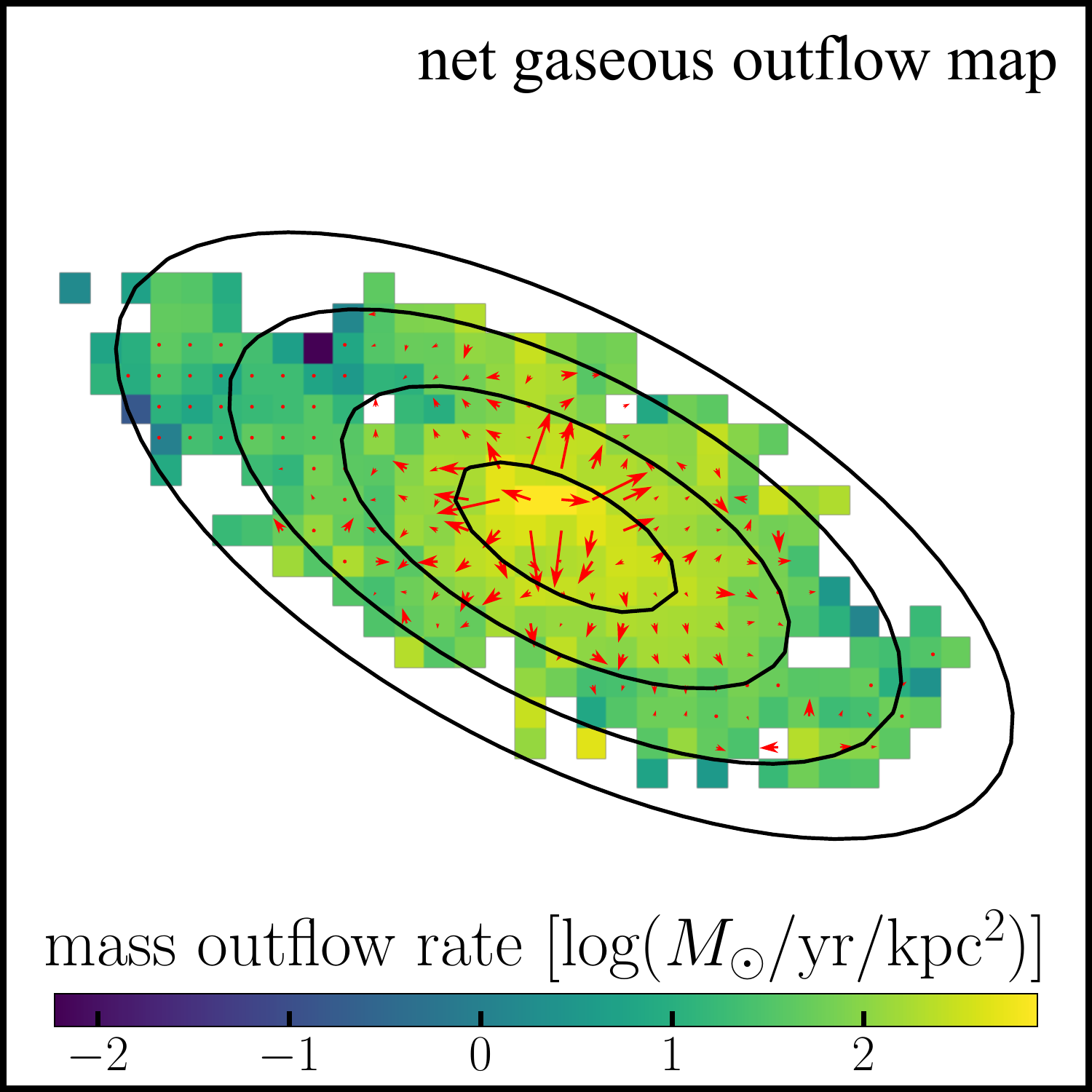}
    \includegraphics[width=.49\textwidth]{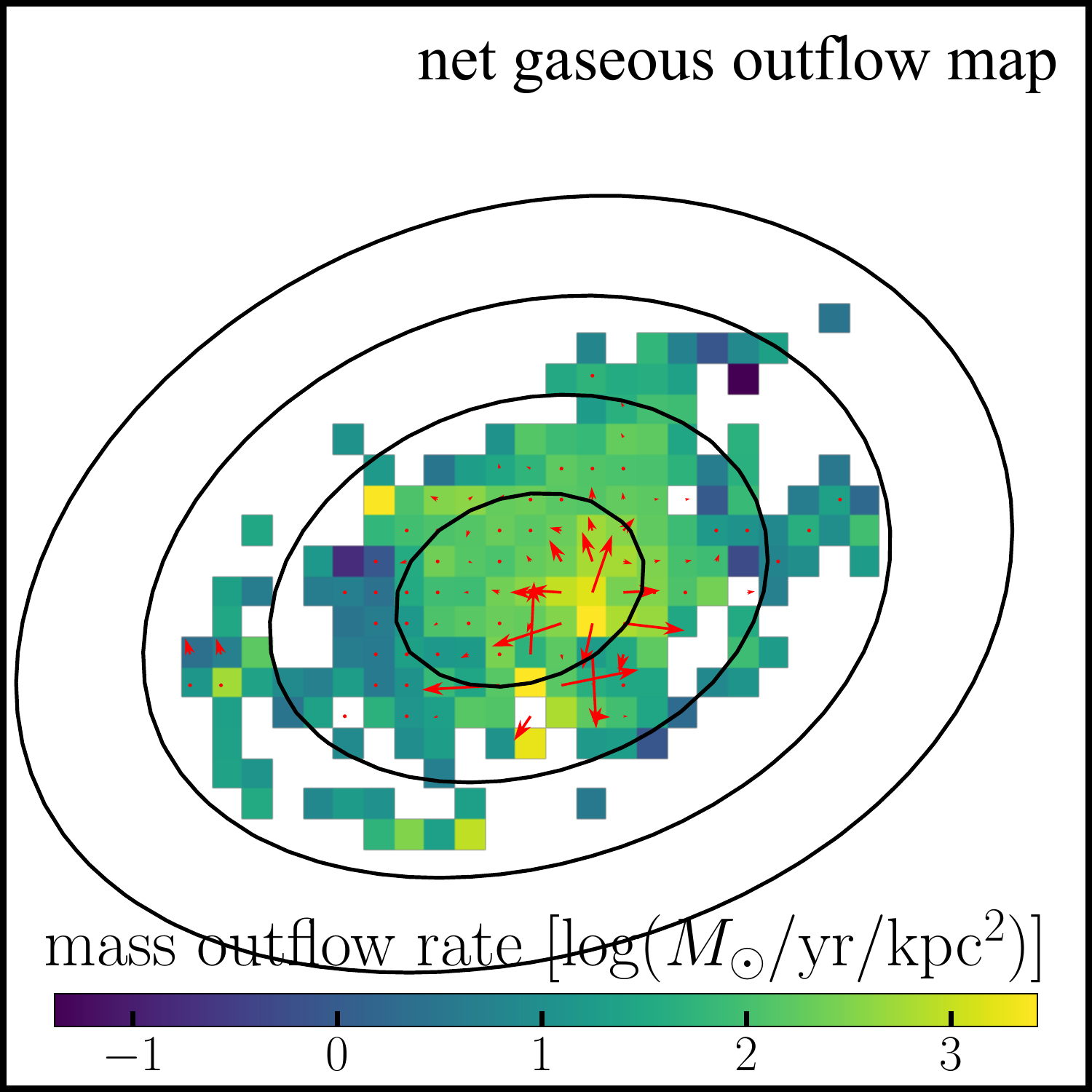}
    \caption{
    Maps of gaseous outflow rates derived from our analysis combining gas regulator models and empirical star-formation laws.
    The spatial extent and orientation follows that in Figure~\ref{fig:combELmap}.
    Red arrows show the net direction and magnitude of the gaseous outflows driven by galactic winds.
    We argue that outflows play a key role in effectively transporting stellar nucleosynthesis yields from the inner regions of 
    these two galaxies to their outskirts.
    \label{fig:outmass}}
\end{figure*}

The application of simple chemical evolution in Section~\ref{sect:physprop} is enlightening but depends on 
strong assumption, such as that the azimuthal variations are negligible and galaxies live in equilibrium.
In reality, these conditions might not be valid, \eg, due to rapid gas flows.
To gain a more precise understanding of the physics of galactic winds and the role of gaseous outflows in 
shaping the observed spatial distribution of metallicity, independent of those assumptions, we can turn to a
more advanced framework for galaxy chemical evolution: the gas regulator model 
\citep{Lilly:2013ko,Peng:2014hn}.
This model provides an informative and coherent view of the full baryon cycle, involving the accretion of underlying DM halos, as 
well as the instantaneous regulation of star formation by a time-variable gas reservoir.
A key feature of this model is that it does not assume that galaxies live in an equilibrium state, where the total amount of gas 
mass remains constant. The non-equilibrium flexibility is especially important for applying this model to spatially resolved 
regions within a galaxy, where gas may be transported radially from one region to another. Chemical evolution within the gas 
regulator model is described by the equations
\begin{align}
    Z_{\rm gas} & = \left[Z_0 +
    y\taueq\epsilon\(1-\exp(-\frac{t}{\taueq})\)\right]\left[1 -
    \exp(\frac{-t/\taueq}{1-\exp(-t/\taueq)})\right],      \\
    \tau_{\rm eq} & = \frac{1}{\epsilon\(1-R+\lambda\)}.    \non
\end{align}
Here we adopt the convention of symbols itemized in Table~1 of \citet{Peng:2014hn}:
\Zgas is the mass fraction of metals in the gas reservoir (determined from the observed \oh as in \citet{Peeples:2011ew}), $t$ is 
the average stellar population age, $\taueq$ is the time scale on which the baryon cycle reaches equilibrium, $\epsilon$ is the 
\sf efficiency (defined as $\epsilon\equiv\SFR/\Mgas=\Sigma_\SFR/\Sgas$), and $\lambda$ is the mass loading factor (defined in 
terms of the mass outflow rate $\Psi$, such that $\lambda=\Psi/\SFR$\footnote{Note that $\lambda$ and $f_o$ in 
Section~\ref{sect:physprop} represent the same quantity but here we are solving for $\lambda$ in a spatially resolved fashion.}).
We adopt a stellar nucleosynthesis yield $y=0.003$ \citep{Dalcanton:2007kc} with $R=0.4$ estimated from BC03 
\citep{Bruzual:2003ck} stellar population models. Finally, we assume that gas inflows are pristine ($Z_0=0$).

For each spatial region where we have estimated the metallicity, \SFR, gas surface density, and age (Figures~\ref{fig:oh12grad}, 
\ref{fig:physmap}), we solve the above equations for the mass loading factor $\lambda$ and subsequently calculate the mass outflow 
rate $\Psi$.
The 2D distribution of $\Psi$ is displayed in Figure~\ref{fig:outmass}.
Taking the gradient field of this gaseous outflow map, we obtain the net direction of the outflowing mass flux on 
sub-galactic scales, projected along the line of sight, denoted by the red arrows in Figure~\ref{fig:outmass}.
The results demonstrate that strong galactic winds transport mass from the center to the outskirts, with the net 
radial transport of heavy elements causing the inverted gradients observed in our targets.

\begin{figure*}
    \centering
    \includegraphics[width=.49\textwidth]{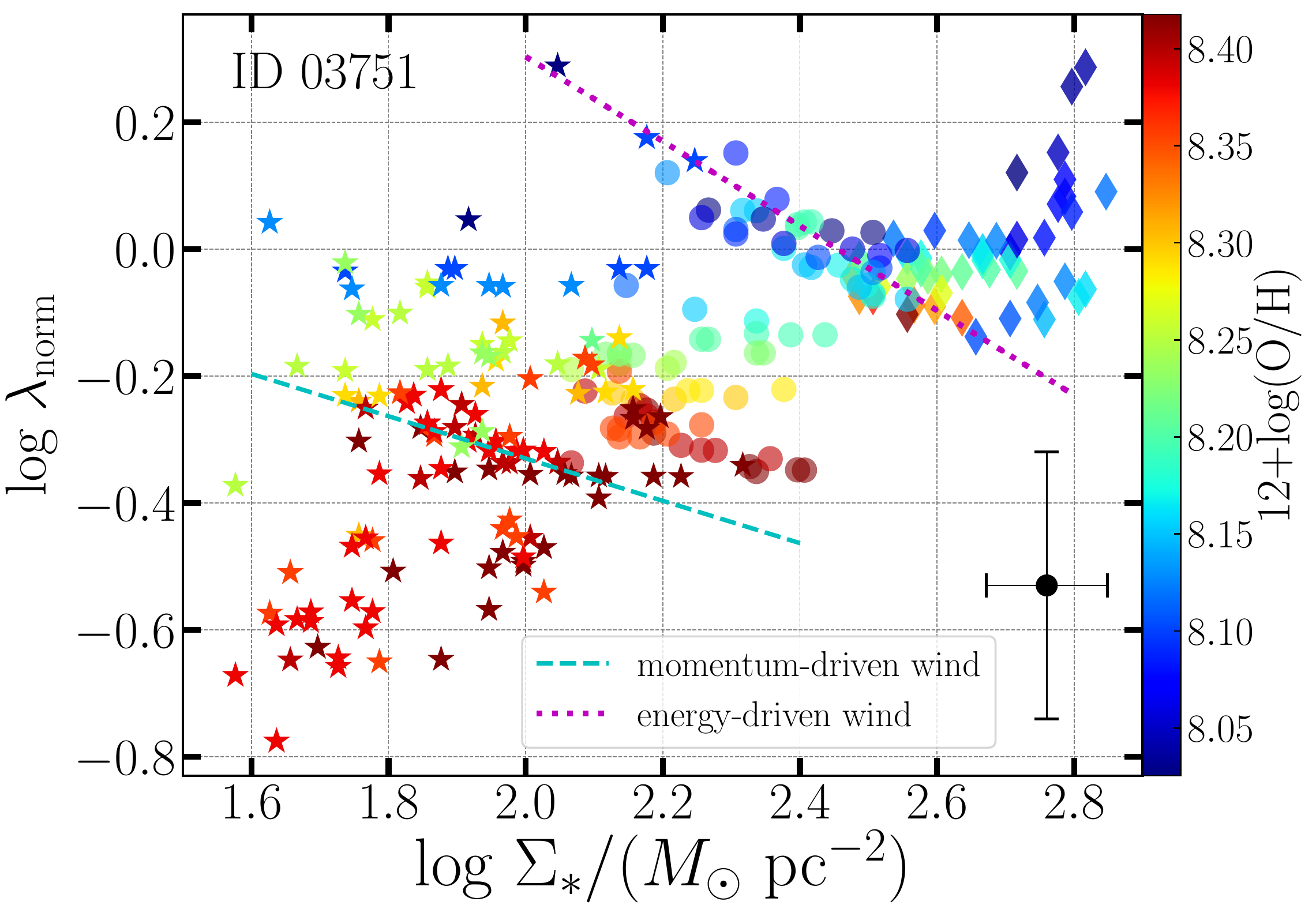}
    \includegraphics[width=.49\textwidth]{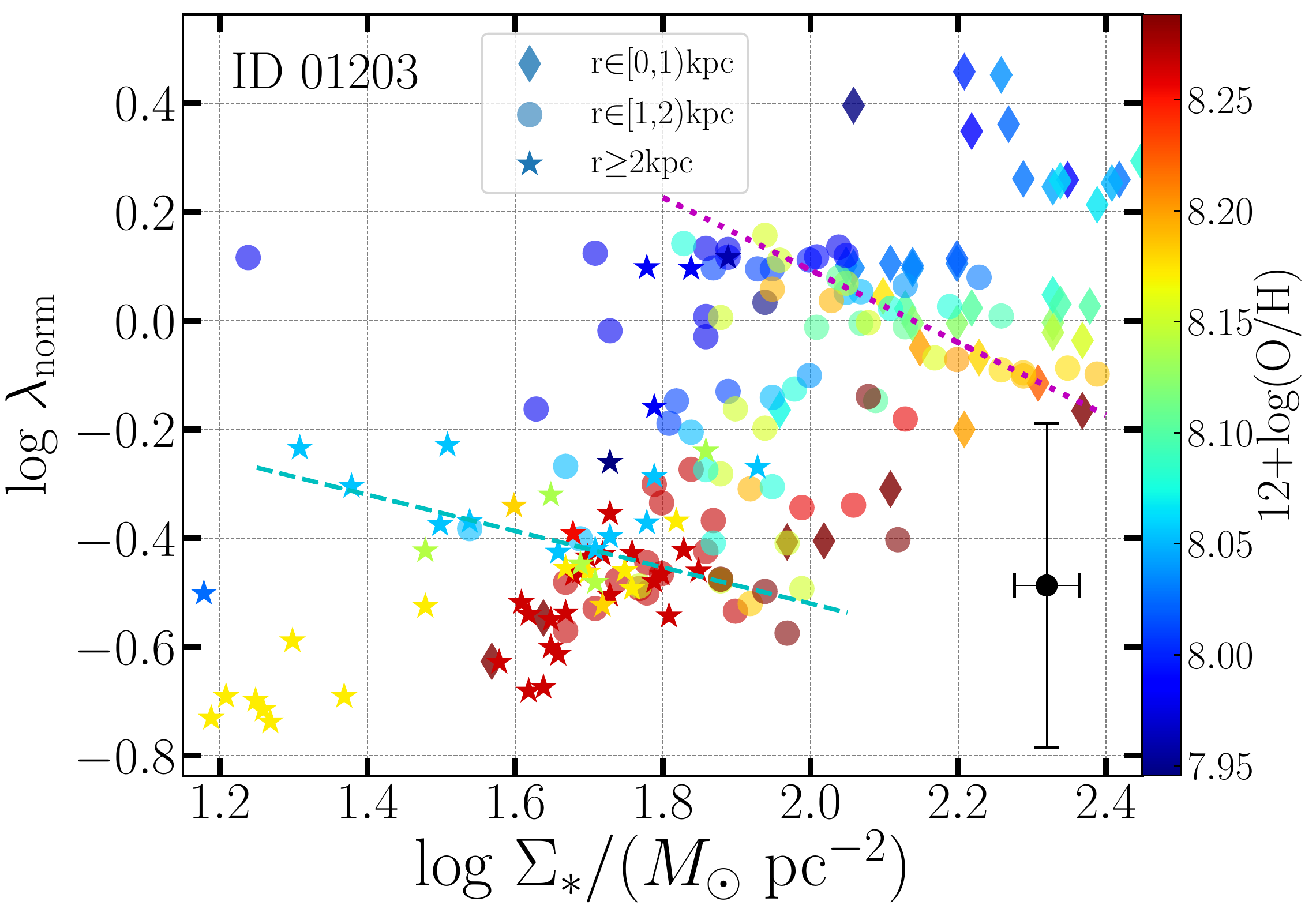}
    \caption{
    Correlation between spatially resolved mass loading factor $\lambda$ (normalized to the value at radius 1
    kpc; see Table~\ref{tab:srcprop}) and stellar surface density $\Sigma_\ast$, color-coded by metallicity.
    As in Figure~\ref{fig:physmap}, the diamond, circle, and star symbols represent measurements derived at a 
    galacto-centric radius
    of $r\in[0,1)$\kpc, $r\in[1,2)$\kpc, and $r\gtrsim2$\kpc, respectively.
    We overlay as an illustration two scaling relations that are commonly assumed to describe integrated measurements:
    $\lambda\propto\Sigma_\ast^{-2/3}$ for an energy-driven wind model marked by magenta dotted lines, and
    $\lambda\propto\Sigma_\ast^{-1/3}$ for a momentum-driven wind model by cyan dashed lines.
    Evidently, a single scaling relation is not sufficient to describe the spatially resolved data, demonstrating the need for a
    more sophisticated approach.
    The black point in the lower right corner in each panel displays the median uncertainties for these measurements.
    \label{fig:lamSDstar}}
\end{figure*}

The distribution of mass loading factors $\lambda$ within each of our targets is also shown in 
Figure~\ref{fig:lamSDstar}, revealing higher $\lambda$ (and therefore a higher fraction of metals lost) in the central regions.
This preferential removal of metals from the center, and subsequent deposition at larger radii, gives rise to the strong 
positively sloped metallicity gradients evident in Figure~\ref{fig:oh12grad}.
The high values of $\lambda$ have important implications for the role of feedback in galaxy formation.
Most fundamentally, our results support feedback as a solution to the ``over-cooling'' problem in galaxy formation, by ejecting
gas and preventing overly condensed baryonic regions at high redshifts \citep{1978MNRAS.183..341W,Dekel:1986cv}.
Such strong outflows are also expected to suppress the formation of stellar bulges from low angular momentum gas 
\citep{Governato:2010ed,Brook:2012gj}.
This is consistent with low bulge fraction in these two galaxies measured from high resolution \hst imaging 
(Table~\ref{tab:srcprop}).

A key feature in the $\lambda$ distribution is that neither of the wind modes, driven by momentum or energy conservation, can 
explain the behavior of the mass dependence of $\lambda$ alone, {\em within} individual galaxies.
Outflows are typically parameterized by either a momentum-driven \citep{Oppenheimer:2006eq,Oppenheimer:2008bu} or an
energy-driven \citep{Springel:2003eg} wind mode, both of which are physically well motivated \citep{Murray:2005jt}.
The energy-driven wind scenario assumes that outflows are launched by the thermal pressure of supernova (SN) explosions and/or
winds from massive stars.
A portion of this thermal energy provides the outflow kinetic energy, \ie, $\Psi\times v^2_{\rm wind} \sim \SFR$, where the wind 
speed $v_{\rm wind}$ can mimic the escape velocity from DM halo, \ie, $v_{\rm esc}\sim M_{\rm h}^{1/3}$ given by the virial 
theorem.
This results in the scaling relation of $\lambda\propto\Mstar^{-2/3}$, assuming the linear correlation between the mass 
constituents of stellar and dark components.
The energy-driven wind model is found successful in explaining the low abundance of satellite galaxies in the Milky Way 
\citep{Okamoto:2010ba}.
The momentum-driven wind model instead relies on the momentum injection deposited by radiation pressure from SN explosions and/or 
massive stars, leading to $\Psi\times v_{\rm wind} \sim \SFR$ and $\lambda\propto\Mstar^{-1/3}$.
In this scenario, $v_{\rm wind}$ is proportional to \Mstar and \SFR, broadly consistent with some observational
results \citep{Martin:2005kx}.
The transition from energy- to momentum-driven winds is typically thought to be a galaxy-wide phenomenon,
resulting in the steepening of the mass-metallicity relation below \Mstar~$\simeq10^{9.3}$\Msun at $z\simeq2$ 
\citep{Henry:2013gx}.
However, our analysis indicates that a single mode is not sufficient to describe spatially resolved data \emph{within} one galaxy 
and it is highly likely that the transition from energy- to momentum-driven winds occurs on sub-galactic scales, governed by local 
gas and star formation properties in addition to the global gravitational potential.

\section{Summary and discussion}\label{sect:conclu}

We present the first robust confirmation of the existence of strongly inverted metallicity radial gradient (\ie $\gtrsim$0.1 
dex/kpc) in star-forming dwarf galaxies ($\Mstar\lesssim10^9\Msun$) at the peak of star formation and chemical enrichment 
($z\sim2$).
Our synergy of the diffraction-limited imaging spectroscopy from \hst NIR grisms and lensing magnification permits exquisite 
spatial sampling, \ie, at the scale of 50-100 pc, to securely resolve our $z\sim2$ galaxies with $\gtrsim$300 
resolution elements (Figures~\ref{fig:combELmap}--\ref{fig:oh12grad}) to deliver precise radial gradient 
measurements.
To understand the physical origin of these strongly inverted gradients, we obtain high resolution 2D maps of star formation rate, 
characteristic stellar age (or equivalently star formation timescale), and gas fraction, from \hst observations of source stellar 
continuum and nebular emission.
These 2D maps show that the galactic disks of our sources are rapidly assembling stellar mass through in-situ star formation, in 
the early phase of inside-out growth (Figure~\ref{fig:physmap}).
By comparing our observations with simple chemical evolution models, we find that gas accretion alone cannot explain these 
strongly inverted gradients in our galaxies (Figures~\ref{fig:Zerb_fgas}).

Using a more advanced gas regulator model, we are able to calculate the spatial distribution of mass loss rates from outflows, 
treating each spaxel as an independent star-forming region, and thus map the macroscopic patterns of net gaseous outflows 
(Figure~\ref{fig:outmass}).
It turns out that the mass loss rates are highest in the central regions of both galaxies, coincident with
the peak star formation surface densities.
A natural explanation is thus that active star formation in galaxy centers gives rise to powerful winds that transport gas and
metals away from the center toward larger radii, forming ``galactic fountains'' \citep{Martin:2002ee}.

Furthermore, our spatially resolved analysis of metals, \SFR, and stellar populations shows that a single type of wind mechanism 
(either energy or momentum driven) cannot explain the entire galaxy (Figure~\ref{fig:lamSDstar}).
A primary physical parameter that has been proposed to set the transition between the two wind dynamics is the gravitational 
potential, often parameterized by velocity dispersion ($\sigma$). There exists a critical scale \scrit 
\citep{Murray:2005jt} such that for galaxies with $\sigma<\scrit$, energy injection by SNe sets a limiting 
\SFR above which interstellar gas is ejected in galactic winds. For galaxies with $\sigma>\scrit$, momentum 
deposition limits the maximum \SFR above which the ISM is likewise ejected.
The presence of both energy- and momentum-driven wind scalings in one galaxy suggests that feedback-triggered winds are connected
to physical properties on sub-galactic scales, \eg, \emph{local} velocity dispersion ($\sigma_{\rm local}$), 
which is sensitive to the optical depth of gas flows, the coupling efficiency between gas clouds and dust 
parcels, \etc.
On sub-galactic scales, there exists a strong correlation among velocity dispersion (not necessarily 
$\sigma_{\rm local}$), surface density and size of molecular clouds \cite[see][and references 
therein]{BallesterosParedes:2011gk}.
It appears that in our galaxies, the wind-launching mechanism transitions from energy- to momentum-driven as 
galacto-centric radius increases.
This gives rise to a hypothesis that $\sigma_{\rm local}$ in our galaxies should increase from inner to outer 
regions.
Our current kinematic data on source ID01203 have high spatial resolution (at $0\farcs05$ plate scale) yet narrow FoV so that it 
is infeasible to map sub-kpc scale velocity dispersion accurately to outer regions at $r\gtrsim2$ \kpc, where momentum-driven wind 
seems to take over.
To test this hypothesis conclusively, more spatially resolved data taken under sufficient spatial sampling will be required to 
robustly derive a full 2D map of velocity dispersion out to the periphery of the galactic disk, using instruments 
with relatively large FoV, \eg, the \jwst NIRSpec IFU \citep{Kalirai:2018gs}.

Physically, the momentum-driven wind scaling applies to ``cool'' ($T \sim 10^4$ K) ambient interstellar gas entrained in outflows,
whereas the energy-driven wind is appropriate when entrained gas is shock heated to temperatures where cooling is inefficient ($T
\sim 10^6$ K). A plausible scenario for our galaxies is that feedback from an intense burst of star formation in the central
regions heats the ejected gas to a highly ionized phase, while gas entrained in outflows from the outer regions remains cool. If
this interpretation is correct, then we expect a distinct signature in the absorption properties of outflowing gas. Outflows from
the central regions should be dominated by highly ionized species (e.g. \ionp{O}{vi}, \ionp{C}{iv}, \ionp{Si}{iv}) whereas
outflows from the outer regions should have relatively more of the low ions characteristic of $T \sim 10^4$ K gas (e.g.
\ionp{Fe}{ii}, \ionp{Mg}{ii}, \ionp{Si}{ii}). Both high and low ion species are commonly observed in outflows from star forming
galaxies at $z\simeq2$ \citep{Berg:2018gd,Du:2018tr}, although their spatial distributions are not yet well known 
\citep[but see][]{James:2018km}.
Our hypothesis suggests a more central concentration of the high ions {\it in the specific cases} where a 
combination of both outflow scalings results in inverted metallicity gradients.
This prediction can be directly tested with spatially resolved spectroscopy of rest-frame
ultraviolet absorption lines using instruments such as \keck/KCWI or VLT/MUSE.

\acknowledgements
We thank ApJ and especially the editor Prof. Brad Gibson for a thoughtful and constructive review process that 
improved the quality of our manuscript.
This study makes use of the Hubble Space Telescope data collected by the \glass program.
We gratefully acknowledge support by NASA through HST grant HST-GO-13459.
XW is supported by the UCLA Graduate Division Dissertation Year Fellowship.
XW thanks Michele Cappellari, Renyue Cen, Tuan Do, Phil Hopkins, Bethan James, Suoqing Ji, Du{\v s}an Kere{\v s}, 
Claus Leitherer,
Xiangcheng Ma, Roberto Maiolino, Yong Shi, Enci Wang, Pieter van Dokkum, Anita Zanella, and Dong Zhang for 
helpful discussions.
YP acknowledges support from the National Key Program for Science and Technology Research and Development under grant number 
2016YFA0400702, and the NSFC grant no. 11773001.

{\it Software:} Astropy \citep{astropy:2018ti}, APLpy \citep{Robitaille:2012wl}, \adriz 
\citep{Gonzaga:2012tj}, \emc \citep{ForemanMackey:2013io}, \fast \citep{Kriek:2009eo}, Galfit 
\citep{Peng:2002di}, \sex \citep{Bertin:1996hf}, VorBin \citep{Cappellari:2003eu}.

\begin{appendix}

\section{Extracting and fitting 1D and 2D \protect\hst grism spectra}\label{sect:grismspec}

As briefly mentioned in Section~\ref{subsect:grism_reduce}, we employed the Grism Redshift and Line analysis 
software \grzl to reduce the \hst WFC3/NIR grism data from raw exposures acquired by the \glass program.
Our primary goal is to obtain the spatially resolved emission line intensities after removing the contribution 
from source continuum.
In terms of modeling the continuum spectrum, \grzl first produces a simple flat (in $F_{\lambda}$)
spectral model for all sources within the WFC3 FoV with \H-band magnitude brighter than 26 ABmag.
The normalization is determined to match the flux in the corresponding reference image (in our cases, F105W as 
the reference to G102, and F140W to G141, ascribed to similar wavelength coverage).
Then second-order polynomial functions are fitted to the sources whose \H-band magnitude is brighter than 24 
ABmag.
This process is done iteratively, until a convergence point where the residual in the grism exposures after
subtracting the fitted continuum models becomes negligible.

While the polynomially fitted continuua serve as good enough models for contamination subtraction
associated with neighboring objects, this polynomial functional form is clearly not physically representative of
the actual SED of the underlying stellar continuum for our sources of interest.
To facilitate a more accurate continuum subtraction, we further refine the source continuum model by considering 
primarily four template continuum spectra in a range of characteristic ages for stellar populations:
\begin{enumerate}
    \item a low-metallicity Lyman-break galaxy (Q2343-BX418) showing very young, blue continuum 
    \citep{Erb:2010iy},
    \item an intermediate-age composite SED with moderate Balmer break and 4000 \AA{} break, synthesized in 
    \citet{Brammer:2008gn} following the method of \citet{KCorrectionsandFi:WJFCKqrW},
    \item a post-starburst SED showing prominent Balmer break and 4000 \AA{} break from the UltraVISTA survey 
    \citep{Muzzin:2013is},
    \item a single stellar population SED with a 13.5 Gyr age and solar metallicity \citep{Conroy:2012bl}.
\end{enumerate}
This combination of both empirical and synthetic SED templates constitutes an optimized set appropriate for 
redshift fitting and continuum subtraction under our situation.
As discussed in \citet{Brammer:2008gn}, there is a trade-off between the number of templates used in SED fitting 
and numerical efficiency, and they find that the improvement is negligible if the number of templates is 
increased to above 5.
For a sanity check, we also run the template fitting procedures using a more complete template library built from 
the Flexible Stellar Population Synthesis (FSPS) models \citep{Conroy:2009ks,Conroy:2010ja,Conroy:2010bi} and 
found no noticeable changes in the spectroscopic redshift determinations nor continuum subtractions.

In addition to fitting stellar continuum, we model the intrinsic nebular emission lines in 1D spectra as Gaussian
functions. The amplitudes and flux ratios between most of the line species are allowed to vary (except for some
certain line complexes, \eg, $f_{\OIII~5008}/f_{\OIII~4960}$ = 3:1).
Given the relatively low instrument resolution of \hst grisms, the dynamic motion of gas and stellar components 
leave no effect on the observed profiles (both in 1D and 2D) of line emission/absorption features.
However, for spatially extended sources, the effective spectral resolution is lowered by morphological broadening 
\citep{vanDokkum:2011cq}, which usually varies with respective to the light-dispersion direction, \ie, the 
position angle (P.A.).
We explicitly take the source morphology into account via convolving the model spectra (stellar continuum + 
nebular emission) with the direct image in reference frames averaged along light-dispersion directions.

\begin{figure*}
    \includegraphics[width=\textwidth]{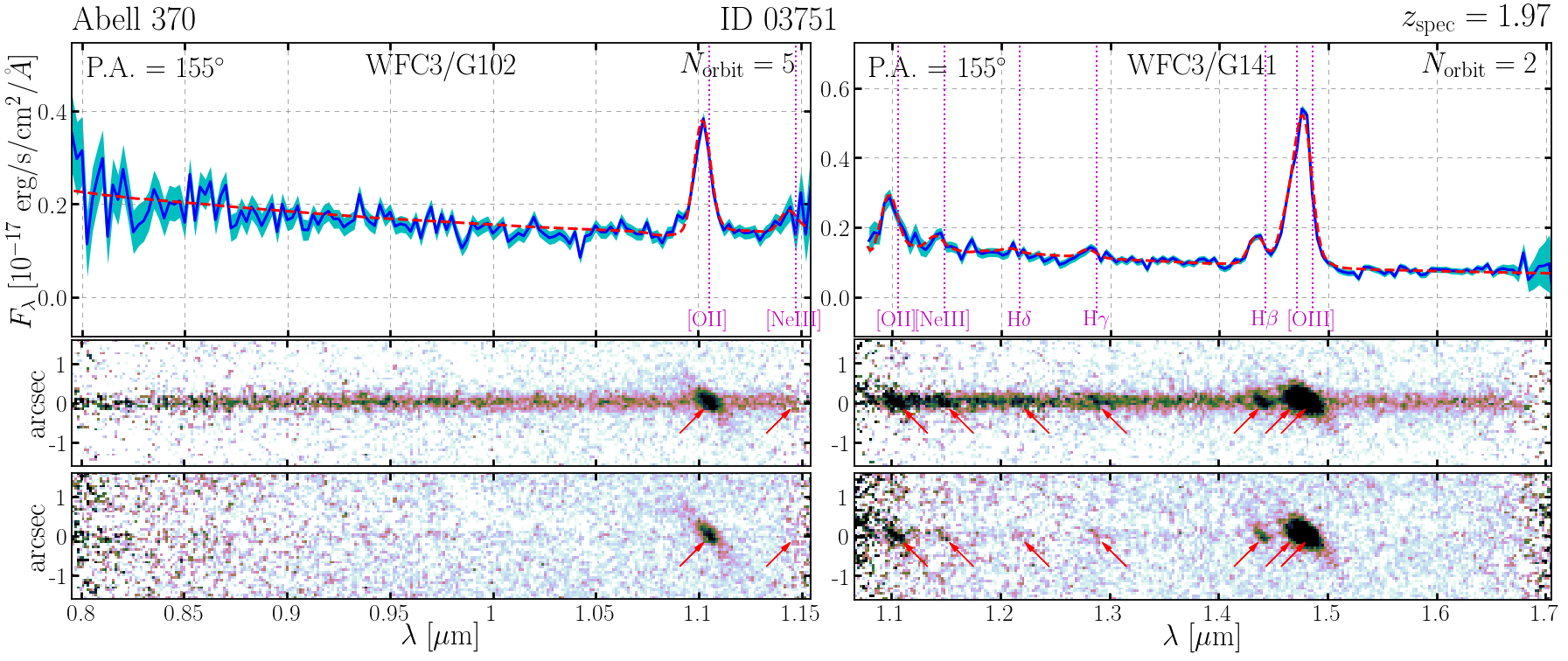}
    \includegraphics[width=\textwidth]{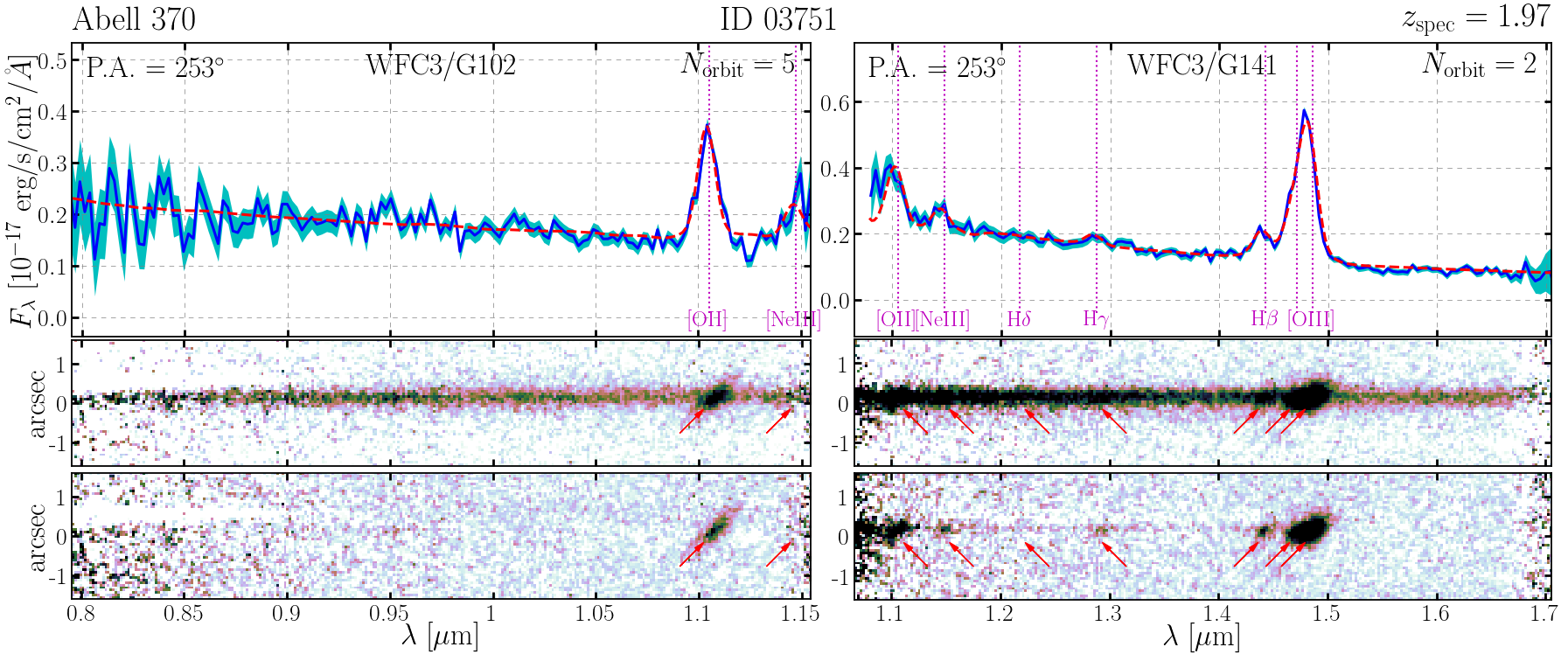}
    \caption{The \hst grism spectra for source ID03751 in the field of \clsan taken by the \glass program.
    The total science integration is equally distributed into two separate P.A.s,
    reaching 5 orbits of G102 exposures and 2 orbits of G141 exposures per P.A., shown in two sub-figures.
    In each sub-figure, from top to bottom, we show the optimally extracted 1D spectra and the full 2D spectra 
    before and after source continuum subtraction, for both grism elements.
    On the 1D spectra, the observed flux is represented by the blue solid line with 1-$\sigma$ noise level 
    denoted by the cyan shaded band, and the 1D model spectrum (source continuum + nebular emission) is 
    represented by the red dashed curve.
    The observed locations of emission features are highlighted by vertical dotted lines in magenta and arrows in 
    red, in 1D and 2D spectra respectively.
    Note that due to low spectral resolution, \NeIII is blended with \ionp{He}{i}$\lambda3889$+H8 on the red side 
    and H9 on the blue side.
    }
    \label{fig:3751spec}
\end{figure*}

\begin{figure*}
    \includegraphics[width=\textwidth]{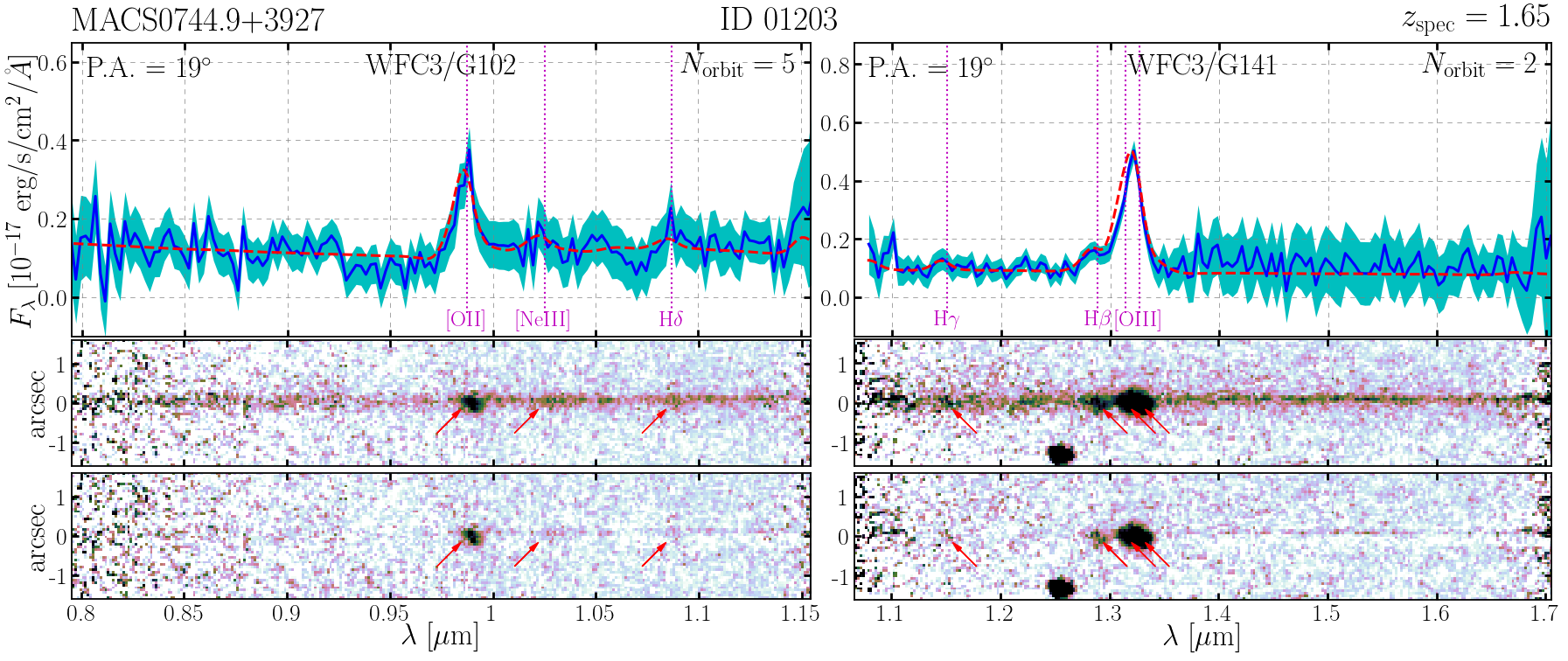}
    \includegraphics[width=\textwidth]{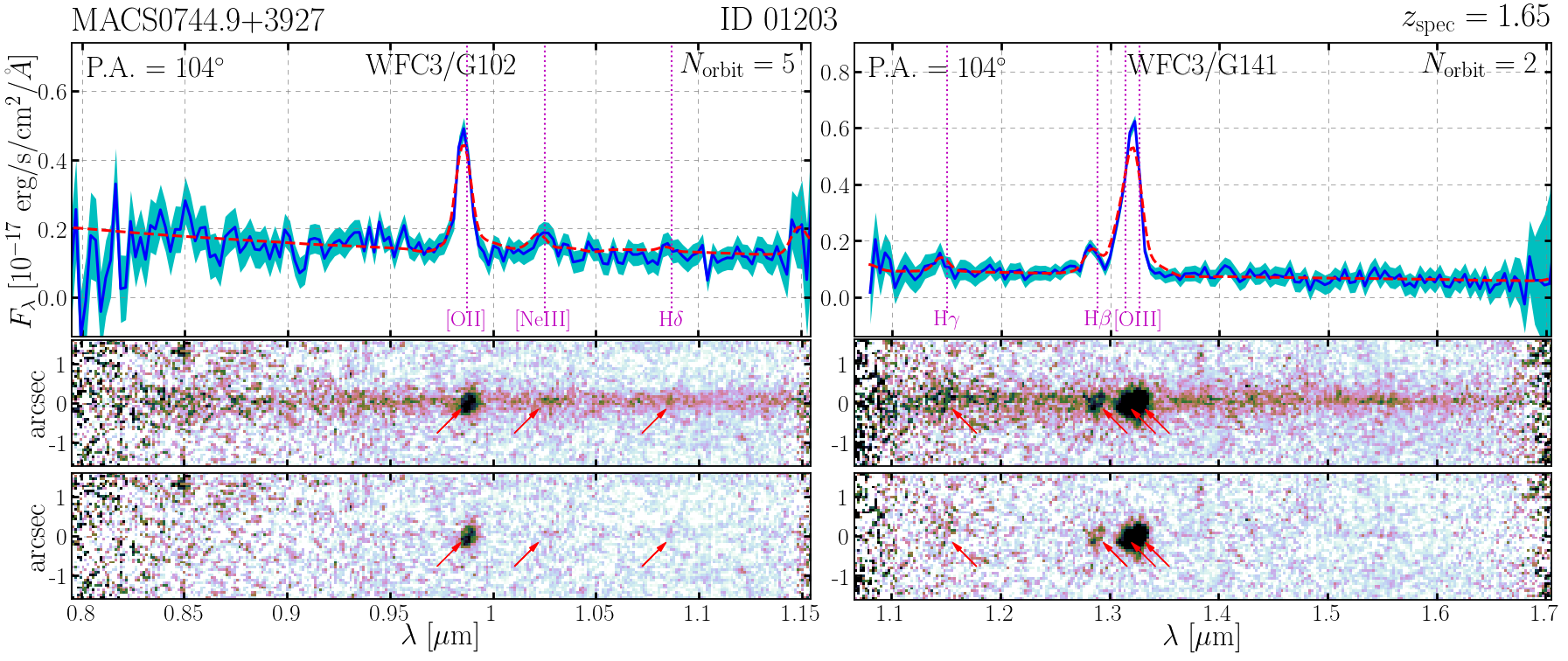}
    \caption{Same as Figure~\ref{fig:3751spec}, except that source ID01203 in the field of \clba is shown.}
    \label{fig:1203spec}
\end{figure*}

As a result, in Figures~\ref{fig:3751spec} and \ref{fig:1203spec}, we show the observed and fitted grism spectra 
for both of our sources at separate P.A.s.
Albeit slightly different in shape and slope, the red curves in 1D spectra comes from the same best-fit spectral 
model for each source and the difference is due to slightly varying morphological broadening.
We also see that the 2D continuum-subtracted spectra are sufficiently clean, preserving only the nebular emission
features that we later combined to get the spatially resolved emission line maps shown in 
Figure~\ref{fig:combELmap}.

\section{Gas kinematics from \keck \osiris observations}\label{sect:kinem}

Kinematics of \HII regions are of interest both for determining whether rotating gaseous disks are present, and the overall scale 
of velocity dispersion which is thought to correlate with the mode of feedback.  We have obtained kinematic maps from \Ha emission 
for source ID01203 as part of a GLASS followup campaign with the \osiris integral field spectrograph 
\citep{Larkin:2006jd} on the \keck I telescope. Full details of the observations and analysis are presented 
elsewhere \citep{Hirtenstein:2018tn}; here we give a brief summary.  Data were obtained on 2016 October 21 using 
the Hn5 filter, 50 milliarcsecond scale, and laser guide star AO, which provides the excellent spatial sampling 
needed to resolve velocity structure on the relevant $\sim$0\farcs1 scales. We obtained 3 exposures of 900 
seconds each.  The \osiris Data Reduction Pipeline was used to process the data, following the standard methods 
adopted in our previous work \citep{2013ApJ...765...48J}. We fit \Ha line emission in each spaxel with a Gaussian 
function, requiring $\geq5\sigma$ significance for acceptable fits. Gas rotation velocity ($V$) and velocity 
dispersion ($\sigma$) are determined from the Gaussian centroid and width. We correct velocity dispersions for 
the effects of instrument resolution and beam smearing by subtracting these terms in quadrature from the best-fit 
Gaussian dispersion. The median beam smearing correction is a 7\% reduction in $\sigma$. 

\begin{figure*}
    \centering
    \includegraphics[height=1.7in]{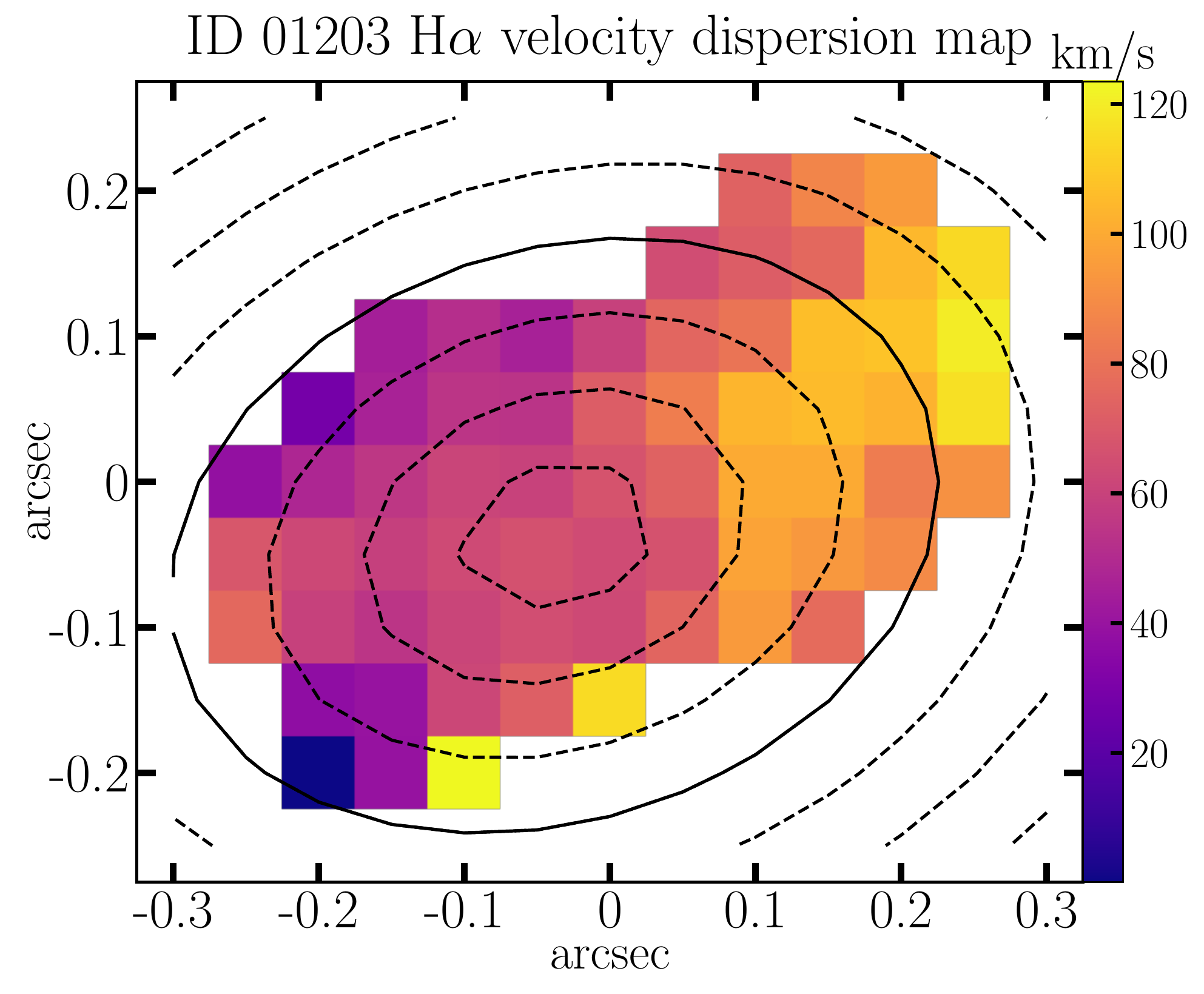}
    \includegraphics[height=1.7in]{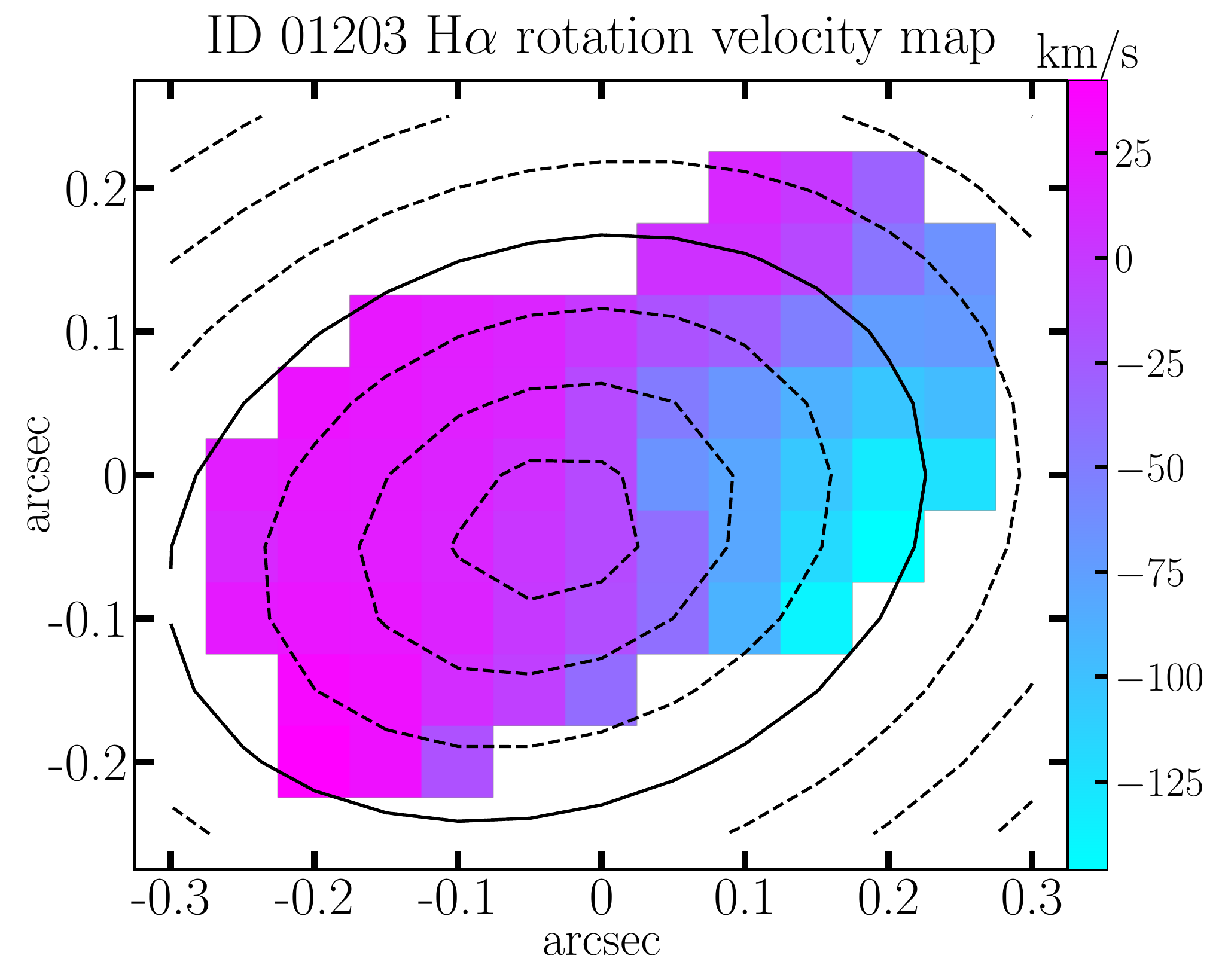}
    \includegraphics[height=1.7in]{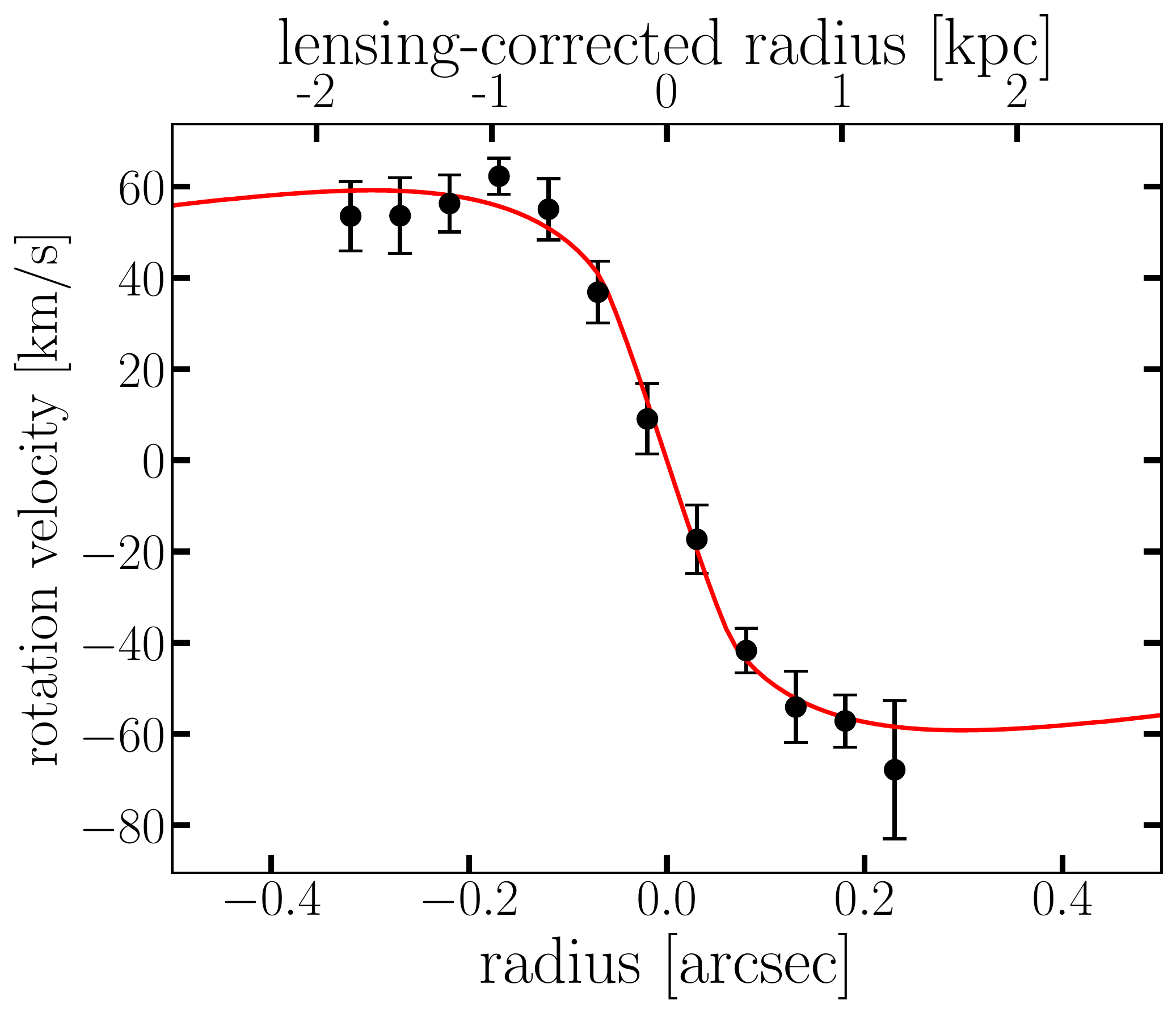}
    \caption{\Ha emission kinematics of ID01203 from our \osiris AO-assisted observations showing the velocity 
    dispersion ({\it
    left}), the rotation velocity ({\it center}), and rotation curve extracted along the major axis ({\it right}).
    The overlaid contours in the left and center panels also represent the source plane de-projected galacto-centric radii, but in
    0.25 kpc interval. To give relative scale with respect to our \hst stamps, the 1 kpc radius contour is still in solid.
    The red curve in the right panel is the best-fit rotation curve for an exponential disk mass distribution, giving a maximum
    line-of-sight velocity $V\sin i = 94 \pm 7$ \kms.
    The data are in good agreement with a thick disk rotation curve despite a high level of turbulence.
    \label{fig:kinem}}
\end{figure*}

Resulting maps of $V$ and $\sigma$ in Figure~\ref{fig:kinem} reveal a sheared velocity field with high local velocity dispersion
($\gtrsim50$ \kms), common among disk galaxies at similar redshift. To quantify the degree of rotational support, we extract a 1D
velocity profile along the kinematic major axis. We fit this with the circular rotation curve of an exponential
disk mass profile. The disk rotation curve is in good agreement with the data, with maximum velocity $V\sin i = 94\pm7$ \kms.
Here $i$ is the disk inclination angle relative to the line-of-sight. The pixel-averaged $\sigma = 73\pm3$ \kms 
such that we derive $V/\sigma = (1.3\pm0.1) / \sin i$ indicating orderly rotation in spite of a high level of ISM 
turbulence.
This $V/\sigma$ ratio is typical of the galaxy population harboring thick disks
at similar mass and redshift \citep{2015ApJ...799..209W,2015arXiv150901279L}.

\section{A comparative study of metallicity gradients derived using different strong line 
calibrations}\label{sect:compare_calib}

Generally speaking, three approaches exist in deriving metallicity information in extragalactic sources from
spectroscopic observations: the metal recombination line method, the electron temperature $T_{\rm e}$
(``direct'') method, and the strong emission line calibration method \citep[see \eg,][for recent 
reviews]{Blanc:2015hl,Bianco:2015hn,Maiolino:2018vq}.

The first method is believed to be the most direct measurement of the chemical abundances, as the emissivities of 
the permitted recombination lines of metal ions are only weakly dependent on $T_{\rm e}$ and the electron density 
$n_{\rm e}$ of the ionized media.
Yet these recombination lines are extremely weak, usually $\sim10^{-3}-10^{-4}$ fainter than the hydrogen 
recombination lines, even for the most abundant metal elements like carbon and oxygen.
So the first method is only feasible not far beyond the local group \citep[see recent results by
\eg,][]{GarciaRojas:2007er,LopezSanchez:2007hv,Bresolin:2009hh,Esteban:2009ew}.

The second method resorts to the CELs of metal ions, given that their emissivities depend strongly on $T_{\rm e}$ 
of the ionized gas, hence this method is usually coined the $T_{\rm e}$ method.
The flux ratios of the temperature-sensitive auroral to nebular CELs (\eg $\OIII\lambda4363$/$\OIII\lambda5008$) 
are often adopted to estimate $T_{\rm e}$.
Yet the auroral lines are intrinsically too faint --- usually at the percent level of the strong nebular lines 
--- to be observable in individual galaxies at high redshifts and high metallicity (corresponding to low $T_{\rm 
e}$).
There are only a handful of such detections to date 
\citep{Christensen:2012bg,2013MNRAS.436.1040S,James:2014gg,Sanders:2016uo,Gburek:2019vb}.

To solve the limitations of the $T_{\rm e}$ method, and entail efficient determinations of chemical abundances in 
faint \HII regions in the high-$z$ universe,
numerous authors rely on the calibrations of flux ratios involving bright nebular CELs (\OIII, \OII, \NII, \etc) 
and Balmer lines (\Ha, \Hb) as a proxy for metallicity (\ie the strong line calibration method).
These strong lines are among the most accessible species of nebular line emission at high redshifts, thus 
rendering this third method the most viable approach to estimate metallicity at extragalactic distances.
The line flux ratios can be calibrated against direct $T_{\rm e}$ measurements in galactic \HII regions and 
nearby individual or stacked galaxy spectra 
\citep{Pilyugin:2000vk,Pilyugin:2001kh,2004MNRAS.348L..59P,Pilyugin:2005kn,Pilyugin:2010bx,Pilyugin:2012ib,2015ApJ...813..126J,Brown:2016bg,Gebhardt:2015ui,
2016MNRAS.457.3678P,Curti:2017fn,Bian:2018km}.
These calibrations are often referred to as empirical calibrations.
Another kind of calibrations is based on the theoretical predictions given by photoionization models 
\citep{Edmunds:1984hd,McCall:1985bl,McGaugh:1991bi,Zaritsky:1994ch,Kewley:2002ep,Kobulnicky:2004kr,Nagao:2011cj,Dopita:2013bj,Strom:2017wl},
and therefore known as theoretical calibrations.
In addition, several authors devise a ``hybrid'' calibration via combining both kinds of line ratio results 
\citep{Denicolo:2002ft,2006A&A...459...85N,2008A&A...488..463M}.

Albeit the strong line calibration method is most widely used,
we must warn the readers of some potential systematics associated with it.
For instance, \citet{Kewley:2008be} show that different calibrations produce offsets in the absolute scale of the 
mass-metallicity relation as large as 0.7 dex \citep[also see \eg,][for similar 
conclusions]{Moustakas:2010ke,LopezSanchez:2012df}.
This primarily originates from the different treatments and assumptions of some secondary parameters that also 
affect the brightness of strong nebular CELs, \eg, the ionization parameter, hardness of ionizing spectrum, the 
nitrogen vs. oxygen abundance ratio, the cosmic evolutions of $T_{\rm e}$ and $n_{\rm e}$, \etc.

In particular, the flux ratio of $\NII\lambda6583$ and \Ha is one of the most frequently calibrated metallicity 
diagnostics.
But it essentially traces the relative abundance of nitrogen, instead of oxygen, with respect to hydrogen.
There have been reported large offset (0.2-0.4 dex) between the loci of star-forming galaxies locally from SDSS 
and at high-$z$ in the BPT diagram \citep{2014ApJ...795..165S,2015ApJ...801...88S,Strom:2016vn}.
This indicates that extending the locally calibrated strong line diagnostics involving nitrogen to high-$z$ can 
be potentially problematic, due to the evolving ionization conditions in the ISM \citep{Strom:2017wl}.

On the other hand, the flux ratios between oxygen CELs and \Hb are also popular diagnostics, particularly
R$_{23}$=$(f_{\OIII\lambda5008}+f_{\OIII\lambda4960}+f_{\OII})/f_{\Hb}$.
However, due to its bi-modality, one often has to pre-determine the locations of metallicity branches before 
applying it \citep{LopezSanchez:2010bi,Guo:2016wk}.
This has presented some challenges especially when the source of interest is believed to be located in the 
transition regions between the two branches.
This dilemma can be effectively alleviated when the ionization and excitation states of the ionized gas are taken 
into account \citep{Kobulnicky:2004kr,Pilyugin:2005kn,Jiang:2019hi}.

\begin{deluxetable}{cccccclcccccc}
    \tablecolumns{13}
    \tablewidth{0pt}
    \tablecaption{Coefficients for the strong line metallicity calibrations used and tested in this work.
    \label{tab:calibr}}
\tablehead{
    \colhead{Group\tablenotemark{a}} &
    \colhead{$\log R$} &
    \colhead{$c_0$} &
    \colhead{$c_1$} &
    \colhead{$c_2$} &
    \colhead{$c_3$} &
    \colhead{$c_4$} &
    \colhead{$\sigma_{\log R}$\tablenotemark{b} [dex]}
}
    \startdata
    \multicolumn{8}{c}{\citet{2008A&A...488..463M} calibrations} \\
    \multirow{2}{*}{O3Hb-O2Hb} & \OIII/\Hb   &  0.1549 & -1.5031 & -0.9790 & -0.0297 & \nodata & 0.1 \\
    & \OII/\Hb    &  0.5603 &  0.0450 & -1.8017 & -1.8434 & -0.6549 & 0.15 \\
    \multirow{2}{*}{R23-O3O2}  &  R$_{23}$    &  0.7462 & -0.7149 & -0.9401 & -0.6154 & -0.2524 & 0.05 \\
    & \OIII/\OII\tablenotemark{c}  & -0.2839 & -1.3881 & -0.3172 & \nodata & \nodata & 0.22 \\
    \hline\noalign{\smallskip}
    \multicolumn{8}{c}{\citet{2015ApJ...813..126J} calibrations} \\
    \multirow{2}{*}{O3Hb-O2Hb} & \OIII/\Hb   &  -88.4378  & 22.7529 & -1.4501 & \nodata & \nodata & 0.1 \\
    & \OII/\Hb    &  -154.9571 & 36.9128 & -2.1921 & \nodata & \nodata & 0.15 \\
    \multirow{2}{*}{R23-O3O2}  &  R$_{23}$    &  -54.1003  & 13.9083 & -0.8782 & \nodata & \nodata & 0.06 \\
    & \OIII/\OII\tablenotemark{c}  &  17.9828   & -2.1552 & \nodata & \nodata & \nodata & 0.23 \\
    \hline\noalign{\smallskip}
    \multicolumn{8}{c}{\citet{Curti:2017fn} calibrations} \\
    \multirow{2}{*}{O3Hb-O2Hb}  & \OIII/\Hb   & -0.277 &  -3.549 & -3.593 & -0.981 & \nodata & 0.09 \\
    & \OII/\Hb    &  0.418 &  -0.961 & -3.505 & -1.949 & \nodata & 0.11 \\
    \multirow{2}{*}{R23-O3O2}  &  R$_{23}$    &  0.527 &  -1.569 & -1.652 & -0.421 & \nodata & 0.06 \\
    & \OIII/\OII\tablenotemark{c}  & -0.691 &  -2.944 & -1.308 & \nodata & \nodata & 0.15
    \enddata
    \tablecomments{The value of the corresponding EL flux ratio is calculated by the polynomial functional form,
    \ie, $\log{R} = \sum_{i} c_i\cdot x^i$, where $x=\oh-8.69$ for the \citet{2008A&A...488..463M} and
    \citet{Curti:2017fn} calibrations, and $x=\oh$ for the \citet{2015ApJ...813..126J} calibrations.  }
    \tablenotetext{a}{To avoid using the same pieces of information repeatedly, we separate the four strong line
    calibrations into two groups: ``O3Hb-O2Hb'' where we combine the flux ratios of \OIII/\Hb and \OII/\Hb,
    and ``R23-O3O2'' where we combine the flux ratios of R$_{23}$ and \OIII/\OII instead.}
    \tablenotetext{b}{The instrinsic scatter of the calibration quantified in the corresponding reference. This
    scatter has been included in our Bayesian analysis (see Eq.~\ref{eq:calibr}).}
    \tablenotetext{c}{In all three compilations of strong line calibrations, \OIII/\OII refers to the flux ratio of
    $\OIII\lambda5008$ and $\OII\lambda\lambda3727,3730$, \ie, a factor of 3/4 smaller than O$_{32}$ frequently
    quoted in the ``blue'' diagnostic diagrams (see \eg, Fig.~\ref{fig:bluediagram}).}
\end{deluxetable}


\begin{figure}
    \centering
    \includegraphics[width=.4\textwidth]{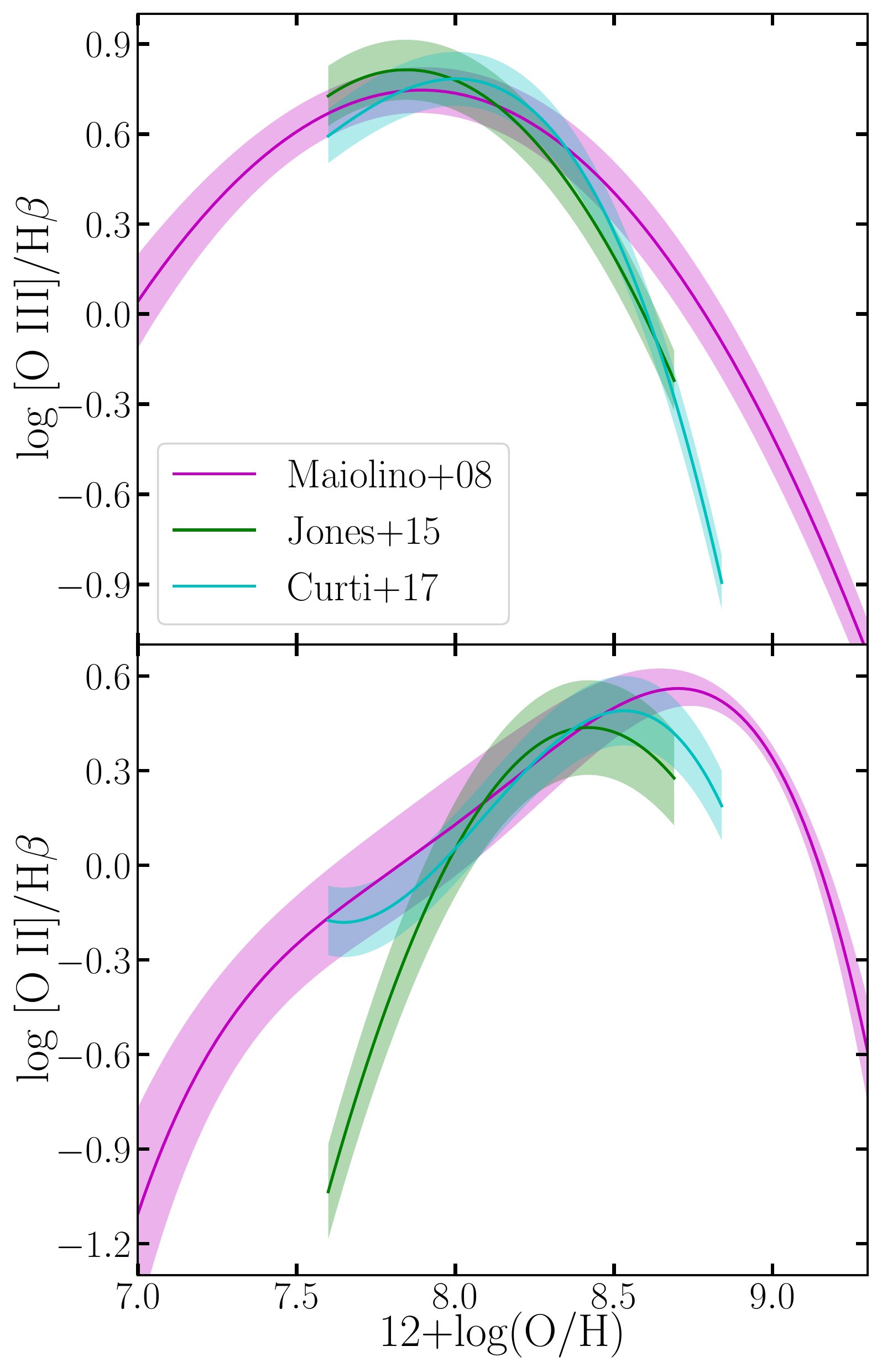}
    \includegraphics[width=.4\textwidth]{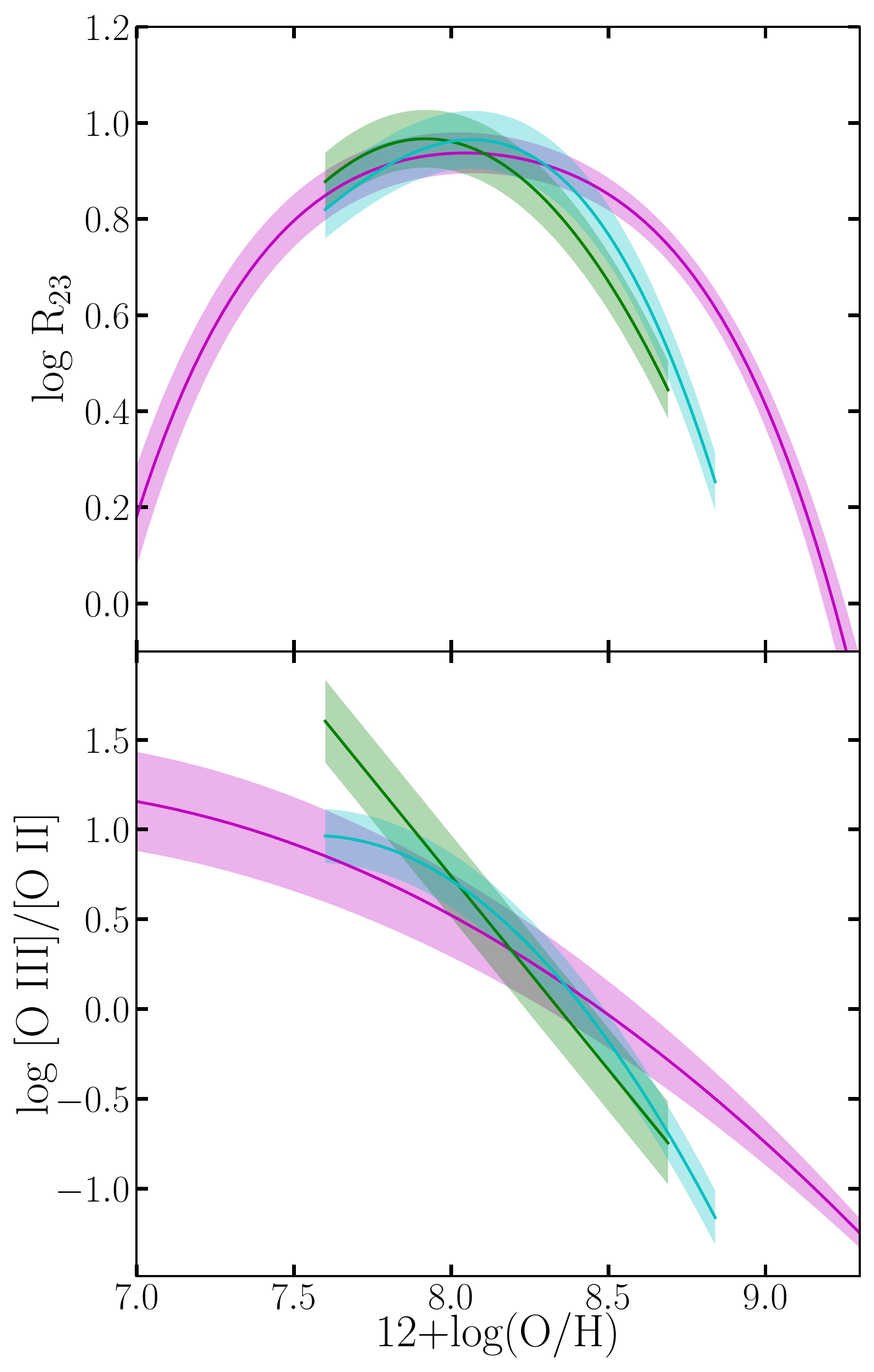}
    \caption{The strong line metallicity calibrations used and tested in this work.
    The group of ``O3Hb-O2Hb'' calibrations is shown on the left, whereas the ``R23-O3O2'' group on the right.
    The corresponding coefficients of these calibrations from all three compilations are presented in 
    Table~\ref{tab:calibr}.}
    \label{fig:compare_calibr}
\end{figure}

While a detailed quantitative comparison of metallicity gradient measurements from different strong line 
calibrations is well beyond the scope of this work,
we want to verify that our results are not significantly altered by the choice of metallicity diagnostics.
For this purpose, we employ two of the most up-to-date pure empirical (\ie calibrated against the actual $T_{\rm 
e}$ metallicities only) strong line calibrations.
\citet{Curti:2017fn} stacked a large number of SDSS galaxy spectra in the redshift range of $0.027<z<0.25$ to
entail high SNR detection of the auroral \OIII$\lambda$4363 lines in the high-metallicity branch, and therefore
homogeneously calibrate the strong line flux ratios over wide metallicity range.
\citet{2015ApJ...813..126J} first extended the strong line calibrations to $z\sim0.8$ using a sample of galaxies 
with \OIII$\lambda$4363 detected in the DEEP2 Galaxy Redshift Survey.
Apart from testing the metallicity diagnostics of $f_{\OIII}/f_{\Hb}$ and $f_{\OII}/f_{\Hb}$ (\ie group 
``O3Hb-O2Hb'') from these independently derived calibration sets,
we also employed an alternative group of calibrations (combining R$_{23}$ and $f_{\OIII}/f_{\OII}$, \ie, group
``R23-O3O2'') in replacement of the ``O3Hb-O2Hb'' ratio diagnostics from the same calibration frameworks.
The coefficients of these calibrations are given in Table~\ref{tab:calibr}, with their behaviors shown in 
Figure~\ref{fig:compare_calibr}.
We note that the results of \citet{2015ApJ...813..126J} demonstrate that oxygen strong line diagnostics based on 
the $T_{\rm e}$ method remain reliable at moderate to high redshifts, thus supporting the conclusions of this 
paper.

\begin{figure*}
    \includegraphics[width=\textwidth]{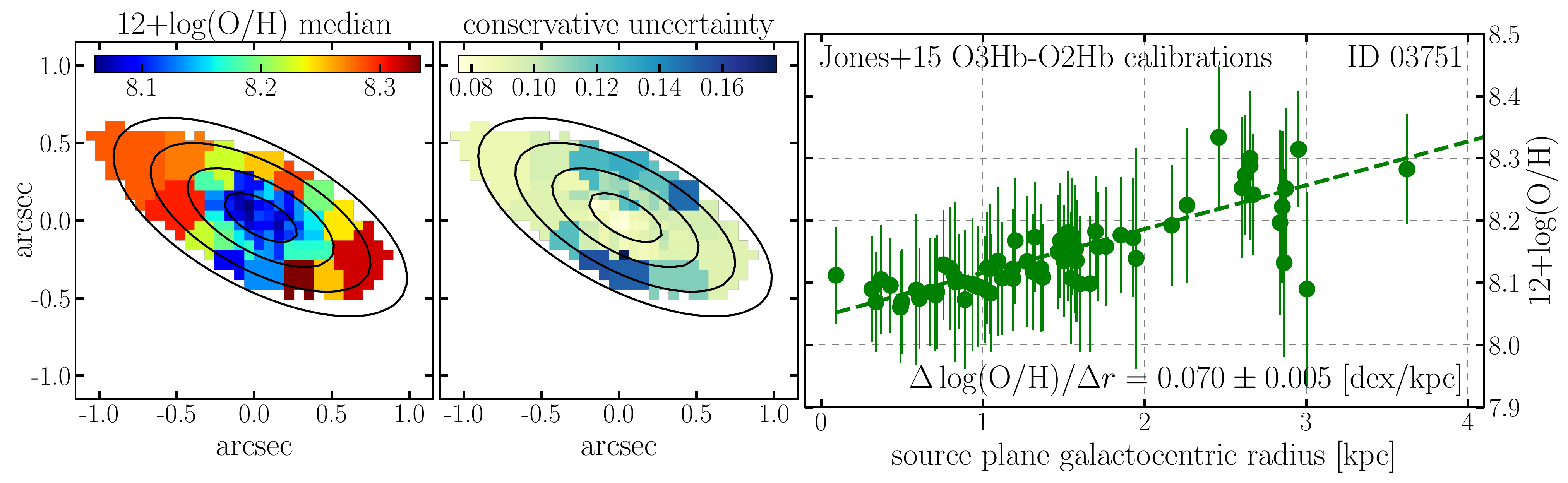}
    \includegraphics[width=\textwidth]{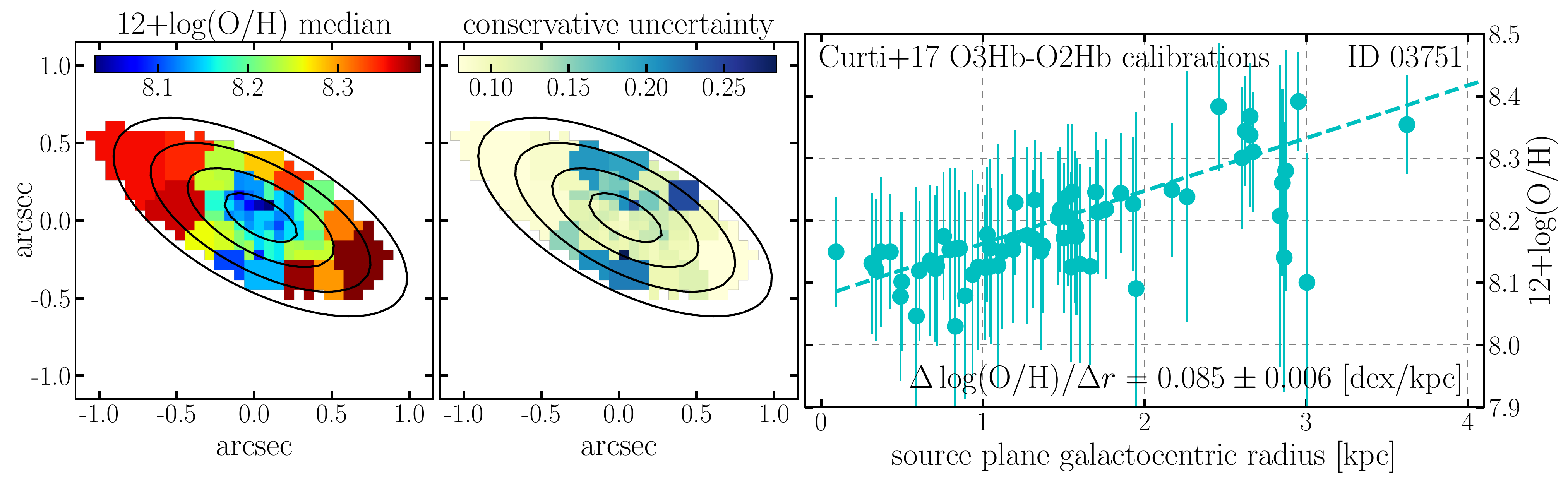}
    \caption{Metallicity maps and radial gradients of source ID03751, determined assuming the 
    \citet{2015ApJ...813..126J} (top) and \citet{Curti:2017fn} (bottom) ``O3Hb-O2Hb'' group of calibrations.
    As in Figure~\ref{fig:oh12grad}, in each row, the left panel shows the median value estimates of metallicity 
    in individual Voronoi bins and the central panel shows the corresponding conservative uncertainties. In the 
    right panel, these metallicity estimates are plotted against their source-plane de-projected galactocentric 
    radii. The dashed line represents the best-fit linear regression with the radial gradient slope denoted on 
    the bottom.
    Albeit different in the exact values of the slope, the gradients are always determined to be strongly 
    inverted, with 0.070$\pm$0.005 (14.0$\sigma$) assuming the \citet{2015ApJ...813..126J} ``O3Hb-O2Hb'' 
    calibrations and 0.085$\pm$0.006 (14.2$\sigma$) for the \citet{Curti:2017fn} ``O3Hb-O2Hb'' calibrations.
    }
    \label{fig:3751calibr}
\end{figure*}

\begin{figure*}
    \includegraphics[width=\textwidth]{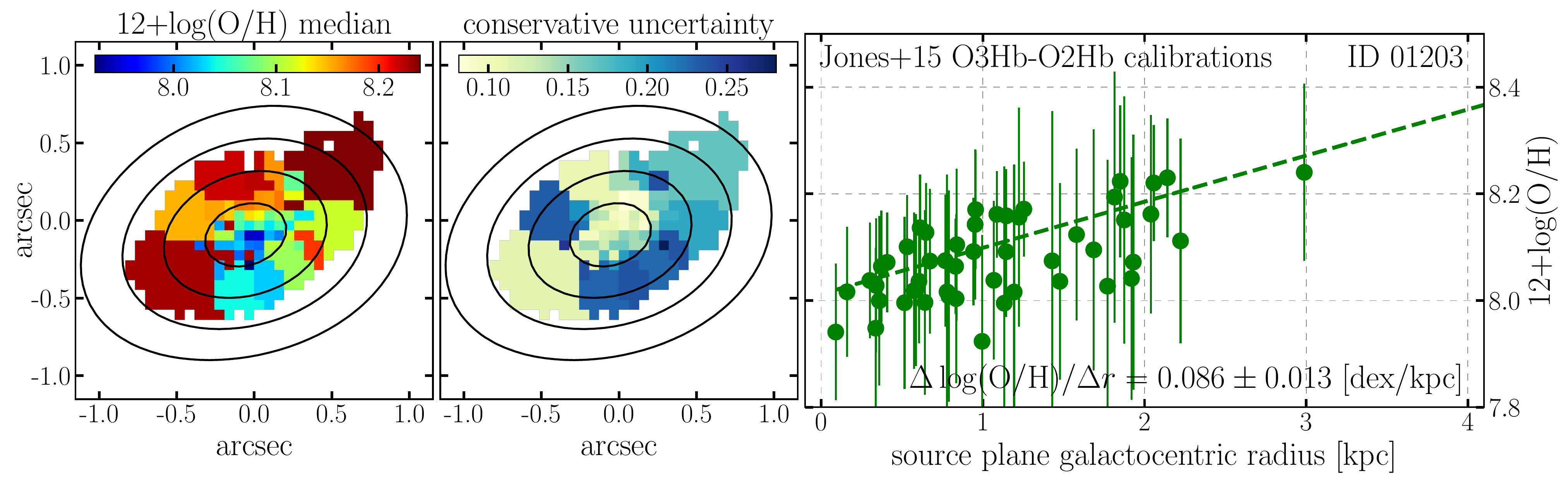}
    \includegraphics[width=\textwidth]{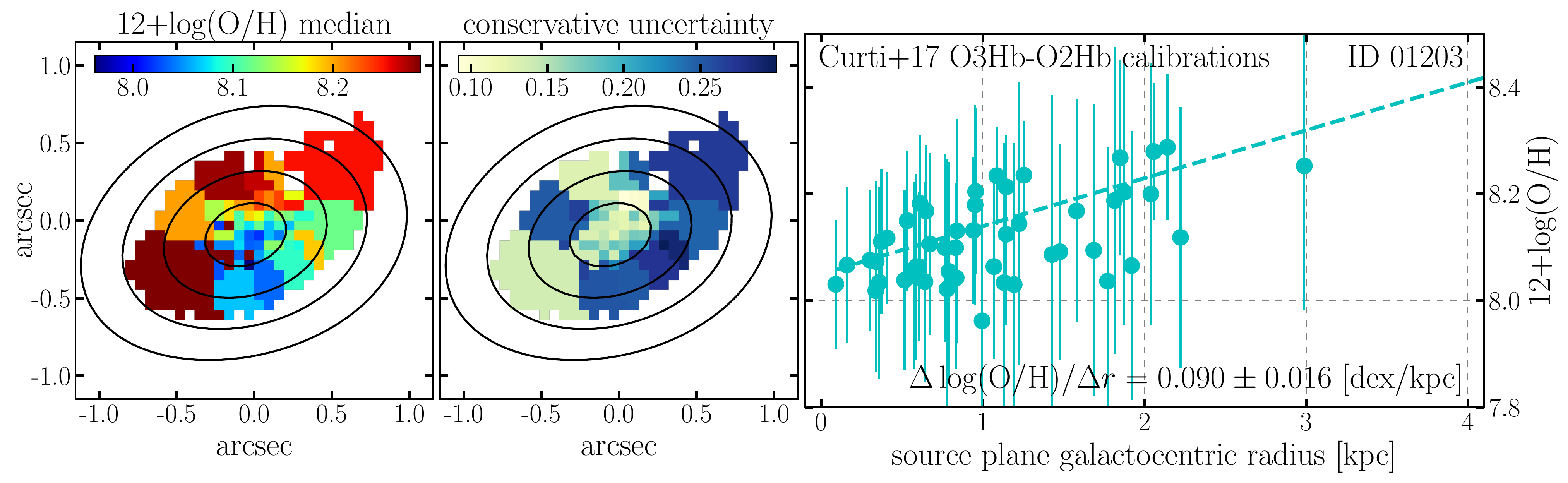}
    \caption{Same as Figure~\ref{fig:3751calibr}, except that source ID01203 is shown.
    Again we see that strongly inverted gradients emerge from our measurements, \ie,
    0.086$\pm$0.013 (6.6$\sigma$) and 0.090$\pm$0.016 (5.6$\sigma$) for the \citet{2015ApJ...813..126J} and 
    \citet{Curti:2017fn} ``O3Hb-O2Hb'' calibrations, respectively.
    }
    \label{fig:1203calibr}
\end{figure*}

Following similar procedures detailed in Section~\ref{sect:rslt}, we measure the radial metallicity gradients of 
our sources using these different groups of strong line diagnostics from independent calibration sets.
The measurement results are summarized in Table~\ref{tab:metalgrad}.
We find that although some systematic differences --- as large as 0.06 dex/kpc --- can be seen in the absolute 
values of the gradient slope, a positive radial gradient is always derived at high statistical significance for 
both of our galaxies, i.e., $\gtrsim12\sigma$ for ID03751 and $\gtrsim5\sigma$ for ID01203, regardless of which 
group of diagnostics from which calibration works is adopted.
Figures~\ref{fig:3751calibr} and \ref{fig:1203calibr} display the metallicity maps and radial gradients 
determined using the ``O3Hb-O2Hb'' group of \citet{2015ApJ...813..126J} and \citet{Curti:2017fn} calibrations for 
galaxies ID03751 and ID01203, respectively.

\begin{deluxetable}{ccccccccccccc}
    \tablecolumns{13}
    \tablewidth{0pt}
    \tablecaption{Inverted radial metallicity gradients measured from two different groups of
    metallicity diagnostics of three independent strong line calibration frameworks.
    \label{tab:metalgrad}}
\tablehead{
    \colhead{Galaxy}  &
    \colhead{Calibration reference} &
    \colhead{Group} &
    \colhead{$\Delta\log({\rm O/H})/\Delta r$}  &
    \colhead{Significance of detection} \\
    & & & [dex/kpc] & [\# of $\sigma$]
}
    \startdata
    \multirow{6}{*}{ID 03751}  
    &\multirow{2}{*}{\citet{2008A&A...488..463M}} &  O3Hb-O2Hb   &  0.122$\pm$0.008\tablenotemark{a}  & 15.2 \\
    &                                             &  R23-O3O2    &  0.087$\pm$0.007  & 12.4 \\
    &\multirow{2}{*}{\citet{2015ApJ...813..126J}} &  O3Hb-O2Hb   &  0.070$\pm$0.005  & 14.0 \\
    &                                             &  R23-O3O2    &  0.060$\pm$0.005  & 12.0 \\
    &\multirow{2}{*}{\citet{Curti:2017fn}}        &  O3Hb-O2Hb   &  0.085$\pm$0.006  & 14.2 \\
    &                                             &  R23-O3O2    &  0.083$\pm$0.006  & 13.8 \\
    \hline\noalign{\smallskip}
    \multirow{6}{*}{ID 01203}  
    & \multirow{2}{*}{\citet{2008A&A...488..463M}} &  O3Hb-O2Hb   &  0.111$\pm$0.017\tablenotemark{a}  & 6.5 \\
    &                                              &  R23-O3O2    &  0.085$\pm$0.011  & 7.7 \\
    & \multirow{2}{*}{\citet{2015ApJ...813..126J}} &  O3Hb-O2Hb   &  0.086$\pm$0.013  & 6.6 \\
    &                                              &  R23-O3O2    &  0.094$\pm$0.010  & 9.4 \\
    & \multirow{2}{*}{\citet{Curti:2017fn}}        &  O3Hb-O2Hb   &  0.090$\pm$0.016  & 5.6 \\
    &                                              &  R23-O3O2    &  0.100$\pm$0.010  & 10.0
    \enddata
    \tablenotetext{a}{The default results quoted in the main body of the paper (see \eg
    Table~\ref{tab:srcprop}).}
\end{deluxetable}


We also perform a simple test using the theoretical calibrations by \citet{Strom:2017wl}, based on the latest 
results of the photoionization models incorporating the hard ionizing radiation from massive star binary systems 
as clearly revealed in the KBSS-MOSFIRE dataset \citep{2014ApJ...795..165S,Strom:2016vn}.
Since \citet{Strom:2017wl} did not prescribe the ``O3Hb-O2Hb'' calibrations as a function of \oh explicitly, we 
infer the metallicities in the radial bins of our sources using the metallicity indicator ``$X_{\rm O32-R23}$'' 
(see their Eq.~8).
Using the dust-corrected line flux ratios shown in Figure~\ref{fig:bluediagram}, we obtain the metallicity 
estimates of 8.39(8.34) and 8.43(8.46) in the most inner and outer radial annuli of source ID03751(ID01203), 
respectively, again confirming the inverted nature of their radial gradient slopes\footnote{We caution the 
readers that the scatter of the metallicity indicator ``$X_{\rm O32-R23}$'' increases dramatically in the range 
of \oh$\lesssim8.6$.}.

For the sake of consistency with our previous and ongoing metallicity gradient analyses \citep[][and Wang \etal 
in prep]{2015AJ....149..107J,Wang:2016um}, we decide to keep the gradient measurements derived with the 
\citet{2008A&A...488..463M} O3Hb-O2Hb calibrations as our default results.

\end{appendix}

\bibliographystyle{apj}
\bibliography{bibtexlib}

\end{document}